\newif\ifanon
\anonfalse
\documentclass[11pt]{article}
\usepackage[letterpaper,margin=1in]{geometry}
\usepackage{amsmath,amsthm,amssymb,mathtools,thmtools,cite,enumitem,framed,graphicx,subfigure,xcolor,tikz}
\usetikzlibrary{backgrounds}
\definecolor{myblue}{HTML}{0088cc}
\definecolor{myorange}{HTML}{f26924}
\definecolor{mygreen}{HTML}{3ec636}
\usepackage[colorlinks,citecolor=myblue,urlcolor=myblue,linkcolor=black]{hyperref}
\usepackage[capitalize]{cleveref}
\DeclareUrlCommand{\path}{}
\newtheorem{theorem}{Theorem}[section]
\newtheorem{lemma}[theorem]{Lemma}
\newtheorem{corollary}[theorem]{Corollary}
\newtheorem{observation}[theorem]{Observation}
\crefname{observation}{Observation}{Observations}

\theoremstyle{definition}

\newtheorem{question}[theorem]{Question}
\newtheorem{definition}[theorem]{Definition}

\DeclareMathOperator{\re}{\mathcal R}
\DeclareMathOperator{\rere}{\overline{\mathcal R}}
\DeclareMathOperator{\rest}{\mathcal R ^{\star}}

\newcommand{\fA}{\mathcal{A}}

\newcommand{\fC}{\mathcal{C}}

\newcommand{\fE}{\mathcal{E}}
\newcommand{\fF}{\mathcal{F}}
\newcommand{\fG}{\mathcal{G}}
\newcommand{\fI}{\mathcal{I}}
\newcommand{\fJ}{\mathcal{J}}

\newcommand{\fL}{\mathcal{L}}
\newcommand{\fP}{\mathcal{P}}
\newcommand{\fS}{\mathbb{S}}
\newcommand{\fQ}{\mathcal{Q}}

\newcommand{\A}{\mathsf{A}}
\newcommand{\B}{\mathsf{B}}
\newcommand{\C}{\mathsf{C}}
\newcommand{\D}{\mathsf{D}}

\newcommand{\lI}{\mathsf{I}}
\renewcommand{\L}{\mathsf{L}}
\renewcommand{\S}{\mathsf{S}}

\newcommand{\Z}{\mathsf{Z}}
\newcommand{\lO}{\mathsf{O}}
\newcommand{\lR}{\mathsf{R}}
\newcommand{\lC}{\mathsf{C}}
\newcommand{\lX}{\mathsf{X}}
\newcommand{\lo}{\mathsf{o}}
\newcommand{\lr}{\mathsf{r}}
\newcommand{\lx}{\mathsf{x}}
\newcommand{\lc}{\mathsf{c}}
\newcommand{\lnine}[2]{(#1,#2)}

\newcommand{\one}{\Sigma_1}
\newcommand{\two}{\Sigma_2}

\newcommand{\mybox}[1]{\mspace{2mu}{\setlength{\fboxsep}{1.5pt}\color{darkgray}\boxed{\color{black} #1}}\mspace{2mu}}

\newcommand{\nodeconst}{\ensuremath{\mathcal{N}}}
\newcommand{\edgeconst}{\ensuremath{\mathcal{E}}}

\newcommand{\spi}{\Sigma_{\Pi}}
\newcommand{\npi}{\nodeconst_{\Pi}}
\newcommand{\epi}{\edgeconst_{\Pi}}
\newcommand{\repi}{\re(\Pi)}
\newcommand{\srepi}{\Sigma_{\re(\Pi)}}
\newcommand{\nrepi}{\nodeconst_{\re(\Pi)}}
\newcommand{\erepi}{\edgeconst_{\re(\Pi)}}
\newcommand{\rerepi}{\rere(\Pi)}
\newcommand{\srerepi}{\Sigma_{\rere(\Pi)}}
\newcommand{\nrerepi}{\nodeconst_{\rere(\Pi)}}
\newcommand{\ererepi}{\edgeconst_{\rere(\Pi)}}
\newcommand{\sig}[1]{\Sigma_{#1}}
\newcommand{\nod}[1]{\nodeconst_{#1}}
\newcommand{\edg}[1]{\edgeconst_{#1}}

\newcommand{\zrr}{\xrightarrow{\scriptscriptstyle \smash 0}}
\newcommand{\nzrr}{\overset{\scriptscriptstyle 0}{\nrightarrow}}

\newcommand{\s}{~}
\newcommand{\ninecol}{\Pi^{\Delta}}
\newcommand{\orcx}{\Pi^\mathsf{orcx}}

\newcommand{\namedref}[2]{\hyperref[#2]{#1~\ref*{#2}}}

\newcommand{\enumerateref}[2]{part \ref{#2} of \Cref{#1}}

\newenvironment{myabstract}%
{\list{}{\listparindent 1.5em
		\itemindent    \listparindent
		\leftmargin    1cm
		\rightmargin   1cm
		\parsep        0pt}%
	\item\relax}%
{\endlist}

\newenvironment{mycover}%
{\list{}{\listparindent 0pt
		\itemindent    \listparindent
		\leftmargin    1cm
		\rightmargin   1cm
		\parsep        0pt}%
	\raggedright
	\item\relax}%
{\endlist}

\newcommand{\myaff}[1]{\,$\cdot$\, {\small #1}\par\smallskip}
\newcommand{\institution}[1]{#1, }
\newcommand{\country}[1]{#1}

\begin{document}

\begin{mycover}
	{\huge\bfseries On the Universality of Round Elimination Fixed Points \par}

\bigskip
\bigskip

\ifanon
\textbf{Anonymous authors}
\else
	\textbf{Alkida Balliu} \myaff{\institution{Gran Sasso Science Institute} \country{Italy}}

	\textbf{Sebastian Brandt} \myaff{\institution{CISPA Helmholtz Center for Information Security} \country{Germany}}

	\textbf{Ole Gabsdil} \myaff{\institution{CISPA Helmholtz Center for Information Security} \country{Germany}}

	\textbf{Dennis Olivetti} \myaff{\institution{Gran Sasso Science Institute} \country{Italy}}

	\textbf{Jukka Suomela} \myaff{\institution{Aalto University} \country{Finland}}
\fi

\end{mycover}
\begin{myabstract}
\noindent\textbf{Abstract.} Recent work on distributed graph algorithms [e.g.\ STOC 2022, ITCS 2022, PODC 2020] has drawn attention to the following open question: \emph{are round elimination fixed points a universal technique for proving lower bounds?} That is, given a locally checkable problem $\Pi$ that requires at least $\Omega(\log n)$ rounds in the deterministic LOCAL model, can we always find a relaxation $\Pi'$ of $\Pi$ that is a \emph{nontrivial fixed point} for the round elimination technique [see STOC 2016, PODC 2019]? If yes, then a key part of distributed computational complexity would be also \emph{decidable}.

The key obstacle so far has been a certain family of homomorphism problems [ITCS 2022], which require $\Omega(\log n)$ rounds, but the only known proof is based on Marks' technique [J.AMS 2016].

We develop a new technique for constructing round elimination lower bounds systematically. Using so-called \emph{tripotent inputs} we show that the aforementioned homomorphism problems indeed admit a lower bound proof that is based on round elimination fixed points. Hence we eliminate the only known obstacle for the universality of round elimination.

Yet we also present a new obstacle: we show that there are some problems \emph{with inputs} that require $\Omega(\log n)$ rounds, yet there is no proof that is based on relaxations to nontrivial round elimination fixed points. Hence round elimination cannot be a universal technique for problems with inputs (but it might be universal for problems without inputs).

We also prove the first fully general lower bound theorem that is applicable to any problem, with or without inputs, that is a fixed point in round elimination. Prior results of this form were only able to handle certain very restricted inputs.
\end{myabstract}

\thispagestyle{empty}
\setcounter{page}{0}
\newpage

\section{Introduction}\label{sec:intro}

Can we always systematically distinguish between easy and hard problems in distributed graph algorithms? We study this question in the context of locally checkable labeling problems (LCLs) \cite{naor-stockmeyer-1995-what-can-be-computed-locally} in trees; these are constraint satisfaction problems defined on bounded-degree graphs that can be conveniently specified by listing a finite set of valid labeled neighborhoods. Numerous problems familiar from the theory of distributed graph algorithms such as vertex and edge coloring, maximal independent set, maximal matching, sinkless orientation, and many other splitting and orientation problems are examples of LCLs in bounded-degree graphs.

\subsection{Classification of LCL problems}

As a result of a long research program \cite{
	cole-vishkin-1986-deterministic-coin-tossing-with,
	naor-1991-a-lower-bound-on-probabilistic-algorithms-for,
	linial-1992-locality-in-distributed-graph-algorithms,
	naor-stockmeyer-1995-what-can-be-computed-locally,
	brandt-fischer-etal-2016-a-lower-bound-for-the,
	fischer-ghaffari-2017-sublogarithmic-distributed,
	ghaffari-harris-kuhn-2018-on-derandomizing-local,
	balliu-hirvonen-etal-2018-new-classes-of-distributed,
	chang-pettie-2019-a-time-hierarchy-theorem-for-the,
	chang-kopelowitz-pettie-2019-an-exponential-separation,
	rozhon-ghaffari-2020-polylogarithmic-time-deterministic,
	balliu-brandt-etal-2020-how-much-does-randomness-help,
	balliu-brandt-etal-2021-almost-global-problems-in-the,
	balliu-censor-hillel-etal-2021-locally-checkable
}, the landscape of LCL problems is now well understood \cite{suomela-2020-landscape-of-locality-invited-talk}, especially in the case of trees \cite{chang-2020-the-complexity-landscape-of-distributed,balliu-brandt-etal-2021-almost-global-problems-in-the,grunau-rozhon-brandt-2022-the-landscape-of-distributed}. For our purposes the most interesting part is the following dichotomy; any LCL problem in trees falls in one of these complexity classes:
\begin{enumerate}
	\item \textbf{Easy problems:} the round complexity is $O(\log^* n)$ in the deterministic and randomized LOCAL models \cite{linial-1992-locality-in-distributed-graph-algorithms,peleg-2000-distributed-computing-a-locality-sensitive} and $O(1)$ in the deterministic and randomized SLOCAL models \cite{ghaffari-kuhn-maus-2017-on-the-complexity-of-local}. A canonical example of such a problem is \emph{$3$-coloring of paths}, for which all of these complexities are tight \cite{cole-vishkin-1986-deterministic-coin-tossing-with,goldberg-plotkin-shannon-1988-parallel-symmetry,linial-1992-locality-in-distributed-graph-algorithms}.
	\item \textbf{Intermediate (or harder) problems:} the round complexity is $\Omega(\log n)$ in deterministic LOCAL, $\Omega(\log\log n)$ in randomized LOCAL, $\Omega(\log \log n)$ in deterministic SLOCAL, and $\Omega(\log \log \log n)$ in randomized SLOCAL. A canonical example of such a problem is \emph{sinkless orientation}, for which all of these complexities are tight \cite{brandt-fischer-etal-2016-a-lower-bound-for-the,chang-kopelowitz-pettie-2019-an-exponential-separation,ghaffari-harris-kuhn-2018-on-derandomizing-local,ghaffari-su-2017-distributed-degree-splitting-edge,balliu-korhonen-etal-2023-sinkless-orientation-made}.
\end{enumerate}
But given some unknown problem $\Pi$, how can we tell whether it is easy or (at least) intermediate?

\paragraph{How to prove that a given problem is easy?}

One half of this is known; being easy is semi-decidable. More precisely, for any easy problem $\Pi$ there exists a natural number $k$ such that one can map a distance-$k$ coloring to a valid solution of~$\Pi$ \cite{chang-kopelowitz-pettie-2019-an-exponential-separation,brandt-hirvonen-etal-2017-lcl-problems-on-grids}. For any fixed $k$, there are only finitely many candidate functions to check, so at least in principle one could systematically check $k = 1, 2, \dotsc$ and stop once a suitable function is found. However, the process will never terminate for harder problems.

\paragraph{How to prove that a given problem is intermediate-hard?}

What is missing is a systematic way to detect if a problem is at least intermediate-hard. There are many different proof techniques used to show that a problem is intermediate or harder, but especially when more elementary techniques have failed, by far the most successful proof technique has been round elimination, which we will discuss in more depth next.

\pagebreak
\subsection{Round elimination}

Round elimination  \cite{brandt-2019-an-automatic-speedup-theorem-for,brandt-fischer-etal-2016-a-lower-bound-for-the} is an explicit function $\fQ$ that maps an LCL problem $\Pi$ to another problem $\fQ(\Pi)$. The key property is this:
\begin{quote}
If $\Pi$ is an easy but nontrivial problem, then $\fQ(\Pi)$ is exactly one round easier to solve than $\Pi$.
\end{quote}
So if $\fQ(\Pi)$ is \emph{not} one round easier than $\Pi$, and $\Pi$ is \emph{not} trivial (it requires at least one round to solve), the only possible explanation was that $\Pi$ was at least intermediate-hard.

In general, it is hard to compare $\fQ(\Pi)$ and $\Pi$. However, if we are lucky enough to find a \emph{fixed point}, i.e., $\fQ(\Pi) = \Pi$, then clearly $\fQ(\Pi)$ cannot be one round easier than $\Pi$, and we can conclude that $\Pi$ must be intermediate-hard (or trivial).

In this work we slightly extend (and abuse) the notion of fixed points to also cover cases where $\Pi$ is a \emph{relaxation} of $\fQ(\Pi)$. We write $\fQ(\Pi) \zrr \Pi$ to denote that a solution of $\fQ(\Pi)$ can be turned in zero rounds into a solution of $\Pi$; clearly in this case $\fQ(\Pi)$ cannot be one round easier than $\Pi$, and hence $\Pi$ has to be at least intermediate-hard.

Fixed points are rare, but for many intermediate problems $\Pi$ the following proof idea does the trick: find a relaxation $\Pi \zrr \Pi'$, show that $\Pi'$ is nontrivial, show that $\fQ(\Pi') \zrr \Pi'$, conclude that $\Pi'$ has to be at least intermediate-hard, and hence is the original problem $\Pi$.

\subsection{Is round elimination universal?}\label{ssec:intro-is-re-univ}

If this scheme worked for \emph{every} intermediate problem, then we would have a procedure for proving that a problem is at least intermediate-hard: simply start to enumerate the countable set of possible relaxations $\Pi'$ of $\Pi$, and apply the $\fQ$ function to each of them, and stop if we find a nontrivial fixed point. If in parallel with this we also apply the procedure that searches for a proof that $\Pi$ is easy, one of these procedures would always terminate. This would mean that distinguishing easy vs.\ intermediate-hard problems would be decidable, resolving a major open question of this field, see e.g.\ \cite{balliu-brandt-etal-2022-efficient-classification-of}.

Inspired by this, the following open question was stated e.g.\ in \cite[Open Problem 1]{balliu-brandt-etal-2022-distributed-delta-coloring} and
\cite[Open Problem 6.3]{brandt-olivetti-2020-truly-tight-in-delta-bounds-for}:
\begin{question}\label{q:universality}
	Do all intermediate problems relax to a nontrivial round elimination fixed point?
\end{question}

So far there is only one known family of problems that is known to be intermediate-hard, yet it is not known if it admits a proof through a relaxation to a round elimination fixed point, namely the graph homomorphism problems introduced by \cite{brandt-chang-etal-2022-local-problems-on-trees-from-the}. The simplest example is the task that we will call \emph{restricted $9$-coloring}; the task is to find a homomorphism from the input graph $G$ to the following $9$-node graph $H$:
\begin{equation}\label{eq:graph}
	\raisebox{-0.5\height}{\includegraphics[page=1]{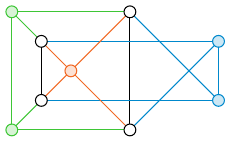}}
\end{equation}
This problem is known to be intermediate-hard \cite{brandt-chang-etal-2022-local-problems-on-trees-from-the}, and the proof uses Marks' technique \cite{marks-2016-a-determinacy-approach-to-borel}, familiar from the context of measurable combinatorics.

If the family of homomorphism problems is indeed outside the scope of round elimination fixed points, then it would mean that round elimination fixed points are not a universal technique. Furthermore, it would mean that round elimination and Marks' technique are orthogonal lower bound techniques, and one does not subsume the other.
In \cite{brandt-chang-etal-2022-local-problems-on-trees-from-the} obtaining the same lower bounds for the homomorphism problems that Marks' technique achieves is stated as an ``exciting open problem.''

\subsection{Our contributions}

\paragraph{Contribution 1: homomorphism problems admit nontrivial round elimination fixed points (\cref{sec:homom}).}
We show that round elimination fixed points can be indeed used to show that the homomorphism problems studied in \cite{brandt-chang-etal-2022-local-problems-on-trees-from-the} are intermediate-hard. We show that for each of these problems $\Pi$, it is possible to find a relaxation $\Pi \zrr \Pi'$ such that $\fQ(\Pi') \zrr \Pi'$.

In particular, we remove all known obstacles to \cref{q:universality}; after this work, there is \emph{no candidate problem} that would serve as a counterexample for the conjecture that round elimination fixed points are indeed a universal technique for proving that a problem is intermediate-hard. Furthermore, these were the only known examples in which Marks' technique was previously necessary to prove lower bounds in the LOCAL model; our work is giving evidence that it might indeed be the case that Marks' technique can be always replaced by round elimination in this context.

\paragraph{Contribution 2: problems with inputs do not admit nontrivial round elimination fixed points (\cref{sec:sso-with-so}).}
However, everything that we discussed above holds only for problems without inputs. We show that \emph{problems with inputs}, i.e., problems where the input graph comes together with an arbitrary solution for some fixed \emph{input LCL}, can be much more challenging. In particular, we prove that \emph{sinkless and sourceless orientation given a sinkless orientation} (SSO-SO) is a problem that is intermediate-hard, yet it admits no relaxation to a round elimination fixed point.

Hence the answer to \cref{q:universality} in full generality is negative, but it could be still true for problems without inputs.

\paragraph{Contribution 3: auxiliary inputs can be eliminated (\cref{sec:tripotent}).}
We will next zoom into the role of inputs. One trick that was used already in the very first applications of round elimination \cite{brandt-fischer-etal-2016-a-lower-bound-for-the} may at first look counterintuitive: we want to show that $\Pi$ is intermediate-hard, and we make our life seemingly harder and try to argue that $\Pi$ is intermediate-hard \emph{even if we are given some friendly auxiliary input $\fI$}. For example, to show that $3$-vertex-coloring is intermediate-hard, one can show that $3$-vertex-coloring is intermediate-hard even if we are given a $3$-edge-coloring as input. Why this makes sense is that it may enable one to find a useful relaxation: a $3$-vertex-coloring in $3$-edge-colored graphs can be used to find a sinkless orientation, and sinkless orientations are a nontrivial round elimination fixed point that remains nontrivial even if we are given a $3$-edge-coloring.

We show that such a roundabout approach is never needed. If we can find a relaxation to a nontrivial fixed point with the help of auxiliary inputs, we can also eliminate such inputs. We introduce the concept of \emph{tripotent inputs} that allow us to turn fixed points with auxiliary inputs in a mechanical way into fixed points without inputs. We demonstrate the effectiveness of this tool to construct a pure input-free round elimination fixed point for the homomorphism problems (Contribution~1).

Tripotent inputs also open up a genuinely new way for discovering lower bounds: if we can \emph{guess} a fixed point relaxation (say, using one of the well-known fixed points such as sinkless orientation), we can also work backwards to construct a suitable input, and then we can finally verify whether our guess was indeed correct. Hence instead of guessing a suitable input and finding a fixed point (as has been done in the literature so far), we can guess a fixed point and calculate the input.

\paragraph{Contribution 4: round elimination works with inputs (\cref{sec:lifting}).}
Finally, we turn our attention to problems where inputs can play a crucial role; the input is not merely helpful information that the algorithm may choose to use to make the problem easier to solve (e.g.\ SSO-SO), but the input may restrict what are possible valid outputs (e.g.\ list coloring).

It is known from prior work that the existence of a nontrivial fixed point relaxation for a problem $\Pi$ implies that $\Pi$ is intermediate-hard \cite{balliu-brandt-etal-2022-distributed-delta-coloring}. However, this implication is not generally known to be true in the case of LCL problems with inputs (such as list coloring problems). We prove for the first time that we can use round elimination fixed points to derive lower bounds also for problems with general inputs.

\paragraph{Discussion.}
In summary, our work provides new evidence pointing in the following direction:
\begin{enumerate}
	\item For intermediate problems without inputs, round elimination fixed points might be a universal technique, and if this is indeed true, classification of input-free problems to easy vs.\ intermediate would be decidable (at least in principle, if not in practice).
	\item For intermediate problems with inputs, round elimination fixed points are not a universal technique, and automatic classification of such problems seems to require the development of new lower-bound proof techniques.
	\item The distinction between problems with inputs vs.\ problems without inputs is clear-cut: auxiliary inputs that were only introduced as a proof technique in order to find appropriate fixed point relaxations can be always systematically eliminated.
\end{enumerate}

\subsection{Key new ideas and proof techniques}

\paragraph{Tripotent inputs (\cref{sec:tripotent}).}

The key novel tool that we introduce in this work is what we call \emph{tripotent inputs}. This is a function $\tau_{P}$ parameterized by an LCL problem $P$ and it maps any given LCL problem $\Pi$ to another problem $\tau_P(\Pi)$. For us the most interesting case is if $P = \fF$ for some nontrivial round elimination fixed point $\fF$.

The intuition is that $\tau_{\fF}(\Pi)$ would be a good choice of input if we wanted to show that $\Pi$ with some input can be relaxed to $\fF$, a nontrivial fixed point. But the remarkable property of $\tau_{\fF}$ is that as long as $\fF$ is nontrivial given $\tau_{\fF}(\Pi)$, then $\Pi' = \tau_{\fF}(\tau_{\fF}(\Pi))$ will be a nontrivial fixed point relaxation of $\Pi$. Furthermore, as long as there is \emph{some} input $\fI$ such that $\Pi$ with $\fI$ can be relaxed to $\fF$ and $\fF$ remains nontrivial given $\fI$, then $\fF$ is also nontrivial given $\tau_{\fF}(\Pi)$.

Put together, tripotent inputs open up \emph{two} new possibilities for proving hardness with the help of round elimination fixed points:
\begin{enumerate}
	\item We can freely cheat with inputs (e.g.\ assume $3$-edge-coloring) to discover that $\Pi$ with some input $\fI$ can be relaxed to some fixed point $\fF$ that remains nontrivial even given $\fI$. Then we know that $\fF$ is also nontrivial given $\tau_{\fF}(\Pi)$. Then we can in a mechanical way compute $\Pi' = \tau_{\fF}(\tau_{\fF}(\Pi))$, and $\Pi'$ is guaranteed to be a nontrivial input-free fixed-point relaxation of $\Pi$, that is, $\Pi \zrr \Pi'$ and $\fQ(\Pi') \zrr \Pi'$ and $\Pi'$ is nontrivial. We can fully eliminate the input!
	\item We can also \emph{guess} some fixed point candidate $\fF$, compute $\tau_{\fF}(\Pi)$ in a mechanical way, see if $\fF$ remains hard given $\tau_{\fF}(\Pi)$, and if so, we can find a nontrivial input-free fixed point relaxation $\Pi' = \tau_{\fF}(\tau_{\fF}(\Pi))$ of $\Pi$.
\end{enumerate}
We will make use of the first approach to argue that e.g.\ restricted $9$-coloring problem and other homomorphism problems indeed admit an input-free round elimination fixed point. Here it is important to note that while $\Pi$, $\fI$, and $\fF$ can be relatively simple problems with concise descriptions, the description of $\Pi'$ can be much longer, and in particular very hard to guess or systematically discover without the use of our new tool.

It turns out that tripotent inputs satisfy numerous convenient mathematical properties. Among others, as the name suggests, $\tau_{\fF}$ is always a tripotent function in the sense that $\tau_{\fF}(\Pi)$ and $\tau_{\fF}(\tau_{\fF}(\tau_{\fF}(\Pi)))$ are equivalent problems. We expect that further study of this function will shed much more light on the round elimination technique.

\paragraph{Non-existence of round elimination fixed points (\cref{sec:sso-with-so}).}

In \cref{sec:sso-with-so} we show that the SSO-SO problem (find a sinkless and sourceless orientation, given a sinkless orientation as input) does not admit any nontrivial fixed point relaxation, even though it is intermediate-hard.

To prove this result, let $\Pi$ be the SSO problem. First, we analyze the structure of problems $\fQ^k(\Pi)$ that are obtained by applying round elimination $k$ times---we essentially obtain a closed-form characterization of all such problems.

Second, we consider a hypothetical fixed-point relaxation $\hat{\Pi}$ of $\Pi$. Since $\hat{\Pi}$ is a relaxation of $\Pi$, also $\fQ^k(\hat{\Pi}) = \hat{\Pi}$ is a relaxation of $\fQ^k(\Pi)$. We choose a $k$ sufficiently large in comparison with the number of labels in $\hat{\Pi}$, and use the structural characterization of $\fQ^k(\Pi)$ to identify a subproblem of $\hat{\Pi}$ that is easy to solve.

Building on this idea, we then show that given a sinkless orientation, we can find a feasible solution to $\fQ^k(\hat{\Pi}) = \hat{\Pi}$; hence $\hat{\Pi}$ cannot be a relaxation of SSO that is a round elimination fixed point that remains nontrivial given a sinkless orientation.

Finally, we can then use the machinery developed for tripotent inputs to show that there is no such relaxation of SSO-SO, either.

\paragraph{ORCX problem (\cref{sec:homom}).}

To show that homomorphism problems from \cite{brandt-chang-etal-2022-local-problems-on-trees-from-the} admit a fixed-point relaxation, we introduce a family of graph problem that we call the \emph{ORCX problem}.

To better understand the key idea, let us consider the special case of $3$-regular graphs and the restricted $9$-coloring problem introduced in \cref{ssec:intro-is-re-univ}. We can show that in $3$-regular graphs, the ORCX problem has these properties:
\begin{enumerate}
	\item Let $\Pi$ be the restricted $9$-coloring problem and let $\fI$ be the auxiliary input of $3$-edge coloring. Then $\Pi$ with input $\fI$ can be used to solve the ORCX problem.
	\item The ORCX problem is a fixed point in round elimination.
	\item The ORCX problem remains non-trivial even if we are given a $3$-edge coloring.
\end{enumerate}
Now we are ready to apply the machinery of tripotent inputs to turn ORCX together with $3$-coloring into an input-free problem $\Pi'$ that is a nontrivial fixed point relaxation of the $9$-coloring problem.

This idea can be generalized to a broader family of homomorphism problems. In \cref{sec:homom} we will study them in $\Delta$-regular graphs and the input will be a $\Delta$-edge-coloring, but the same construction of the ORCX problem applies also there.

\paragraph{Error probability analysis with inputs (\cref{sec:lifting}).}

To show that round elimination fixed points imply an $\Omega(\log \log n)$ lower bound in the randomized LOCAL model, the key step is to analyze failure probabilities for a single step of round elimination.

While round elimination is a function that maps problems to problems, it also has algorithmic implications: given an algorithm $\fA$ that solves $\Pi$ in $T$ rounds, we can construct an algorithm $\fA'$ that solves $\Pi' = \fQ(\Pi)$ in $T-1$ rounds. However, the catch is that if $\fA$ is a randomized algorithm that fails with probability $p$, then $\fA'$ is a randomized algorithm that fails with probability $p' > p$.

While in prior work there are results that analyze how $p'$ depends on $p$ and the structural properties of $\Pi$, none of these results can handle inputs in full generality. In this work we present the first analysis of error probabilities that is able to take into account also inputs.

We will use our new analysis in \cref{sec:sso-with-so} to show that the SSO-SO problem is indeed at least intermediate-hard, and in \cref{sec:homom} to derive a lower bound for homomorphism problems.

\section{Preliminaries}\label{sec:Preliminaries}

In this section, we provide some of the technical background from the literature and collect definitions and observations that we need to obtain our results.

\paragraph{Graph-theoretic basics.}
Throughout the paper, we will use standard graph notation, such as $G = (V,E)$ for a graph, $\deg(v)$ for the degree of a node $v$, etc.
Moreover, we will make use of the following notion of a \emph{half-edge}:

\begin{definition}[Half-edge]
	Let $G = (V, E)$ be a graph.
	A \emph{half-edge $h$} is a pair $(v, e)$ such that $e$ is incident to $v$.
	We say that the half-edge $(v, e)$ is \emph{incident to $v$} and \emph{belongs} to $e$.
	We denote the set of half-edges of $G$ by $H$, i.e., $H := \{ (v, e) \mid v \in V, e \in E, v \text{ is an endpoint of } e \}$.
\end{definition}

\paragraph{The LOCAL model.}
The LOCAL model is a synchronous message-passing model defined as follows.
A network is represented as a graph $G = (V,E)$, where $V$ is the set of machines and $E$ is the set of communication links.
At the beginning of the computation, each node $v$ is aware of the size $n$ of the graph, the maximum degree $\Delta$ of the graph, its own degree $\deg(v)$, and the input assigned to its incident half-edges. Moreover, in the case of deterministic algorithms, each node is assigned a unique ID in $\{1,\ldots,n\}$, while in the case of randomized algorithms each node is assigned a random bit-string of infinite length.
Then, the computation proceeds in rounds, and at each round each node can send a (possibly different) message to each of its neighbors. Then, as a function of the received messages, a node updates its state and proceeds to the next round. In a $T$-round algorithm, each node is required to output a solution for the considered problem within round $T$. In this model, the size of the messages, and the computational power of each node, are unbounded (messages do not even have to be of finite length: as in this paper we prove lower bounds, considering this unreasonable assumption only makes our results stronger).  In the case of randomized algorithms, it is required that the produced solution is correct with probability at least $1 - 1/n$. For technical reasons, we also assume that each node $v$ comes with a \emph{port numbering}, that is, an arbitrary assignment of the numbers $1,\ldots,\deg(v)$ to the half-edges incident to $v$, such that different half-edges get different numbers. Additionally, an arbitrary orientation is assigned to the edges, or in other words, also edges come with a port numbering, in this case from $\{1,2\}$. The \emph{port numbering} (PN) model is defined similarly as the LOCAL model, and the only difference is that there are no IDs assigned to the nodes.
Since the size of the messages is not restricted, we can see any $T$-round algorithm as a function that maps a given radius-$T$ neighborhood into an output.

\paragraph{Node-edge-checkable problems.}
Next, we introduce the notion of a node-edge-checkable problem \cite{brandt-2019-an-automatic-speedup-theorem-for,balliu-censor-hillel-etal-2021-locally-checkable}.
These are problems that capture the vast majority of problems studied in the LOCAL model, including maximal independent set, maximal matching, and coloring problems. We define such problems in the case of regular graphs. The reason is that all known lower bounds that are obtained by the lower bound techniques relevant to our work---round elimination~\cite{brandt-fischer-etal-2016-a-lower-bound-for-the,brandt-2019-an-automatic-speedup-theorem-for} and Marks' technique~\cite{marks-2016-a-determinacy-approach-to-borel}---are obtained on regular graphs or regular trees\footnote{A regular tree is simply a tree where every node of degree $\neq 1$ has the same degree.}, and the same holds for the lower bound constructions we develop in this work.
As such, we can restrict attention to $\Delta$-regular graphs/trees and will provide our definitions for this setting.
Consequently, for the remainder of the paper (except for \Cref{sec:lifting}), we will assume that some value $\Delta$ is fixed.

We note that our definitions generalize straightforwardly to the setting of arbitrary graphs though they might be much more cumbersome to write down in that setting. In fact, we will provide a generalized definition in \Cref{sec:lifting}, which will consider problems in graphs that are not necessarily regular, and problems where the validity of the output may depend on some given input.

\begin{definition}[Node-edge-checkable problem]
	A \emph{node-edge-checkable problem} $\Pi$ is a tuple $(\Sigma_{\Pi},\allowbreak \nodeconst_{\Pi},\allowbreak \edgeconst_{\Pi})$ where
	\begin{enumerate}[noitemsep]
		\item $\Sigma_{\Pi}$ is a finite set,
		\item $\nodeconst_{\Pi}$ is a set of cardinality-$\Delta$ multisets of elements from $\Sigma_{\Pi}$, and
		\item $\epi$ is a set of cardinality-$2$ multisets of elements from $\spi$.
	\end{enumerate}
	The set $\Sigma_{\Pi}$ is called the \emph{(output) label set of $\Pi$} and the elements of $\spi$ are called \emph{output labels (of $\Pi$)}.
	The set $\nodeconst_{\Pi}$ is called the \emph{node constraint of $\Pi$} and $\edgeconst_{\Pi}$ is called the \emph{edge constraint of $\Pi$}.
	A multiset of elements from $\Sigma_{\Pi}$ is called a \emph{configuration}; in particular, the elements of $\epi$ and of $\npi$ are configurations.
	A correct solution for $\Pi$ on a $\Delta$-regular graph is an assignment of labels from $\spi$ to the half-edges of the input graph such that, for each node $v$, the cardinality-$\Delta$ multiset of labels assigned to the half-edges incident to $v$ is an element of $\npi$ and, for each edge $e$, the cardinality-$2$ multisets of labels assigned to the half-edges belonging to $e$ is an element of $\epi$. In the case of $\Delta$-regular trees, the same requirements apply, but only on nodes of degree $\Delta$, that is, leaves are unconstrained.
	When considering a distributed algorithm outputting a solution to some node-edge-checkable problem $\Pi$, we will assume that, for each half-edge $h = (v, e)$, the node deciding which output label is assigned to $h$ is the incident node $v$.
	When considering a configuration that consists of the labels assigned to the half-edges incident to some node, resp.\ edge (or when considering a configuration in $\npi$, resp.\ in $\epi$), we may refer to it as a \emph{node configuration}, resp.\ \emph{edge configuration}.

	We will write a configuration containing labels $\L_1, \dots, \L_j$ as $\L_1\s\dots\s\L_j$.
	Note that the contained labels do not have to be pairwise distinct, as a configuration is a multiset.
	Note further that, for the same reason, any displayed order of the labels represents the same configuration, e.g., $\L_1\s\dots\s\L_j = \L_j\s\dots\s\L_1$.
\end{definition}

Note that all the known lower bounds obtained via round elimination or via the Marks' technique hold on regular \emph{trees}, and the lower bounds are already achieved on the ``inner part'' of the tree, i.e., which configurations are admissible on the leaf nodes does not affect the lower bounds (thereby ensuring that the presented standard definition is indeed all we need for our paper).

\paragraph{Condensed configurations.}
For writing down the configurations in a node or edge constraint, it is often convenient to use so-called condensed configurations, which provide a shorthand for capturing many configurations at once.

\begin{definition}[Condensed configuration]
	Let $\Pi$ be a node-edge-checkable problem with output label set $\Sigma_{\Pi}$.
	Formally, a \emph{condensed configuration} is a configuration $\S_1\s\dots\s\S_j$ (for some positive integer $j$) where each $\S_i$ is a collection of labels from $\Sigma_{\Pi}$.
	However, such a condensed configuration is to be interpreted as the collection of all configurations $\L_1\s\dots\s\L_j$ of labels from $\Sigma_{\Pi}$ such that $\L_i \in \S_i$ for each $1 \leq i \leq j$.
	To support this interpretation notationally, we will write the sets $\S_i$ with square brackets, e.g., $[\L_1 \L_2]\s[\L_1 \L_3]$ represents the collection containing the configurations $\L_1\s\L_1$, $\L_1\s\L_2$, $\L_1\s\L_3$, and $\L_2\s\L_3$. 
\end{definition}

\paragraph{Example: sinkless orientation.}
We provide an example of a node-edge checkable problem. Consider the problem of orienting the edges of a $3$-regular graph such that no node is a \emph{sink}, i.e., no node has all edges oriented incoming. This problem is called sinkless orientation, and can be defined as a node-edge checkable problem $\Pi = (\spi,\npi,\epi)$ as follows.
\begin{itemize}
	\item $\spi = \{\lI,\lO\}$, where $\lI$ stands for \emph{incoming} and $\lO$ stands for \emph{outgoing}.
	\item $\epi = \{ \lI \s \lO \}$, that is, on the edges the only allowed configuration is $\lI \s \lO$. In other words, if a node claims that an edge is outgoing, then the other endpoint must claim that such an edge is incoming, and vice versa.
	\item $\npi$ contains all the configurations given by the condensed configuration $[\lO] \s [\lI\s\lO] \s [\lI\s\lO]$, that is, $\lO \s \lI \s \lI$, $\lO \s \lO \s \lI$, and $\lO \s \lO \s \lO$. (Recall that the order of labels does not matter.) In other words, for each node, at least one incident half-edge must be labeled as outgoing.
\end{itemize}

\paragraph{Relaxations.}
In the following, we define the notion of a \emph{relaxation} of a node-edge-checkable problem (introduced in \cite{brandt-olivetti-2020-truly-tight-in-delta-bounds-for} with a slightly different terminology).
Intuitively, a relaxation of a problem $\Pi$ is a problem $\hat{\Pi}$ such that $\hat{\Pi}$ can be solved in $0$ rounds given a solution to $\Pi$.
Formally, this can be expressed as follows, making use of the definition of a node-edge-checkable problem.

\begin{definition}[Relaxation]\label{def:relax}
	Let $\Pi = (\Sigma_{\Pi}, \nodeconst_{\Pi}, \edgeconst_{\Pi})$ and $\hat{\Pi} = (\Sigma_{\hat{\Pi}}, \nodeconst_{\hat{\Pi}}, \edgeconst_{\hat{\Pi}})$ be two node-edge-checkable problems.
	Let $z$ denote the number of configurations in $\nodeconst_{\Pi}$ and let $\L_{i1}~\dots~\L_{i\Delta}$, where $1 \leq i \leq z$, denote the $z$ configurations.
	Then $\hat{\Pi}$ is called a \emph{relaxation} of $\Pi$ if there exists a mapping\footnote{We slightly abuse notation here by considering the $\L_{ij}$ simultaneously as two different objects: entries in configurations and labels from $\Sigma_{\Pi}$. In particular, please note that the function $f$ may map the same label $\L$ from $\Sigma_{\Pi}$ to two different labels in $\Sigma_{\hat{\Pi}}$ if $\L$ occurs in two different configurations of $\nodeconst_{\Pi}$ (or if $\L$ occurs more than once in the same configuration).} $f \colon \{\L_{ij} \mid 1 \leq i \leq z, 1 \leq j \leq \Delta\} \rightarrow \Sigma_{\hat{\Pi}}$ such that
	\begin{enumerate}[noitemsep]
		\item\label{item:node-based1} for every configuration $\L_{i1}~\dots~\L_{i\Delta} \in \nodeconst_{\Pi}$, we have $f(\L_{i1})~\dots~f(\L_{i\Delta}) \in \nodeconst_{\hat{\Pi}}$, and
		\item\label{item:node-based2} for every configuration $\L_{ij}~\L_{i'j'} \in \edgeconst_{\Pi}$, we have $f(\L_{ij})~f(\L_{i'j'}) \in \edgeconst_{\hat{\Pi}}$.
	\end{enumerate}
	We call such a mapping $f$ a \emph{relaxation function from $\Pi$ to $\hat{\Pi}$}.
	Moreover, if $\hat{\Pi}$ is a relaxation of $\Pi$, then we write $\Pi \zrr \hat{\Pi}$, symbolizing the fact that a solution to $\hat{\Pi}$ can be obtained from a solution to $\Pi$ by a $0$-round algorithm.
	If $\Pi \zrr \hat{\Pi}$ does not hold, we write $\Pi \nzrr \hat{\Pi}$.
\end{definition}
This definition captures the aforementioned intuition in the following sense:
The only information that a node $v$ executing a $0$-round algorithm has about a solution to $\Pi$ is the output labels of that solution on the half-edges incident to $v$.
Hence, in order to infer from said solution a (local) solution to $\hat{\Pi}$ (without making use of any further assumptions about the input graph), node $v$ must change each of the labels on its incident half-edges to an output label for $\hat{\Pi}$ in a way that guarantees that the new assignment satisfies the node and edge constraints of the new problem $\hat{\Pi}$.
As, for each edge $e = \{ u, v \}$ incident to $v$, the only guarantee that $v$ has about the label assigned to half-edge $(u, e)$ in the solution to $\Pi$ is that together with the label assigned to half-edge $(v, e)$ it forms a configuration contained in $\edgeconst_{\Pi}$, we arrive at the definition given above.

A second kind of relaxation we will make use of in our work is what we call \emph{port-local relaxations}.
Differently from the regular relaxations defined above in which \emph{occurrences of labels in node configurations} are mapped (i.e., where the mapping takes entire node configurations into account), a port-local relaxation is a direct mapping from the output label set of a problem to an output label set of a second problem.
Formally, we define port-local relaxations as follows.
\begin{definition}[Port-local relaxation]\label{def:pl-relax}
	Let $\Pi = (\Sigma_{\Pi}, \nodeconst_{\Pi}, \edgeconst_{\Pi})$ and $\hat{\Pi} = (\Sigma_{\hat{\Pi}}, \nodeconst_{\hat{\Pi}}, \edgeconst_{\hat{\Pi}})$ be two node-edge-checkable problems.
	Then we say that $\hat{\Pi}$ is a \emph{port-local relaxation} of $\Pi$ (and, equivalently, that $\Pi$ is \emph{port-locally relaxable} to $\hat{\Pi}$) if there exists a function $f: \sig{\Pi} \rightarrow \sig{\hat{\Pi}}$ such that
	\begin{enumerate}[noitemsep]
		\item for each $L_1~\dots,~L_{\Delta} \in \nod{\Pi}$, we have $f(L_1)~\dots~f(L_{\Delta}) \in \nod{\hat{\Pi}}$, and
		\item for each $L_1~L_2 \in \edg{\Pi}$, we have $f(L_1) \s f(L_2) \in \edg{\hat{\Pi}}$.
	\end{enumerate}
\end{definition}

From \Cref{def:relax,def:pl-relax}, it follows that a port-local relaxation is always a relaxation.
\begin{observation}\label{obs:plrisr}
	Let $\Pi$ and $\hat{\Pi}$ be two node-edge-checkable problems such that $\hat{\Pi}$ is a port-local relaxation of $\Pi$.
	Then $\Pi \zrr \hat{\Pi}$.
\end{observation}

Moreover, from the definition of a relaxation it follows immediately that a relaxation of a relaxation of some problem $\Pi$ is a relaxation of $\Pi$ (and analogously for port-local relaxations), which we capture in the following observation.
\begin{observation}\label{obs:combinerelax}
	For any three node-edge-checkable problems $\Pi, \Pi', \Pi''$, it holds that if $\Pi \zrr \Pi'$ and $\Pi' \zrr \Pi''$, then also $\Pi \zrr \Pi''$.
	Moreover, if $\Pi'$ is a port-local relaxation of $\Pi$ and $\Pi''$ a port-local relaxation of $\Pi'$, then $\Pi''$ is a port-local relaxation of $\Pi$.
\end{observation}

\paragraph{Equivalent problems.}
Based on the notion of a relaxation, we now formalize the concept of \emph{equivalent} problems, which we will encounter in several places throughout the paper.

\begin{definition}[Equivalent problems]\label{def:eq-problems}
	We call two node-edge-checkable problems $\Pi, \Pi'$ \emph{equivalent} if $\Pi$ is a relaxation of $\Pi'$ and $\Pi'$ is a relaxation of $\Pi$.
\end{definition}
A simple case of equivalent problems that we will encounter frequently are two problems $\Pi, \Pi'$ where one is obtained from the other by renaming labels.
As, in this case, these two problems are \emph{identical} for essentially all purposes, for simplicity we will treat problems obtained from each other via renaming of labels as \emph{the same} problem and write $\Pi = \Pi'$.

\paragraph{Round elimination.}
The lower bounds we obtain in our work make use of the round elimination technique, which we introduce in the following.
The round elimination framework is based on two functions $\re(\cdot)$ and $\rere(\cdot)$ that take a node-edge-checkable problem as input and return a node-edge-checkable problem.
For the definition of the two functions, we need the notion of a \emph{maximal} configuration of sets.

\begin{definition}[Maximal and dominated configurations]
	Let $k$ be a positive integer, $\Sigma$ a finite set of sets and $\fC$ a collection of cardinality-$k$ multisets with elements from $\Sigma$ (i.e., each such multiset consists of $k$ sets).
	Then a configuration $\S_1~\dots~\S_k \in \fC$ is called \emph{maximal (in $\fC$)} if there is no configuration $\S'_1~\dots~\S'_k \in \fC$ such that there exists a permutation $\rho: \{ 1, \dots, k \} \rightarrow \{ 1, \dots, k \}$ satisfying 1) $\S_i \subseteq \S'_{\rho(i)}$ for all $1 \leq i \leq k$ and 2) $\S_i \subsetneq \S'_{\rho(i)}$ for at least one $1 \leq i \leq k$. Otherwise, we say that $\S_1~\dots~\S_k$ is non-maximal and that it is \emph{dominated} by $\S'_1~\dots~\S'_k$.
\end{definition}

Now we are set to define $\re(\cdot)$ and $\rere(\cdot)$.
Let $\Pi$ be a node-edge-checkable problem.
The node-edge-checkable problem $\re(\Pi)$ is defined as follows.

Define $\Sigma'$ as the set of all nonempty subsets of $\spi$, i.e., $\Sigma' := 2^{\spi} \setminus \{ \emptyset \}$.
We will call the elements of $\Sigma'$ labels but keep in mind that, formally, they are sets.
For defining the edge constraint $\erepi$ of $\repi$, first define $\mathcal{C}'$ as the collection of all configurations $\S_1~\S_2$ consisting of labels from $\Sigma'$ (i.e., consisting of sets of labels from $\spi$) such that, for each pair $(\L_1, \L_2) \in \S_1 \times \S_2$, we have $\L_1~\L_2 \in \epi$.
Then, $\erepi$ is simply defined as the set of all maximal configurations in $\mathcal{C}'$.

Next, define the output label set $\srepi$ of $\repi$ as the subset of $\Sigma'$ containing precisely those labels that appear in at least one configuration in $\erepi$.
Finally, define the node constraint $\nrepi$ of $\repi$ as the set of all configurations $\S_1~\dots~\S_{\Delta}$ consisting of labels from $\srepi$ such that there exists a tuple $(\L_1, \dots, \L_{\Delta}) \in \S_1 \times \dots \times \S_{\Delta}$ satisfying $\L_1~\dots~\L_{\Delta} \in \npi$.
This concludes the definition of $\re(\Pi)$.

The node-edge-checkable problem $\rere(\Pi)$ is defined dually (where the application of the universal and existential quantifiers is reversed):
Define $\Sigma''$ as the set of all nonempty subsets of $\spi$, i.e., $\Sigma'' := 2^{\spi} \setminus \{ \emptyset \}$.
For defining the node constraint $\nrerepi$ of $\rerepi$, first define $\mathcal{C}''$ as the collection of all configurations $\S_1~\dots~\S_{\Delta}$ consisting of labels from $\Sigma''$ such that, for each tuple $(\L_1, \dots, \L_{\Delta}) \in \S_1 \times \dots \times \S_{\Delta}$, we have $\L_1~\dots~\L_{\Delta} \in \npi$.
Then, $\nrerepi$ is simply defined as the set of all maximal configurations in $\mathcal{C}''$.

Next, define the output label set $\srerepi$ of $\rerepi$ as the subset of $\Sigma''$ containing precisely those labels that appear in at least one configuration in $\nrerepi$.
Finally, define the edge constraint $\ererepi$ of $\rerepi$ as the set of all configurations $\S_1~\S_2$ consisting of labels from $\srerepi$ such that there exists a pair $(\L_1, \L_2) \in \S_1 \times \S_2$ satisfying $\L_1~\L_2 \in \epi$.
This concludes the definition of $\rerepi$.

We note that, while the function $\rere(\cdot)$ can be applied to any node-edge-checkable problem, in this work we will only apply it to node-edge-checkable problems obtained by the application of the function $\re(\cdot)$, i.e., to problems of the form $\re(\Pi)$.
Furthermore, to simplify notation, we define $\fQ(\cdot) := \rere(\re(\cdot))$.

Moreover, when computing $\edgeconst_{\re(\Pi)}$, resp.\ $\nodeconst_{\rere(\Pi)}$, we will refer to configurations in the above collections $\mathcal{C}'$, resp.\ $\mathcal{C}''$, as configurations that \emph{satisfy the universal quantifier}.

\paragraph{\boldmath An easy way to compute $\nodeconst_{\re(\Pi)}$ and $\edgeconst_{\rere(\Pi)}$.}
It has already been noticed in previous works \cite{brandt-2019-an-automatic-speedup-theorem-for} that there is an easy way to compute $\nodeconst_{\re(\Pi)}$ and $\edgeconst_{\rere(\Pi)}$:

\begin{observation}\label{obs:easy-exists}
The constraint $\nodeconst_{\re(\Pi)}$ (resp.\ $\edgeconst_{\rere(\Pi)}$) can be computed as follows: for each configuration $\L_1\s\ldots\s\L_\Delta\in \npi$ (resp.\ $\L_1\s\L_2\in \epi$), add to the constraint $\nodeconst_{\re(\Pi)}$ (resp.\ $\edgeconst_{\rere(\Pi)}$) all the configurations that can be picked from the condensed configuration obtained by replacing each $\L_i$ with the set containing all the labels in $\Sigma_{\re(\Pi)}$ (resp.\ $\Sigma_{\rere(\Pi)}$) that are supersets of $\L_i$.
\end{observation}

\paragraph{\boldmath An easy way to compute $\edgeconst_{\re(\Pi)}$ and $\nodeconst_{\rere(\Pi)}$.}
The hard part of computing $\re(\Pi)$ and $\rere(\Pi)$ is computing the constraints $\edgeconst_{\re(\Pi)}$ and $\nodeconst_{\rere(\Pi)}$. A recent work developed an algorithm that makes it easier to compute such constraints \cite{balliu-brandt-etal-2025-towards-fully-automatic}. Such an algorithm is based on an iterative application of a simple operation, that we now describe.

Let $\fC = \S_1~\dots~\S_k$ and $\fC' = \S'_1~\dots~\S'_k$ be two configurations of sets of labels.
Let $1 \leq u \leq k$, and let $\sigma \colon \{ 1, \dots, k\} \rightarrow \{ 1, \dots, k\}$ be a permutation.
The \emph{combination of $\fC$ and $\fC'$ w.r.t.\ $u$ and $\sigma$} is defined as the configuration $\fC'' = \S''_1~\dots~\S''_k$ satisfying the following:
\begin{itemize}[noitemsep]
	\item $\S''_j = \S_j \cap \S'_{\sigma(j)}$ for each $j \in \{ 1, \dots, k\} \setminus \{ u \}$, and
	\item $\S''_u = \S_u \cup \S'_{\sigma(u)}$.
\end{itemize}

We now describe an algorithm $\mathcal{A}$ that takes as input a constraint $\fC$ described as a collection of condensed configurations and returns a new constraint $\fC'$ of configurations. If $\fC$ is not given as a set of condensed configurations, it is straightforward to convert it into a set of condensed configurations, that is, by replacing each label with the singleton set containing that label. 

The algorithm $\mathcal{A}$ works as follows. At first, $\fC'$ is initialized as $\fC':= \fC$. Then, remove non-maximal configurations from $\fC'$. Choose two arbitrary (possibly the same) configurations $\C_1$ and $\C_2$ from $\fC'$, and combine them w.r.t\ some arbitrarily chosen $u$ and $\sigma$. Add the obtained configuration to $\fC'$ if it is not dominated by some configuration already present in $\fC'$, and remove non-maximal configurations from $\fC'$. 
Repeat this operation for all possible choices of $\C_1$, $\C_2$, $u$, and $\sigma$, where $\C_1$ and $\C_2$ can also be configurations that have been added to $\fC'$ in the previous steps, until no new configuration is added to $\fC'$.

\begin{observation}[Theorem 4.1 of \cite{balliu-brandt-etal-2025-towards-fully-automatic}]\label{obs:newre}
	The result of applying $\mathcal{A}$ on $\edgeconst_{\Pi}$ is $\edgeconst_{\re(\Pi)}$, and the result of applying $\mathcal{A}$ on $\nodeconst_{\Pi}$ is $\nodeconst_{\rere(\Pi)}$.
\end{observation}

We say that two sets $S$ and $S'$ are \emph{comparable} if $S\subseteq S'$ or $S'\subseteq S$. Otherwise, we say that they are non-comparable.
In \cite{balliu-brandt-etal-2025-distributed-quantum-advantage}, the authors made the following observation.

\begin{observation}\label{obs:union-non-comparable}
	Let $\mathcal{A'}$ be defined similarly as $\mathcal{A}$, with the only difference that two configurations $\C_1=\S_1\s\ldots\s\S_k$ and $\C_2=\S'_1\s\ldots\s\S'_k$ are combined w.r.t.\ $u$ and $\sigma$ only if $S_u$ and $S'_{\sigma(u)}$ are non-comparable.
	Then, the output of $\mathcal{A'}$ is the same as the one of $\mathcal{A}$.
\end{observation}

\paragraph{Obtaining lower bounds via round elimination.}
The way the round elimination framework (and in particular the two functions $\re(\cdot)$ and $\rere(\cdot)$) are used to obtain lower bounds is based on the round elimination theorem~\cite{brandt-2019-an-automatic-speedup-theorem-for}, which, very informally speaking, states that, for any node-edge-checkable problem $\Pi$, the complexity of $\fQ(\Pi)\coloneqq \rere(\re(\Pi))$ is precisely one round less than the complexity of $\Pi$ (or $0$ in case also $\Pi$ is $0$-round-solvable).\footnote{This comes with a lot of fine print; for a modern overview containing all the technical details, see, e.g., \cite[Appendix A, arxiv version]{balliu-brandt-etal-2022-distributed-delta-coloring} or~\Cref{sec:lifting}.}
Applying this theorem recursively to the sequence $\Pi, \fQ(\Pi), \fQ^2(\Pi), \dots$ of problems then yields that the complexity of $\fQ^i(\Pi)$ is $i$ rounds less than the complexity of $\Pi$.
Now, to obtain a lower bound of $k$ rounds for the complexity of $\Pi$, we merely need to show that problem $\fQ^{k - 1}(\Pi)$ cannot be solved in $0$ rounds (which is much easier to check than $j$-round solvability for $j >0$).

We emphasize that this overview is highly informal and hides a lot of details.
In particular, those hidden details guarantee that if none of the problems in the sequence $\Pi, \fQ(\Pi), \fQ^2(\Pi), \dots$ is $0$-round solvable, then one obtains a deterministic lower bound of $\Omega(\log n)$ rounds and a randomized lower bound of $\Omega(\log \log n)$ rounds.\footnote{Again, some details are hidden here; for a formal statement, see~\cite[Theorem 7.1]{balliu-brandt-etal-2022-distributed-delta-coloring}, or \Cref{thm:newlifting} for a more general version.}

\paragraph{Fixed points.}
One beautiful scenario in which the above approach yields the desired lower bounds without much effort is that $\Pi$ satisfies $\fQ(\Pi) = \Pi$, which implies that the aforementioned sequence reduces to $\Pi, \Pi, \Pi, \dots$ and that therefore none of the problems in the sequence can be solved in $0$ rounds, assuming that $\Pi$ cannot be solved in $0$ rounds.
A problem $\Pi$ satisfying $\fQ(\Pi) = \Pi$ is called a \emph{fixed point}.
We slightly extend the definition of a fixed point in a way that guarantees the same lower bounds but, for technical reasons, may capture more problems.

\begin{definition}[Fixed point]
	A \emph{fixed point} is a node-edge-checkable problem $\Pi$ satisfying $\fQ(\Pi) \zrr \Pi$.
	We call a fixed point \emph{nontrivial} if it cannot be solved in $0$ rounds, and trivial otherwise.
	Moreover, for two node-edge-checkable problems $\Pi, \hat{\Pi}$, we call $\hat{\Pi}$ a \emph{(trivial/nontrivial) fixed point relaxation} of $\Pi$ if $\hat{\Pi}$ is a (trivial/nontrivial) fixed point and $\Pi \zrr \hat{\Pi}$. 
\end{definition}

If $\Pi$ is a nontrivial fixed point, then by combining the above argumentation with the fact that $\Pi$ can be solved in $0$ rounds given a solution to $\fQ(\Pi)$, we indeed again obtain the deterministic and randomized lower bounds of $\Omega(\log n)$ and $\Omega(\log \log n)$, respectively.

An example of a problem that is a nontrivial fixed point is the sinkless orientation problem \cite{brandt-fischer-etal-2016-a-lower-bound-for-the}. An example of a problem that is not a fixed point but can be relaxed to a nontrivial fixed point is the $\Delta$-coloring problem \cite{balliu-brandt-etal-2022-distributed-delta-coloring}.

\paragraph{Inputs.}
As discussed in~\Cref{sec:intro}, one interesting aspect that our work explores are settings with inputs.
In this setting, each half-edge is labeled with an input label from some input label set $\Sigma_{\mathrm{in}}$ and each node is aware of the input labels on its incident half-edges at the beginning of an algorithm.

An important distinction in the setting with inputs is between problems where the correctness of the output depends on the input and problems where the correctness of the output is independent of the input (and the input is merely additional information that the algorithm can use).
Apart from~\Cref{sec:lifting}, where we will develop a generic lifting theorem that will allow for problems of the former kind (to capture problems like list coloring), throughout the paper we will focus on the latter kind as it is sufficient for our purposes.

One nice property of round elimination is that \emph{in the deterministic port numbering model, i.e., when nodes do not have access to unique identifiers or random bits}, round elimination is known to work also in the setting with input (of the latter kind) as long as the input (or, more specifically, the considered input-labeled graph class) satisfies some independence property~\cite{brandt-2019-an-automatic-speedup-theorem-for}.
This property, informally speaking, can be phrased as follows: for any sufficiently small $T$ and any $T$-hop neighborhood $N$ of any node $v$, ``the input that $v$ would see when extending its view one hop beyond $N$ in any fixed direction does not reveal any information about the input $v$ would see when extending its view one hop in any other direction''.\footnote{For the formal definition of this independence property, see the definition of \emph{$t$-independence} in~\cite{brandt-2019-an-automatic-speedup-theorem-for}.}
In particular, this independence property is satisfied when the input labels assigned to the half-edges of the input graph constitute a correct solution to some fixed node-edge-checkable problem.
As such, we will allow precisely those inputs: when studying the setting with inputs, a fixed node-edge-checkable problem $\fI$ specifies which input configurations can appear, namely, precisely those that satisfy the constraints of problem $\fI$.
In other words, the input graph comes with an arbitrary solution to $\fI$, and this solution can be used by the nodes while performing their communication and computation to solve some given problem $\Pi$ of interest.

As mentioned above, in the setting with such inputs, round elimination is only known to work in the port numbering model.
In \Cref{sec:lifting}, we show that round elimination works in the input setting also in the LOCAL model.

To conclude the discussion of inputs, let us clarify what we mean when we say that round elimination ``works'' in the setting with inputs: the definition and use of the functions $\re(\cdot)$ and $\rere(\cdot)$ stays precisely the same; the only difference that the input makes is that a problem is considered to be $0$-round solvable if it can be solved in $0$ rounds \emph{given the input}, i.e., in a potential $0$-round algorithm, each node $v$ may make use of the information about the input on its incident half-edges (which is all that $v$ knows about the input).

\paragraph{Relaxations using inputs.}
A natural idea to make use of fixed points to obtain lower bounds is via relaxations: finding, for a given problem $\Pi$, a nontrivial fixed point $\fF$ satisfying $\Pi \zrr \fF$ directly transfers the aforementioned $\Omega(\log n)$- and $\Omega(\log \log n)$-round lower bounds from $\fF$ to $\Pi$.
Now, based on our result in~\Cref{sec:lifting} that round elimination works in the setting with inputs, we can extend this idea to the setting with inputs: in particular, we obtain these lower bounds already whenever we can find just a single node-edge-checkable ``input'' problem $\fI$ such that
\begin{enumerate}[noitemsep]
	\item\label{item:inputrelax} we can relax $\Pi$ to some fixed point $\fF$ given (a solution to) $\fI$ as input, and
	\item $\fF$ is nontrivial given input $\fI$, i.e., $\fI \nzrr \fF$.
\end{enumerate}

Formally, for defining what we mean by a relaxation given some input, we introduce the notion of the product of two problems.

\begin{definition}[Product of problems]
	Let $\Pi, \Pi'$ be two node-edge-checkable problems.
	Then the \emph{product $\Pi \times \Pi' = (\Sigma_{\Pi \times \Pi'}, \nodeconst_{\Pi \times \Pi'}, \edgeconst_{\Pi \times \Pi'})$ of $\Pi$ and $\Pi'$} is defined by setting
	\begin{align*}
		\Sigma_{\Pi \times \Pi'} &:= \Sigma_{\Pi} \times \Sigma_{\Pi'},\\
		\nodeconst_{\Pi \times \Pi'} &:= \{ (\L_1, \L'_1)\s\dots\s(\L_{\Delta}, \L'_{\Delta}) \mid \L_1\s\dots\s\L_{\Delta} \in \nodeconst_{\Pi} \textrm{ and } \L'_1\s\dots\s\L'_{\Delta} \in \nodeconst_{\Pi'} \},\textrm{ and}\\
		\edgeconst_{\Pi \times \Pi'} &:= \{ (\L_1, \L'_1)\s(\L_2, \L'_2) \mid \L_1\s\L_2 \in \edgeconst_{\Pi} \textrm{ and } \L'_1\s\L'_2 \in \edgeconst_{\Pi'} \}.
	\end{align*}
\end{definition}

Now, we can formally restate point~\ref{item:inputrelax} from above as $\Pi \times \fI \zrr \fF$.

Similarly as for relaxations, in the setting with inputs, we can also strengthen the notion of a fixed point (in the sense that it captures more problems while providing the same complexity implications) by making use of the available input.
\begin{definition}[Generalized fixed point]
	Assume that we are in the setting with inputs, where the input is given by some node-edge-checkable problem $\fI$.
	Then we call a node-edge-checkable problem $\Pi$ a \emph{generalized fixed point} if $\fQ(\Pi) \times \fI \zrr \Pi$.
	Moreover, we call a generalized fixed point $\Pi$ \emph{nontrivial} if $\Pi$ cannot be solved in $0$ rounds given $\fI$, i.e., if $\fI \nzrr \Pi$.
\end{definition}
Given the definition of a generalized fixed point and using the insights from the above discussion, we can now formulate a strong version of the lower bound guarantees provided by fixed point relaxations in the setting with input $\fI$: if $\fF$ is a nontrivial generalized fixed point and $\Pi$ a node-edge-checkable problem satisfying $\Pi \times \fI \zrr \fF$, then $\Pi$ requires $\Omega(\log n)$ rounds to be solved deterministically and $\Omega(\log \log n)$ rounds to be solved randomized.
Note that the fact that these lower bounds hold in the setting with input also implies that the lower bounds hold in the (harder) setting without input.

\section{The tripotent input}\label{sec:tripotent}

In this section we will investigate the technique of proving lower bounds for node-edge-checkable problems via fixed point relaxations with input.
Suppose we are given a node-edge checkable problem $\Pi$ for which we want to prove hardness.
As we explained in \Cref{sec:Preliminaries}, the hardness follows if we can find a nontrivial fixed point $\fF$ and an input $\fI$ (which are both node-edge-checkable problems themselves) such that
\begin{equation*}
	\Pi \times \fI \zrr \fF, \quad \quad \fI \nzrr \fF.
\end{equation*}
The first condition means that $\Pi$ can be relaxed to $\fF$ in the setting with input $\fI$, the second ensures that $\fF$ is nontrivial even given a solution to $\fI$.
For the moment, suppose that we already chose the fixed point $\fF$ we want to use.
We are then left to find an input $\fI$ ensuring the above properties, where we focus on the first property $\Pi \times \fI \zrr \fF$.
In the following, we provide a generic construction of such a suitable input problem.
More than that, we will show that our construction in fact leads to \emph{the easiest} node-edge-checkable problem $\fI$ satisfying this property.
By ``easiest'', we mean that any other input problem $\fJ$ allowing a relaxation from $\Pi$ to $\fF$ can itself be relaxed to $\fI$.
Note that the existence of such an easiest input is highly nontrivial, but at the same time optimal for our purposes: since we want to also ensure the second property $\fI \nzrr \fF$, the input $\fI$ should contain as little information as possible, i.e., be as easy as possible.

Later in this section we connect this new construction to the existing round elimination framework.
Rather surprisingly, we will prove that under some conditions, the constructed inputs themselves are again fixed points.
More than that, we will encounter cases in which we can find nontrivial fixed point relaxations (without input) for problems of interest using this construction.

\subsection{Constructing an input problem}

Suppose we are given two node-edge-checkable problems, $\Pi$ and $P$, where $\Pi$ is a problem of interest and $P$ is some problem we chose in order to prove a lower bound or $\Pi$.
One can think of $P$ as being a fixed point, however for now this is not an assumption we make.
Before we can define our generic input problem $\tau_P(\Pi)$, we need to introduce another problem, $\rest(P)$, which is pretty similar to $\re(P)$.
The only difference in the definition is that we also allow non-maximal edge configurations, $\Sigma$ and $\nodeconst$ stay the same.
Formally, $\rest(\cdot)$ is defined as follows:
\begin{align*}
	\sig{\rest(P)} & := 2^{\sig{P}} \setminus \{\emptyset\},\\
	\nod{\rest(P)} & := \{ \S_1 \s \dots \s \S_{\Delta} | \exists \L_1 \in \S_1, \dots, \L_{\Delta} \in \S_{\Delta} \colon \L_1 \s \dots \s \L_{\Delta} \in \nod{P} \},\\
	\edg{\rest(P)} & := \{ \S_1 \s \S_2 | \forall \L_1 \in \S_1, \L_2 \in \S_2 \colon \L_1 \s \L_2 \in \edg{P} \}.
\end{align*}
Although the definitions are similar, the reason we need $\rest(P)$ has nothing to do with round elimination.
Instead, we will later make use of its definition in \Cref{lem:R(P)-allows-plr}, stating that $\rest(P)$ behaves nicely w.r.t.\ port-local relaxations.

We are now ready to define our generic input:
\begin{definition}\label{def:tau}
	Let $\Pi$ and $P$ be node-edge-checkable problems.
	The \emph{tripotent input} $\tau_{P}(\Pi)$ of $\Pi$ w.r.t.\ $P$ is defined as follows:
	\begin{align*}
		\sig{\tau_P(\Pi)} & := \{ f\,|\,f \colon \Sigma_{\Pi} \rightarrow \Sigma_{\rest(P)} \},\\
		\nod{\tau_P(\Pi)} & := \{ f_1 \s \dots \s f_{\Delta} | \forall \L_1 \s \dots \s \L_{\Delta} \in \nod{\Pi} \colon f_1(\L_1) \s \dots \s f_{\Delta}(\L_{\Delta}) \in \nod{\rest(P)} \},\\
		\edg{\tau_P(\Pi)} & := \{ f_1 \s f_2 | \forall \L_1 \s \L_2 \in \edg{\Pi} \colon f_1(\L_1) \s f_2(\L_2) \in \edg{\rest(P)} \}.
	\end{align*}
	We call $\tau$ the \emph{tripotent input generator}.
\end{definition}
The intuition behind this definition is that it provides an easy relaxation $\Pi \times \tau_P(\Pi) \zrr \rest(P)$, meaning that $\Pi$ can be relaxed to $\rest(P)$ given $\tau_P(\Pi)$ as input.
A half-edge that is assigned labels $\L$ and $f$ by $\Pi$ and $\tau_P(\Pi)$ respectively can choose its new label as $f(\L) \in \sig{\rest(P)}$.
The node and edge constraints of $\tau_P(\Pi)$ ensure that this indeed leads to a valid solution of $\rest(P)$.
In fact, we prove the following stronger statement, proving that $\tau_P(\Pi)$ can be chosen as input when one wants to solve $P$ (instead of $\rest(P)$) given a solution to $\Pi$.
\begin{lemma}\label{lem:tau-is-input}
	Two node-edge-checkable problems $P$ and $\Pi$ always satisfy $\Pi \times \tau_P(\Pi) \zrr P$.
\end{lemma}
\begin{proof}
	We will prove $\Pi \times \tau_P(\Pi) \zrr \rest(P) \zrr P$, our claim $\Pi \times \tau_P(\Pi) \zrr P$ then follows by \Cref{obs:combinerelax}.
	
	We start with the relaxation $\Pi \times \tau_P(\Pi) \zrr \rest(P)$, which we perform port-locally as described above: each node replaces all its labels $(\L, f) \in \sig{\Pi \times \tau_P(\Pi)}$ by $f(\L) \in \sig{\rest(P)}$.
	We need to prove that this is indeed a (port-local) relaxation, therefore consider an edge that was assigned labels $(\L_1, f_1) \s (\L_2, f_2) \in \edg{\Pi \times \tau_P(\Pi)}$.
	By the definition of the product of two problems, this implies $\L_1 \s \L_2 \in \edg{\Pi}$ and $f_1 \s f_2 \in \edg{\tau_P(\Pi)}$.
	\Cref{def:tau} now immediately yields $f_1(\L_1) \s f_2(\L_2) \in \edg{\rest(P)}$ which is what we needed to prove.
	The definition of $\nod{\tau}$ is analogous to $\edg{\tau}$, thus one can use the exact same argument to conclude that $(\L_1, f_1) \s \dots \s (\L_{\Delta}, f_{\Delta}) \in \nod{\Pi \times \tau_P(\Pi)}$ implies $f_1(\L_1) \s \dots \s f_{\Delta}(\L_{\Delta}) \in \nod{\rest(P)}$.
	This proves the above relaxation to be correct.
	
	It is left to provide the second relaxation $\rest(P) \zrr P$.
	Therefore, assume that a node is assigned labels $\S_1 \s \dots \s \S_{\Delta}$ by $\rest(\Pi)$.
	By definition of $\nod{\rest(\Pi)}$ that node can pick $\L_1 \in \S_1, \dots, \L_{\Delta} \in \S_{\Delta}$ such that $\L_1 \s \dots \s \L_{\Delta} \in \nod{\Pi}$.
	If each node proceeds like this, the definition of $\edg{\rest(\Pi)}$ ensures that we end up with a solution of $\Pi$.
\end{proof}

As we already mentioned, we will now prove that $\tau_P(\Pi)$ is not only a ``valid input'' for $\Pi$ in the sense of \Cref{lem:tau-is-input} but is in fact the easiest such problem.
Formally, we will prove that each input $\fI$ allowing a relaxation from $\Pi$ to $P$ can itself be relaxed to $\tau_P(\Pi)$, as stated in the following theorem:
\begin{theorem}\label{thm:easiest-input}
	Let $\Pi, P$ and $\fI$ be node-edge-checkable problems.
	Then $\Pi \times \fI \zrr P$ holds if and only if $\fI \zrr \tau_P(\Pi)$.
\end{theorem}
To prove \Cref{thm:easiest-input}, we first need to introduce the notion of a problem property related to port-local relaxations.
\begin{definition}\label{def:pl-relax2}
	We say that a node-edge-checkable problem $P$ \emph{allows port-local relaxations} if for all node-edge-checkable problems $\Pi$ satisfying $\Pi \zrr P$, $\Pi$ is also port-locally relaxable to $P$.
\end{definition}
Recall that port-local relaxations are a restriction of general relaxations.
Hence it may seem weird to introduce the ``allowing port-local relaxations''-property for node-edge-checkable problems we want to relax to.
However, the following lemma shows that the $\rest(\cdot)$-operator constructs problems with exactly this property, which is why we needed $\rest(\cdot)$ in the first place:
\begin{lemma}\label{lem:R(P)-allows-plr}
	For any node-edge-checkable problem $P$, the problem $\rest(P)$ allows port-local relaxations.
\end{lemma}
\begin{proof}
	Let $\Pi$ be some node-edge checkable problem that can be relaxed to $\rest(P)$, i.e., satisfying $\Pi \zrr \rest(P)$.
	According to \Cref{def:pl-relax2}, we need to prove that there exists a port-local relaxation from $\Pi$ to $\rest(P)$.
	Let $f$ be the given relaxation $\Pi \zrr \rest(P)$.
	$f$ maps an $\L \in \sig{\Pi}$ to one of finitely many $\S_1, \dots, \S_k \in \sig{\rest(P)}$, depending on the other labels that node is assigned.
	Recalling that the $\S_1, \dots, \S_k$ are actually sets of labels of $P$, we can define $g(\L) = \bigcup_{i=1}^k \S_i$.
	This can be done for every $\L \in \sig{\Pi}$ (where $k$ of course depends on $\L$), which yields a function $g \colon \sig{\Pi} \rightarrow \sig{\rest(P)}$.
	We claim that $g$ is a port-local relaxation in the sense of \Cref{def:pl-relax}.
	
	For the first condition, we choose an arbitrary $\L_1 \s \dots \s \L_{\Delta} \in \nod{\Pi}$ which is mapped to $\S_1 \s \dots \s \S_{\Delta}$ by $f$.
	Since $f$ is a correct relaxation, we know $\S_1 \s \dots \s \S_{\Delta} \in \nod{\rest(P)}$.
	According to the definition of $\nod{\rest(P)}$ this means that there exist $\L_1 \in \S_1, \dots, \L_{\Delta} \in \S_{\Delta}$ such that $\L_1 \s \dots \s \L_{\Delta} \in \nod{P}$.
	Note that by definition of $g$, we must have $\S_i \subseteq g(\L_i)$ and thus $\L_i \in g(\L_i)$ for all $i = 1, \dots, \Delta$.
	This implies $g(\L_1) \s \dots \s g(\L_{\Delta}) \in \nod{\rest(P)}$.
	
	For the second condition, we choose an arbitrary edge that is assigned labels $\L$ and $\L'$ by $\Pi$.
	Let $\S_1, \dots, \S_k \in \sig{\rest(P)}$ be the sets $f$ might map $\L$ to.
	Similarly, we define $\S'_1, \dots, \S'_{k'}$ for $\L'$.
	Since $f$ always maps $\L$ and $L'$ to a valid edge configuration of $\rest(P)$, we can apply the definition of $\edg{\rest(P)}$ to all pairs $\S_i, \S'_{i'}$ where $1 \leq i \leq k$ and $1 \leq i' \leq k'$.
	This means we have $s \s s' \in \edg{P}$ for all $s \in \S_i, s' \in \S'_{i'}$.
	Since this holds for all pairs $i,i'$ we also get $s \s s' \in \edg{P}$ for all $s \in g(\L), s' \in g(\L')$.
	Thus we have $g(\L) \s g(\L') \in \edg{\rest(P)}$ which finishes the proof.
\end{proof}
\noindent With this we are ready to prove \Cref{thm:easiest-input}.
\begin{proof}[Proof of \Cref{thm:easiest-input}]
	If we suppose $\fI \zrr \tau_P(\Pi)$, we clearly have
	\begin{equation*}
		\Pi \times \fI \zrr \Pi \times \tau_P(\Pi) \zrr P,
	\end{equation*}
	where the last step follows from \Cref{lem:tau-is-input}.
	We can combine these relaxations as explained in \Cref{obs:combinerelax} to obtain $\Pi \times \fI \zrr P$.
	
	For the other direction we are given a relaxation from $\Pi \times \fI$ to $P$.
	Further, $P$ is (port-locally) relaxable to $\rest(P)$ by just mapping every label $l \in \sig{\Pi}$ to $\{l\} \in \sig{\rest(\Pi)}$.
	Again using \Cref{obs:combinerelax}, we can combine these relaxations $\Pi \times \fI \zrr P$ and $P \zrr \rest(P)$ to obtain $\Pi \times \fI \zrr \rest(P)$.
	Using \Cref{lem:R(P)-allows-plr}, $\Pi \times \fI$ must even be port-locally relaxable to $\rest(P)$.
	Unpacking \Cref{def:pl-relax} this means there exists an $f \colon \sig{\Pi \times \fI} \rightarrow \sig{\rest(P)}$ such that
	\begin{align*}
		\L_1 \s \dots \s \L_{\Delta} \in \nod{\Pi}, i_1 \s \dots \s i_{\Delta} \in \nod{\fI},  & \implies f(\L_1,i_1) \s \dots \s f(\L_{\Delta},i_{\Delta}) \in \nod{\rest(P)} \quad \quad \text{and}\\
		\L_1 \s \L_2 \in \edg{\Pi}, i_1 \s i_2 \in \edg{\fI} & \implies f(\L_1,i_1) \s f(\L_2,i_2) \in \edg{\rest(P)}.
	\end{align*}
	This allows us to define a port-local relaxation from $\fI$ to $\tau_P(\Pi)$ as follows:
	\begin{align*}
		g \colon \sig{\fI} & \rightarrow \sig{\tau_P(\Pi)},\\
		i & \mapsto f(\cdot, i).
	\end{align*}
	We need to prove that $g$ is indeed a port-local relaxation.
	Observe that for each $i_1 \s \dots \s i_{\Delta} \in \nod{\fI}$ and $\L_1 \s \dots \s \L_{\Delta} \in \nod{\Pi}$ we have
	\begin{equation*}
		g(i_1)(\L_1) \s \dots \s g(i_{\Delta})(\L_{\Delta}) = f(\L_1,i_1) \s \dots \s f(\L_{\Delta},i_{\Delta}) \in \nod{\rest(P)}.
	\end{equation*}
	Thus by definition of $\tau_P(\Pi)$ for each $i_1 \s \dots \s i_{\Delta} \in \nod{\fI}$ we have
	\begin{equation*}
		g(i_1) \s \dots \s g(i_{\Delta}) \in \nod{\tau_P(\Pi)}.
	\end{equation*}
	The latter ensures that $g$ maps each of $\fI$'s valid node configurations to a valid node configuration of $\tau_P(\Pi)$.
	$\fE$ behaves exactly the same, proving that $g$ is a port-local relaxation from $\fI$ to $\tau_P(\Pi)$.
\end{proof}
Note that the above proof actually constructs the (port-local) relaxation $\fI \zrr \tau_P(\Pi)$,
so whenever we use \Cref{thm:easiest-input} to infer some relaxation of this type, the proof tells us how to perform that relaxation.

As a direct consequence, we can use \Cref{thm:easiest-input} to see what happens if one repeatedly applies $\tau_P(\cdot)$ to a problem.
We prove that $\tau_P(\tau_P(\Pi))$ is always a relaxation of $\Pi$ and that $\tau_P$ is indeed tripotent, justifying its name.
\begin{corollary}
	For two node-edge-checkable problems $\Pi$ and $P$ the following statements hold:
	\begin{enumerate}[noitemsep]
		\item $\Pi \zrr \tau_P(\tau_P(\Pi))$
		\item $\tau_P(\Pi)$ and $\tau_P(\tau_P(\tau_P(\Pi)))$ are equivalent.
	\end{enumerate}
\end{corollary}
\begin{proof}
	We start with the first statement.
	By \Cref{lem:tau-is-input} we have $\tau_P(\Pi) \times \Pi \zrr P$.
	This allows us to apply \Cref{thm:easiest-input}, where $\Pi$ is the input problem $\fI$ and $\tau_P(\Pi)$ takes the role of $\Pi$.
	That yields $\Pi \zrr \tau_P(\tau_P(\Pi))$ as claimed.
	
	For the second statement, $\tau_P(\Pi) \zrr \tau_P(\tau_P(\tau_P(\Pi)))$ follows immediately from the first.
	The other direction's proof is quite similar.
	Observe that using the first statement, we get:
	\begin{equation*}
		\Pi \times \tau_P(\tau_P(\tau_P(\Pi))) \zrr \tau_P(\tau_P(\Pi)) \times \tau_P(\tau_P(\tau_P(\Pi))) \zrr P
	\end{equation*}
	Here the latter step follows from \Cref{lem:tau-is-input}, and we obtain $\Pi \times \tau_P(\tau_P(\tau_P(\Pi))) \zrr P$ by \Cref{obs:combinerelax}.
	We can now apply \Cref{thm:easiest-input} directly to $\Pi \times \tau_P(\tau_P(\tau_P(\Pi))) \zrr P$, where $\tau_P(\tau_P(\tau_P(\Pi)))$ takes the role of $\fI$, in order to get $\tau_P(\tau_P(\tau_P(\Pi))) \zrr \tau_P(\Pi)$.
	This is everything we needed.
\end{proof}

\subsection{Inputs in the RE framework}

So far we introduced the notion of a generic input $\tau_P(\Pi)$ for two node-edge-checkable problems $\Pi$ and $P$ and proved a couple of nice results coming solely from the definition of $\tau$.
However, for proving lower bounds, we also rely on the round elimination technique, which is not yet really connected to $\tau$.
Broadly speaking, the goal of this section is to prove statements about $\tau$'s behavior in the round elimination framework.

For this purpose, we first need a lemma showing how relaxations behave w.r.t.\ round elimination.
To this end, we need the notion of an \emph{edge-based relaxation}, which can be seen as the dual to the standard relaxation notion where the role of nodes and edges has been reversed.

\begin{definition}[Edge-based relaxation]\label{def:edgerelax}
	Let $\Pi = (\Sigma_{\Pi}, \nodeconst_{\Pi}, \edgeconst_{\Pi})$ and $\hat{\Pi} = (\Sigma_{\hat{\Pi}}, \nodeconst_{\hat{\Pi}}, \edgeconst_{\hat{\Pi}})$ be two node-edge-checkable problems.
	Let $z$ denote the number of configurations in $\edgeconst_{\Pi}$ and let $\L_{i1}\s\L_{i2}$, where $1 \leq i \leq z$, denote the $z$ configurations.
	Then $\hat{\Pi}$ is called an \emph{edge-based relaxation} of $\Pi$ if there exists a mapping $f \colon \{\L_{ij} \mid 1 \leq i \leq z, 1 \leq j \leq 2\} \rightarrow \Sigma_{\hat{\Pi}}$ such that
	\begin{enumerate}[noitemsep]
		\item\label{item:edge-based1} for every configuration $\L_{i1}\s\L_{i2} \in \edgeconst_{\Pi}$, we have $f(\L_{i1})\s f(\L_{i2}) \in \edgeconst_{\hat{\Pi}}$, and
		\item\label{item:edge-based2} for every configuration $\L_{i_1j_1}\s\dots\s\L_{i_{\Delta}j_{\Delta}} \in \nodeconst_{\Pi}$, we have $f(\L_{i_1j_1})\s\dots\s f(\L_{i_{\Delta}j_{\Delta}}) \in \nodeconst_{\hat{\Pi}}$.
	\end{enumerate}
	We call such a mapping $f$ an \emph{edge-based relaxation function from $\Pi$ to $\hat{\Pi}$}.
\end{definition}

Note that unlike port-local relaxations, $f$ is not simply a function mapping a label to a label.
Instead, $f$ may take the indices $i$ and $j$ of the given edge configuration into account.
This means that $f$ ``knows'' about the configuration assigned to the respective edge when it has to output a new label at a port.

We are now ready to prove a statement that, informally speaking, shows that the two notions of relaxing a problem and applying the function $\rere(\re(\cdot))$ commutate.
\begin{lemma}[RE commutativity]\label{lem:commutative}
	Let $\Pi$ and $\hat{\Pi}$ be two node-edge-checkable problems. Then all of these hold:
	\begin{enumerate}[noitemsep]
		\item If $\hat{\Pi}$ is a relaxation of $\Pi$, then there exists an edge-based relaxation from $\re(\Pi)$ to $\re(\hat{\Pi})$. \label{item:cummutative-1}
		\item If $\hat{\Pi}$ is an edge-based relaxation of $\Pi$, then there is a relaxation from $\rere(\Pi)$ to $\rere(\hat{\Pi})$.\label{item:cummutative-2}
		\item If $\hat{\Pi}$ is a relaxation of $\Pi$, then $\fQ^i(\hat{\Pi})$ is a relaxation of $\fQ^i(\Pi)$ for all integers $i$.\label{item:cummutative-3}
	\end{enumerate}
\end{lemma}
\begin{proof}
	We start with Statement~\ref{item:cummutative-1}, therefore let $f$ be a relaxation function from $\Pi$ to $\hat{\Pi}$.
	Consider an arbitrary edge configuration $\S_1\s\S_2 \in \edgeconst_{\re(\Pi)}$.
	Observe that $\S_1, \S_2$ are sets of labels from $\Sigma_{\Pi}$.
	Now, for each $j \in \{ 1, 2 \}$, do the following.
	
	Let $\L_{j1}, \dots, \L_{jk_j}$ denote the labels contained in $\S_j$.
	For each $1 \leq x \leq k_j$, the label $\L_{jx}$ may occur in different configurations from $\nodeconst_{\Pi}$; let $\fL_{jx}$ denote the set of labels that these different occurrences of $\L_{jx}$ are mapped to by $f$.
	Set further $S'_j := \bigcup_{1 \leq x \leq k_j} \fL_{jx}$.
	
	Observe that, by the definition of $\re(\cdot)$, the edge configuration $\L_{1x_1}\s\L_{2x_2}$ is contained in $\edgeconst_{\Pi}$ for each $1 \leq x_1 \leq k_1$ and $1 \leq x_2 \leq k_2$.
	By the fact that $f$ is a relaxation and the definition of $S'_j$, it follows that $\L'_1\s\L'_2 \in \edgeconst_{\hat{\Pi}}$ for each $\L'_1 \in S'_1$ and $\L'_2 \in S'_2$.
	This implies, again by the definition of $\re(\cdot)$, that there exists a (not necessarily unique) edge configuration $\hat{\S}_1\s\hat{\S}_2 \in \edgeconst_{\re(\hat{\Pi})}$ such that $S'_1 \subseteq \hat{\S}_1$ and $S'_2 \subseteq \hat{\S}_2$. 
	
	Now, for the considered edge configuration $\S_1\s\S_2 \in \edgeconst_{\re(\Pi)}$, set $g(\S_1) := \hat{\S}_1$ and $g(\S_2) := \hat{\S}_2$.
	As the considered edge configuration was chosen arbitrarily from $\edgeconst_{\re(\Pi)}$, this concludes the definition of $g$.
	From the construction of $g$, we immediately obtain that Condition~\ref{item:edge-based1} of~\Cref{def:edgerelax} is satisfied.
	In the following, we show that $g$ also satisfies Condition~\ref{item:edge-based2}.
	
	Consider an arbitrary node configuration $\S_1\s\dots\s\S_{\Delta} \in \nodeconst_{\re(\Pi)}$ and let $\hat{\S}_1, \dots, \hat{\S}_{\Delta}$ be labels that $\S_1, \dots, \S_{\Delta}$, respectively, are mapped to by $g$.
	By the definition of $\re(\cdot)$, we know that there exists some configuration $\L_1\s\dots\s\L_{\Delta} \in \nodeconst_{\Pi}$ such that $\L_y \in \S_y$ for each $1 \leq y \leq \Delta$.
	Let $\L'_1, \dots, \L'_{\Delta}$, respectively, be the labels that the labels $\L_1, \dots, \L_{\Delta}$ of the configuration $\L_1\s\dots\s\L_{\Delta}$ are mapped to by $f$.
	By the definition of $f$, this in particular implies that $\L'_1\s\dots\s\L'_{\Delta} \in \nodeconst_{\hat{\Pi}}$.
	By the construction of $g$, we know that there exist subsets $S'_1, \dots, S'_{\Delta}$ of $\hat{\S}_1, \dots, \hat{\S}_{\Delta}$, respectively, such that $\L'_y \in S'_y$ for each $1 \leq y \leq \Delta$.
	By the definition of $\re(\cdot)$, it follows that $\hat{\S}_1\s\dots\s\hat{\S}_{\Delta} \in \nodeconst_{\re(\hat{\Pi})}$, as desired.
	We conclude that $g$ also satisfies Condition~\ref{item:edge-based2} of~\Cref{def:edgerelax}.
	Hence, $g$ is indeed an edge-based relaxation function, proving our first statement.
	
	Statement~\ref{item:cummutative-2} is dual to statement~\ref{item:cummutative-1}, where the roles of nodes and edges are swapped.
	Here we assume that $g$ is an edge-based relaxation function from $\Pi$ to $\hat{\Pi}$.
	
	Consider an arbitrary node configuration $\Z_1\s\dots\s\Z_{\Delta} \in \nodeconst_{\rere(\Pi)}$.
	For each $y \in \{ 1, \dots, \Delta \}$, do the following.
	
	Let $\S_{y1}, \dots, \S_{yk_y}$ denote the labels contained in $\Z_y$.
	For each $1 \leq x \leq k_y$, the label $\S_{yx}$ may occur in different configurations from $\edgeconst_{\Pi}$; let $\fS_{yx}$ denote the set of labels that these different occurrences of $\S_{yx}$ are mapped to by $g$.
	Set further $Z'_y := \bigcup_{1 \leq x \leq k_y} \fS_{yx}$.
	
	Observe that, by the definition of $\rere(\cdot)$, the edge configuration $\S_{1x_1}\s\dots\s\S_{\Delta x_{\Delta}}$ is contained in $\nodeconst_{\Pi}$ for each $(x_1, \dots x_{\Delta}) \in \{ 1, \dots, k_1 \} \times \dots \times \{ 1, \dots, k_{\Delta} \}$.
	By the fact that $g$ is an edge-based relaxation and the definition of $Z'_y$, it follows that $\S'_1\s\dots\s\S'_{\Delta} \in \nodeconst_{\hat{\Pi}}$ for each $(\S'_1, \dots, \S'_{\Delta}) \in Z'_1 \times \dots \times Z'_{\Delta}$.
	This implies, again by the definition of $\rere(\cdot)$, that there is a node configuration $\hat{\Z}_1\s\dots\s\hat{\Z}_{\Delta} \in \nodeconst_{\rere(\hat{\Pi})}$ such that $Z'_1, \dots, Z'_{\Delta}$ are subsets of $\hat{\Z}_1, \dots, \hat{\Z}_{\Delta}$, respectively. 
	
	Now, for the considered node configuration $\Z_1\s\dots\s\Z_{\Delta} \in \nodeconst_{\rere(\Pi)}$, set $h(\Z_y) := \hat{\Z}_y$ for each $1 \leq y \leq \Delta$.
	As the considered node configuration was chosen arbitrarily from $\nodeconst_{\rere(\Pi)}$, this concludes the definition of $h$.
	From the construction of $h$, we immediately obtain that Condition~\ref{item:node-based1} of~\Cref{def:relax} is satisfied.
	In the following, we show that $h$ also satisfies Condition~\ref{item:node-based2}.
	
	Consider an arbitrary edge configuration $\Z_1\s\Z_2 \in \edgeconst_{\rere(\Pi)}$ and let $\hat{\Z}_1, \hat{\Z}_2$ be labels that $\Z_1, \Z_2$, respectively, are mapped to by $h$.
	By the definition of $\rere(\cdot)$, we know that there exists some configuration $\S_1\s\S_2 \in \edgeconst_{\Pi}$ such that $\S_j \in \Z_j$ for each $1 \leq j \leq 2$.
	Let $\S'_1, \S'_2$, respectively, be the labels that the labels $\S_1, \S_2$ of the configuration $\S_1\s\S_2$ are mapped to by $g$.
	By the definition of $g$, this in particular implies that $\S'_1\s \S'_2 \in \edgeconst_{\hat{\Pi}}$.
	By the construction of $h$, we know that there exist subsets $Z'_1, Z'_2$ of $\hat{\Z}_1, \hat{\Z}_2$, respectively, such that $\S'_j \in Z'_j$ for each $1 \leq j \leq 2$.
	By the definition of $\rere(\cdot)$, it follows that $\hat{\Z}_1\s\hat{\Z}_2 \in \edgeconst_{\rere(\hat{\Pi})}$, as desired.
	We conclude that $h$ also satisfies Condition~\ref{item:node-based2} of~\Cref{def:relax}.
	Hence, $h$ is indeed a relaxation function, which concludes the proof of the second part.
	
	For Statement~\ref{item:cummutative-3}, it suffices to prove the statement for $i=1$, it then holds for all $i$ by induction.
	If $\Pi$ is relaxable to $\hat{\Pi}$, then we know by part~\ref{item:cummutative-1} that there exists an edge-based relaxation from $\re(\Pi)$ to $\re(\hat{\Pi})$.
	This allows to apply part~\ref{item:cummutative-2}, proving that $\rere(\re(\Pi)) = \fQ(\Pi)$ is relaxable to $\rere(\re(\hat{\Pi})) = \fQ(\hat{\Pi})$.
\end{proof}

We are now ready to connect $\tau$ to the round elimination framework.
To be precise, we will now prove that $\fQ(\tau_P(\Pi))$ i.e., the problem obtained by performing one round elimination step on $\tau_P(\Pi)$, can always be relaxed to $\tau_{\fQ(P)}(\Pi)$.
This holds without any further assumptions on $P$ and $\Pi$, however the application to the case where $P$ is a fixed point is particularly important.
Broadly speaking, $P$ being a fixed point allows us to replace the appearing $\fQ(P)$ with $P$, which yields a relaxation $\fQ(\tau_P(\Pi)) \zrr \tau_P(\Pi)$.
But this just means that $\tau_P(\Pi)$ itself is a fixed point, which we will formally see in \Cref{cor:tau-is-FP}.

We will now prove the statement $\fQ(\tau_P(\Pi)) \zrr \tau_{\fQ(P)}(\Pi)$ as mentioned above.
The proof requires delving into the definitions of $\re(\cdot)$ and $\rere(\cdot)$ making it very technical, luckily it is the only time we need to unpack these definitions.
\begin{lemma}\label{lem:tau-and-RE}
	For any two node-edge-checkable problems $P$ and $\Pi$ we have
	\begin{equation*}
		\fQ(\tau_P(\Pi)) \zrr \tau_{\fQ(P)}(\Pi).
	\end{equation*}
\end{lemma}

\begin{proof}
	To get a better overview, we visualize the present node-edge-checkable problems in a small diagram, see \Cref{fig:diag-tau-and-RE}.
	A ``$\rightarrow$'' means that one problem can be relaxed to another.
	Two problems pointing at a $\times$-node mean that their product can be relaxed to the problem the $\times$-node points to.
	
	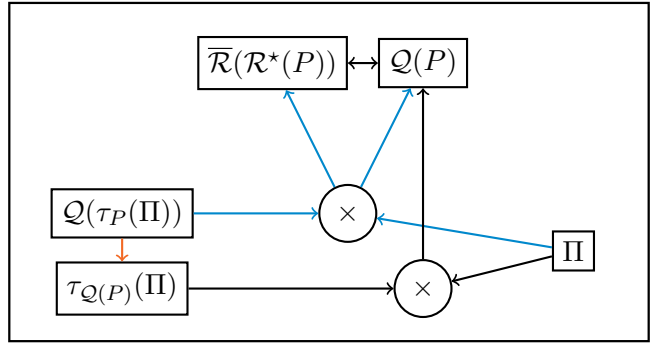
\begin{figure}
		\centering
		\begin{tikzpicture}[thick]
			\useasboundingbox (-4.5,-1.6) rectangle (4,3);
			\node[shape=rectangle,draw=black] (QP) at (1,2) {$\fQ(P)$};
			\node[shape=rectangle,draw=black] (QstarP) at (-1,2) {$\rere(\rest(P))$};
			\node[shape=rectangle,draw=black] (Pi) at (3,-0.5) {$\Pi$};
			\node[shape=rectangle,draw=black] (Qtau) at (-3,0) {$\fQ(\tau_P(\Pi))$};
			\node[shape=rectangle,draw=black] (tauQ) at (-3,-1) {$\tau_{\fQ(P)}(\Pi)$};
			\node[shape=circle,draw=black] (C1) at (0,0) {$\times$};
			\node[shape=circle,draw=black] (C2) at (1,-1) {$\times$};
			\draw[draw = myblue] [->] (Qtau) to (C1);
			\draw[draw = myblue] [->] (Pi) to (C1);
			\draw[draw = myblue] [->] (C1) to (QP);
			\draw[draw = myblue] [->] (C1) to (QstarP);
			\draw [->] (tauQ) to (C2);
			\draw [->] (Pi) to (C2);
			\draw [->] (C2) to (QP);
			\draw[draw = myorange] [->] (Qtau) to (tauQ);
			\draw [<->] (QstarP) to (QP);
			
			\begin{scope}[on background layer]
				\draw[line width=0.3mm, black] (-4.5,-1.7) rectangle (4,2.8);
			\end{scope}
		\end{tikzpicture}
		\caption{Diagram to visualize \Cref{lem:tau-and-RE}} \label{fig:diag-tau-and-RE}
	\end{figure}
	At first we will explain why the black relaxations in \Cref{fig:diag-tau-and-RE} hold.
	Observe that $\tau_{\fQ(P)}(\Pi) \times \Pi \zrr \fQ(P)$ follows directly from \Cref{lem:tau-is-input}.
	Furthermore, $\rest(P)$ can be relaxed to $\re(P)$ using an edge-based relaxation.
	The only thing each edge needs to do is to enlarge its labels (which are sets) so that it ends up with a maximal configuration of $\rest(P)$, which by definition is contained in $\re(P)$.
	Since $\re(P)$ is a more restrictive version of $\rest(P)$, it is clear that, conversely, $\re(P)$ can also be relaxed to $\rest(P)$, using an edge-based relaxation that just does nothing.
	Thus we can apply part~\ref{item:cummutative-2} of \Cref{lem:commutative} to obtain that $\rere(\rest(P))$ and $\rere(\re(P)) = \fQ(P)$ are equivalent in the sense of \Cref{def:eq-problems}.
	This concludes the black relaxations.
	
	The orange relaxation $\fQ(\tau_P(\Pi)) \zrr \tau_{\fQ(P)}(\Pi)$ is the one we want to prove.
	
	Now consider the blue relaxations and observe that $\fQ(\tau_P(\Pi)) \times \Pi \zrr \fQ(P)$ implies $\fQ(\tau_P(\Pi)) \zrr \tau_{\fQ(P)}(\Pi)$ using \Cref{thm:easiest-input}.
	Thus it suffices to prove $\fQ(\tau_P(\Pi)) \times \Pi \zrr \fQ(P)$.
	But since $\fQ(P)$ and $\rere(\rest(P))$ are equivalent, we can instead finish the proof by showing the other blue relaxation
	\begin{equation*}
		\fQ(\tau_P(\Pi)) \times \Pi \zrr \rere(\rest(P))
	\end{equation*}
	which we will do in the following.
	
	Therefore, we first define a function $T \colon \sig{\fQ(\tau_P(\Pi))} \times \sig{\Pi} \rightarrow \sig{\rere(\rest(P))}$ as follows:
	For
	\begin{equation*}
		\S = \{\{f_{11}, \dots, f_{1b_1}\}, \{f_{21}, \dots, f_{2b_2}\}, \dots, \{f_{a1}, \dots, f_{ab_a}\}\} \in \sig{\rere(\re(\tau_P(\Pi)))}
	\end{equation*}
	and some $\L \in \sig{\Pi}$ we set 
	\begin{equation*}
		T(\S,\L) := \left\{ \bigcup_{k=1}^{b_1} f_{1k}(\L), \bigcup_{k=1}^{b_2} f_{2k}(\L), \dots, \bigcup_{k=1}^{b_a} f_{ak}(\L) \right\}.
	\end{equation*}
	One can imagine each node applies this functions to all its label pairs, however that is not quite the relaxation yet.
	For now let $(\S_1, \L_1) \s \dots \s (\S_{\Delta}, \L_{\Delta}) \in \nod{\fQ(\tau_P(\Pi)) \times \Pi}$ be a node configuration.
	Suppose that each $\S_i$ is given by
	\begin{equation*}
		\S_i = \{\{f^{i}_{11}, \dots, f^{i}_{1b_1^{i}}\}, \dots, \{f^{i}_{a^{i}1}, \dots, f^{i}_{a^{i}b_{a^{i}}^{i}}\}\}.
	\end{equation*}
	We have $\S_1 \s \dots \s \S_{\Delta} \in \nod{\rere(\re(\tau_P(\Pi)))}$ and can thus apply the definitions of $\rere(\cdot)$ to obtain 
	\begin{equation*}
		\{f^1_{j^11}, \dots, f^1_{j^1b^1_{j^1}}\} \s \dots \s \{f^{\Delta}_{j^{\Delta}1}, \dots, f^{\Delta}_{j^{\Delta}b^{\Delta}_{j^{\Delta}}}\} \in \re(\tau_P(\Pi))
	\end{equation*}
	for all indices $1 \leq j^{i} \leq a^{i}$.
	We now choose some arbitrary indices $j^1, \dots, j^{\Delta}$.
	By definition of $\re(\cdot)$ we can w.l.o.g.\ assume $f^1_{j^11} \s \dots \s f^{\Delta}_{j^{\Delta}1} \in \nod{\tau_P(\Pi)}$ for these indices $j^{i}$.
	By definition of $\tau$ this implies $f^1_{j^11}(\L_1) \s \dots \s f^{\Delta}_{j^{\Delta}1}(\L_{\Delta}) \in \nod{\rest(P)}$.
	Following the definition of $\rest(P)$ we can enlarge these sets to get
	\begin{equation*}
		\bigcup_{k=1}^{b^1_{j^1}} f^1_{j^1k}(\L_1) \s \dots \s \bigcup_{k=1}^{b^{\Delta}_{j^{\Delta}}} f^{\Delta}_{j^{\Delta}k}(\L_{\Delta}) \in \nod{\rest(P)}.
	\end{equation*}
	Since the indices $1 \leq j^{i} \leq a^{i}$ were arbitrary, the latter relation holds for all possible $j^{i}$.
	We recall that $T(\S_i, \L_i)$ by definition is given by
	\begin{equation*}
		T(\S_i, \L_i) = \left\{ \bigcup_{k=1}^{b^{i}_{j^{i}}} f^{i}_{j^{i}k}(\L_i); 1 \leq j^{i} \leq a^{i} \right\}.
	\end{equation*}
	Hence we can conclude:
	\begin{equation*}
		\forall \S_1' \in T(\S_1, \L_1), \dots, \forall \S_{\Delta}' \in T(\S_{\Delta}, \L_{\Delta}) \colon \S_1' \s \dots \s \S_{\Delta}' \in \nod{\rest(P)}
	\end{equation*}
	Now by definition of $\rere(\cdot)$ there exist supersets $T'(\S_i, \L_i) \supseteq T(\S_i, \L_i)$ for all $1 \leq i \leq \Delta$ such that $T'(\S_1, \L_1) \s \dots \s T'(\S_{\Delta}, \L_{\Delta}) \in \rere(\rest(P))$, which a node knowing the $T(\S_i, \L_i)$ can compute.
	For our relaxation, each node chooses these $T'(\S_i, \L_i)$ as its new labels.
	The definition of $T'$ immediately proves condition~\ref{item:node-based1} of \Cref{def:relax}, so we are left to prove condition~\ref{item:node-based2}.
	Therefore, let $(\S_1, \L_1) \s (\S_2, \L_2) \in \edg{\fQ(\tau_P(\Pi)) \times \Pi}$ be an arbitrary edge configuration.
	Similar to above, we set
	\begin{align*}
		\S_1 &= \{\{f_{11}, \dots, f_{1b_1}\}, \{f_{21}, \dots, f_{2b_2}\}, \dots, \{f_{a1}, \dots, f_{ab_a}\}\},\\
		\S_2 &= \{\{g_{11}, \dots, g_{1d_1}\}, \{g_{21}, \dots, g_{2d_2}\}, \dots, \{g_{c1}, \dots, g_{cd_c}\}\}.
	\end{align*}
	Since we have $\S_1 \s \S_2 \in \edg{\rere(\re(\tau_P(\Pi)))}$, we can use the definition of $\rere(\cdot)$ and w.l.o.g.\ assume \[\{f_{11}, \dots, f_{1b_1}\} \s \{g_{11}, \dots, g_{1d_1}\} \in \edg{\re(\tau_P(\Pi))}.\]
	We then get:
	\begin{align*}
		& \{f_{11}, \dots, f_{1b_1}\} \s \{g_{11}, \dots, g_{1d_1}\} \in \edg{\re(\tau_P(\Pi))}\\
		{} \xRightarrow{\text{Def.\ } \edg{\re}} {} \,\, & \forall 1 \leq i \leq b_1 \forall 1 \leq j \leq d_1 \colon f_{1i} \s g_{1j} \in \edg{\tau_P(\Pi)}\\
		{} \xRightarrow{\text{Def.\ } \edg{\tau}} {} \,\, & \forall 1 \leq i \leq b_1 \forall 1 \leq j \leq d_1 \colon f_{1i}(\L_1) \s g_{1j}(\L_2) \in \edg{\rest(P)}\\
		{} \xRightarrow{\text{Def.\ } \edg{\rest}} {} \,\, & \forall 1 \leq i \leq b_1 \forall 1 \leq j \leq d_1 \forall p_1 \in f_{1i}(\L_1) \forall p_2 \in g_{1j}(\L_2) \colon p_1 \s p_2 \in \edg{P}\\
		{} \Longrightarrow {} \,\, & \forall p_1 \in \bigcup_{i=1}^{b_1} f_{1i}(\L_1) \forall p_2 \in \bigcup_{j=1}^{d_1} g_{1j}(\L_2) \colon p_1 \s p_2 \in \edg{P}\\
		{} \xRightarrow{\text{Def.\ } \edg{\rest}} {} \,\, & \bigcup_{i=1}^{b_1} f_{1i}(\L_1) \s \bigcup_{j=1}^{d_1} g_{1j}(\L_2) \in \edg{\rest(P)}.
	\end{align*}
	Observe that by the definitions of $T$ and $T'$ we now have $\bigcup_{i=1}^{b_1} f_{1i}(\L_1) \in T(\S_1, \L_1) \subseteq T'(\S_1, \L_1)$ and $\bigcup_{j=1}^{d_1} g_{1j}(\L_2) \in T(\S_2, \L_2) \subseteq T'(\S_2, \L_2)$.
	So we finally obtain $T'(\S_1, \L_1) \s T'(\S_2, \L_2) \in \edg{\rere(\rest(P))}$ which is what we needed to finish the proof.
\end{proof}
\Cref{lem:tau-and-RE} has no further assumptions on $P$ and $\Pi$, however as we already mentioned, the case where $P$ is a fixed point is particularly nice.
Informally speaking, it allows us to replace $\fQ(P)$ with $P$, which makes $\tau_P(\Pi)$ a fixed point.
Indeed we have:
\begin{corollary}\label{cor:tau-is-FP}
	Let $P, \fI$ and $\fF$ be node-edge-checkable problems satisfying the following:
	\begin{enumerate}[noitemsep]
		\item $\fQ(\fF) \times \fI \zrr \fF$, meaning that $\fF$ is a generalized fixed point with input $\fI$.
		\item $P \zrr \fI$.
	\end{enumerate}
	Then $\tau_{\fF}(P)$ is a fixed point, even in the setting without input.
\end{corollary}
\begin{proof}
	We immediately get
	\begin{equation*}
		\fQ \left( \tau_{\fF}(P) \right) \times P 
		\underset{\text{\Cref{lem:tau-and-RE}}}{\zrr} \tau_{\fQ(\fF)}(P) \times P
		\underset{\text{\Cref{lem:tau-is-input}}}{\zrr} \fQ(\fF) \times P
		\zrr \fQ(\fF) \times \fI
		\zrr \fF,
	\end{equation*}
	where the last two steps follow from the two assumptions.
	We can combine these relaxations using \Cref{obs:combinerelax} to obtain $\fQ \left( \tau_{\fF}(P) \right) \times P \zrr \fF$.
	This allows us to apply \Cref{thm:easiest-input} to get $\fQ(\tau_{\fF}(P)) \zrr \tau_{\fF}(P)$, proving that $\tau_{\fF}(P)$ is a fixed point.
\end{proof}
What this means is that the fixed point property is carried over from $\fF$ to $\tau_{\fF}(P)$.
We will make use of this in the next section.
Further, it is worth noting that while generating new fixed points using \Cref{cor:tau-is-FP}, we can get rid of a previously required input.
Observe that while $\fF$ is a generalized fixed point requiring $\fI$ as input, $\tau_{\fF}(P)$ is a fixed point in the setting without input.
The input is, of course, hiding in the assumption $P \zrr \fI$, however broadly speaking, this can easily be ensured by replacing $P$ with $P \times \fI$, which does not make $P$ harder in the setting where $\fI$ is a given input.

\subsection{Constructing fixed point relaxations}

We have now collected everything necessary to prove a rather surprising result.
Broadly speaking, it states that whenever a node-edge-checkable problem $\Pi$ allows a relaxation with input to a nontrivial generalized fixed point, there also exists a nontrivial fixed point relaxation of $\Pi$.
\begin{theorem}\label{thm:worthless-input}
	For two node-edge-checkable problems $\Pi$ and $\fI$ the following two statements are equivalent:
	\begin{enumerate}
		\item \label{item:worthless-input-1} There exists a nontrivial generalized fixed point $\fF$ given $\fI$ which is a relaxation of $\Pi$ in the setting with input $\fI$:
		\begin{equation*}
			\Pi \times \fI \zrr \fF, \quad \quad \fI \nzrr \fF, \quad \quad \fQ(\fF) \times \fI \zrr \fF.
		\end{equation*}
		\item \label{item:worthless-input-2} There exists a fixed point relaxation $\fG$ of $\Pi$ in the setting without input, which is nontrivial even given the input $\fI$:
		\begin{equation*}
			\Pi \zrr \fG, \quad \quad \fI \nzrr \fG, \quad \quad \fQ(\fG) \zrr \fG.
		\end{equation*}
	\end{enumerate}
\end{theorem}
\begin{proof}
	Statement 2 clearly implies 1, as we can just choose $\fF := \fG$.
	For the other direction we assume that there exists a problem $\fF$ satisfying the properties listed in statement 1.
	We set $\fG := \tau_{\fF}(\fI)$ and prove that $\fG$ fulfills all required properties.
	First we note that
	\begin{equation*}
		\Pi \times \fI \zrr \fF \,\, \xRightarrow{\text{\Cref{thm:easiest-input}}} \,\, \Pi \zrr \fG,
	\end{equation*}
	where we used $\fF$'s first property.
	Now assume $\fI \zrr \fG$.
	This would imply
	\begin{equation*}
		\fI \zrr \fI \times \fG = \fI \times \tau_{\fF}(\fI) \zrr \fF
	\end{equation*}
	by \Cref{lem:tau-is-input} contradicting $\fI \nzrr \fF$.
	Thus we must have $\fI \nzrr \fG$.
	Lastly, we have $\fQ(\fG) \zrr \fG$ by \Cref{cor:tau-is-FP}, where we choose $P$ to just be $\fI$.
\end{proof}
Here one should think about $\Pi$ as some node-edge-checkable problem for which we want to prove hardness.
$\fI$ is an input problem that may have been introduced artificially to find such a proof.
If we were interested in $\Pi$ in the setting given some input, $\fI$ can be chosen as some node-edge-checkable problem that is 0-round relaxable to this input.

With this interpretation in mind, \enumerateref{thm:worthless-input}{item:worthless-input-1} means that the hardness of $\Pi$ can be proven using a generalized fixed point relaxation with (artificial) input.
Note that the input may be used in the relaxation of $\Pi$ to the fixed point $\fF$ \emph{and} in the relaxation from $\fQ(\fF)$ to $\fF$ making $\fF$ a generalized fixed point.
In contrast, part \ref{item:worthless-input-2} states that the hardness of $\Pi$ can be proven using a fixed point relaxation without input.
Even more, we can ``force'' $\fG$ in strategy \ref{item:worthless-input-2} to not be easier than our artificially introduced input.

Recall that the first is clearly a generalization of the latter, since we can just introduce a trivial input.
However, \Cref{thm:worthless-input} proves that these strategies are equally powerful in the sense that they allow to prove hardness for the same problems $\Pi$.
This may be surprising, but note that it is actually a necessary condition for the correctness of the possible conjecture that there exists a nontrivial fixed point relaxation for each intermediate-hard node-edge-checkable problem.

We further note that given a node-edge-checkable problem $\Pi$ that we want to solve given a fixed input $\fI$, the \enumerateref{thm:worthless-input}{item:worthless-input-1} is nothing else then finding a nontrivial generalized fixed point relaxation in the setting with this given input.
It asks to find a node-edge-checkable problem $\fF$, which is a relaxation of $\Pi$ (here also adding the input $\fI$), nontrivial given $\fI$ and a generalized fixed point in the setting with input $\fI$.

Further, it is worth noticing that the proof of \Cref{thm:worthless-input} is constructive.
When given a node-edge-checkable problem $\Pi$ and a nontrivial generalized fixed point relaxation $\fF$ under a certain input $\fI$, a nontrivial fixed point relaxation of $\Pi$ without input is given by $\tau_{\fF}(\fI)$.

Lastly, we reconsider the setting where a fixed point candidate $\fF$ has already been chosen, and we are left to find a suitable input problem $\fI$ to ensure \enumerateref{thm:worthless-input}{item:worthless-input-1}.
Since we need to ensure $\Pi \times \fI \zrr \fF$ and $\fI \nzrr \fF$, a natural good candidate would be $\fI = \tau_{\fF}(\Pi)$, since it is the easiest problem satisfying $\Pi \times \fI \zrr \fF$ by \Cref{lem:tau-is-input} and \Cref{thm:easiest-input}.
Following the proof of \Cref{thm:worthless-input}, the double-input $\tau_{\fF}(\tau_{\fF}(\Pi))$ would then be a nontrivial fixed point relaxation of $\Pi$ in the setting without input.
This suggests that whenever there exists a fixed point relaxation of $\Pi$ given some input, we can choose that input as $\tau_{\fF}(\Pi)$ and furthermore obtain that $\tau_{\fF}(\tau_{\fF}(\Pi))$ is a nontrivial fixed point relaxation of $\Pi$ without input.
The following corollary formally proves this statement.
\begin{corollary}\label{cor:double-tau-relax}
	For a node-edge-checkable problem $\Pi$ and a fixed point $\fF$, the following three statements are equivalent:
	\begin{enumerate}[noitemsep]
		\item Problem $\tau_{\fF}(\Pi)$ satisfies $\tau_{\fF}(\Pi) \nzrr \fF$.\label{item:tau-as-input}
		\item There exists a node-edge-checkable problem $\fI$ such that $\Pi \times \fI \zrr \fF$ and $\fI \nzrr \fF$.\label{item:general-input}
		\item Problem $\tau_{\fF}(\tau_{\fF}(\Pi))$ is a nontrivial fixed point relaxation of $\Pi$.\label{item:double-tau}
	\end{enumerate}
\end{corollary}
\begin{proof}
	It is easy to see that \ref{item:tau-as-input} implies \ref{item:general-input}, since we can choose $\fI := \tau_{\fF}(\Pi)$ and obtain $\Pi \times \fI \zrr \fF$ directly from \Cref{lem:tau-is-input}.
	
	Conversely, $\Pi \times \fI \zrr \fF$ implies $\fI \zrr \tau_{\fF}(\Pi)$ by \Cref{thm:easiest-input}.
	Thus $\tau_{\fF}(\Pi) \zrr \fF$ would imply $\fI \zrr \fF$ by \Cref{obs:combinerelax}, contradicting the assumption $\fI \nzrr \fF$.
	Hence we must have $\tau_{\fF}(\Pi) \nzrr \fF$, proving that \ref{item:general-input} implies \ref{item:tau-as-input}.
	
	To prove \ref{item:tau-as-input} implies \ref{item:double-tau}, observe that our assumptions $\fQ(\fF) \zrr \fF$ and $\tau_{\fF}(\Pi) \nzrr \fF$ together with \Cref{lem:tau-is-input} correspond to the assumptions in \enumerateref{thm:worthless-input}{item:worthless-input-1}, where $\tau_{\fF}(\Pi)$ takes the role of $\fI$.
	According to the proof of \Cref{thm:worthless-input}, $\tau_{\fF}(\tau_{\fF}(\Pi))$ is a nontrivial fixed point relaxation of~$\Pi$.
	
	Lastly, we prove the contraposition $\neg$ \ref{item:tau-as-input} implies $\neg$ \ref{item:double-tau} and hence assume $\tau_{\fF}(\Pi) \zrr \fF$.
	This immediately yields $\tau_{\fF}(\Pi) \times P \zrr \fF$ for the trivial node-edge-checkable problem $P$, and thus $P \zrr \tau_{\fF}(\tau_{\fF}(\Pi))$ by \Cref{thm:easiest-input}.
	It follows directly that $\tau_{\fF}(\tau_{\fF}(\Pi))$ is also trivial.
\end{proof}

\section{SSO with SO as input}\label{sec:sso-with-so}

In this section, our goal is to prove the following theorem.

\begin{theorem}\label{thm:sso-with-so}
	Let $\Pi$ denote the sinkless and sourceless orientation problem.
	Then, in the setting where a sinkless orientation is given as input, $\Pi$ satisfies the following two properties.
	\begin{enumerate}
		\item\label{prop:lb} Solving $\Pi$ with a deterministic algorithm requires $\Omega(\log n)$ rounds and solving $\Pi$ with a randomized algorithm requires $\Omega(\log \log n)$ rounds.
		\item\label{prop:fp} There is no nontrivial fixed point relaxation of $\Pi$. Moreover, there is no nontrivial generalized fixed point relaxation of $\Pi$.
	\end{enumerate}
\end{theorem}

The sinkless and sourceless orientation problem (SSO) is the problem of orienting the edges such that no node is a sink and no node is a source, i.e., such that each node has at least one outgoing and at least one incoming edge.
For simplicity, in the remainder of this section, we focus on the case of \emph{$3$-regular} SSO.
This already suffices to provide our impossibility results (and generalizes straightforwardly to any $\Delta > 3$).

Analogously to how sinkless orientation can be phrased as a node-edge-checkable problem (see~\Cref{sec:Preliminaries}), SSO can be easily phrased as a node-edge-checkable problem by labeling half-edges with $\lI$ (incoming) and $\lO$ (outgoing), which naturally leads to the following node and edge constraints: the only (condensed) configuration in the node constraint of SSO is $\lI \s \lO \s [\lI\s\lO]$, the only configuration in the edge constraint of SSO is $\lI \s \lO$.

Now we turn our attention towards proving~\Cref{thm:sso-with-so}.
We start with characterizing the problems in the round elimination sequence $\Pi, \fQ(\Pi), \fQ(\fQ(\Pi)), \dots$ for the case of $\Pi$ being SSO.

\begin{lemma}\label{lem:characterization}
	Let $\Pi$ be the problem of sinkless and sourceless orientation and $k$ a nonnegative integer.
	Then the node and edge constraints of $\fQ^k(\Pi)$ are:
	\begin{align*}
		&\text{If $k = 0$:}&
		\nodeconst_{\fQ^k(\Pi)}\colon\quad& \B~\C~[\B\s\C] \\[-1pt]
		&&\edgeconst_{\fQ^k(\Pi)}\colon\quad& \B~\C, \\[3pt]
		&\text{If $k = 1$:}&
		\nodeconst_{\fQ^k(\Pi)}\colon\quad& \B~\C~\D \\[-1pt]
		&&\edgeconst_{\fQ^k(\Pi)}\colon\quad& [\B\s\D]~[\C\s\D], \\[3pt]
		&\text{If $k \geq 2$:}&
		\nodeconst_{\fQ^k(\Pi)}\colon\quad& \B~\C~\D_k\\[-4pt]
		&&&\A_i~\D_i~\D_k \text{ for each } 1 \leq i \leq k - 1\\[0pt]
		&&\edgeconst_{\fQ^k(\Pi)}\colon\quad& [\B\s\D_1\s\D_2\s\dots\s\D_k]~[\C\s\D_1\s\D_2\s\dots\s\D_k]\\[-4pt]
		&&&\A_i~\D_j \text{ for each } 1 \leq i < j \leq k.
	\end{align*}
\end{lemma}
\begin{proof}
	We prove the lemma for the cases $k = 0$, $k = 1$, and $k \geq 2$ in that order.
	The case $k = 0$ follows directly from the definition of sinkless and sourceless orientation by renaming the labels.
	
	Now consider the case $k = 1$.
	It is straightforward to verify that, by applying $\re(\cdot)$ to $\Pi$, we obtain
	\begin{align*}
		\edgeconst_{\re(\Pi)} &= \{ \{\B\}~\{\C\} \},\\
		\Sigma_{\re(\Pi)} &= \{ \{\B\}, \{\C\} \}, \textrm{ and}\\
		\nodeconst_{\re(\Pi)} &= \{ \{\B\}~\{\B\}~\{\C\},\quad\{\B\}~\{\C\}~\{\C\} \}.
	\end{align*}
	Again, it is straightforward to apply $\rere(\cdot)$ to $\re(\Pi)$ (where, for computing $\edgeconst_{\rere(\re(\Pi))}$, we can make use of~\Cref{obs:easy-exists}), and we obtain
	\begin{align*}
		\nodeconst_{\rere(\re(\Pi))} &= \{ \{\{\B\}\}~\{\{\C\}\}~\{\{\B\},\{\C\}\} \},\\
		\Sigma_{\rere(\re(\Pi))} &= \{ \{\{\B\}\}, \{\{\C\}\}, \{\{\B\},\{\C\}\} \}, \textrm{ and}\\
		\edgeconst_{\rere(\re(\Pi))} &= \{ \{\{\B\}\}~\{\{\C\}\}, \quad \{\{\B\}\}~\{\{\B\},\{\C\}\}, \quad \{\{\C\}\}~\{\{\B\},\{\C\}\}, \quad \{\{\B\},\{\C\}\}~\{\{\B\},\{\C\}\} \}.
	\end{align*}
	By performing the renaming
	\begin{align*}
		\{\{\B\}\} &\rightarrow \B\\
		\{\{\C\}\} &\rightarrow \C\\
		\{\{\B\}, \{\C\}\} &\rightarrow \D,
	\end{align*}
	we obtain
		\begin{align*}
		\nodeconst_{\rere(\re(\Pi))} &= \{ \B~\C~\D \} \textrm{ and}\\
		\edgeconst_{\rere(\re(\Pi))} &= \{ \B~\C, \quad \B~\D, \quad \C~\D, \quad \D~\D \},
	\end{align*}
	as desired.

	Finally consider the case $k \geq 2$.
	We will prove this case via induction.
	For the base case, consider that $k = 2$.
	We claim that by applying $\re(\cdot)$ to $\fQ(\Pi)$ (which we already established), we obtain
	\begin{align*}
		\edgeconst_{\re(\fQ(\Pi))} &= \{ \{\B, \D\}~\{\C, \D\}, \quad \{\D\}~\{\B, \C, \D\} \},\\
		\Sigma_{\re(\fQ(\Pi))} &= \{ \{\D\}, \{\B, \D\}, \{\C, \D\}, \{\B, \C, \D\} \}, \textrm{ and}\\
		\nodeconst_{\re(\fQ(\Pi))} &= \{ [\{\B, \D\}\s\{\B, \C, \D\}]~[\{\C, \D\}\s\{\B, \C, \D\}]~[\{\D\}\s\{\B, \D\}\s\{\C, \D\}\s\{\B, \C, \D\}] \}.
	\end{align*}
	It is straightforward to verify that all configurations listed in the claimed $\edgeconst_{\re(\fQ(\Pi))}$ satisfy the universal quantifier.
	Moreover, it is clear that no listed configuration dominates another (different) listed configuration and that each configuration obtainable by taking a configuration from $\edgeconst_{\fQ(\Pi)}$ and replacing each label in the configuration by the singleton set containing the label is dominated by a listed configuration.
	Since $\{\B, \D\}$ and $\{\C, \D\}$ are the only incomparable sets appearing in the claimed $\edgeconst_{\fQ(\Pi)}$ and the configuration $\{\B, \D\} \cup \{\C, \D\} \s \{\C, \D\} \cap \{\B, \D\}$ is identical to $\{\D\}~\{\B, \C, \D\}$ (which is listed above), we obtain by~\Cref{obs:newre} and~\Cref{obs:union-non-comparable} that $\edgeconst_{\re(\fQ(\Pi))}$ is indeed as given above.
	Now the correctness of the claim follows straightforwardly by applying~\Cref{obs:easy-exists}.

	Let us rename the labels in $\Sigma_{\re(\fQ(\Pi))}$ to make the obtained problem $\re(\fQ(\Pi))$ more readable.
	By performing the renaming
	\begin{align*}
		\{\D\} &\rightarrow \D\\
		\{\B, \D\} &\rightarrow \B\\
		\{\C, \D\} &\rightarrow \C\\
		\{\B, \C, \D\} &\rightarrow \A,
	\end{align*}
	we obtain
	\begin{align*}
		\edgeconst_{\re(\fQ(\Pi))} &= \{ \B~\C, \quad \A~\D \} \textrm{ and}\\
		\nodeconst_{\re(\fQ(\Pi))} &= \{ [\A\s\B]~[\A\s\C]~[\A\s\B\s\C\s\D] \}.
	\end{align*}

	We claim that, by applying $\rere(\cdot)$ to $\re(\fQ(\Pi))$, we obtain
	\begin{align*}
		\nodeconst_{\rere(\re(\fQ(\Pi)))} &= \{ \{\A, \B\}~\{\A, \C\}~\{\A, \B, \C, \D\}, \quad \{\A\}~\{\A, \B, \C\}~\{\A, \B, \C, \D\} \},\\
		\Sigma_{\rere(\re(\fQ(\Pi)))} &= \{ \{\A\}, \{\A, \B\}, \{\A, \C\}, \{\A, \B, \C\}, \{\A, \B, \C, \D\} \}, \textrm{ and}\\
		\edgeconst_{\rere(\re(\fQ(\Pi)))} &= \{ [\{\A, \B\}\s\{\A, \B, \C\}\s\{\A, \B, \C, \D\}]~[\{\A, \C\}\s\{\A, \B, \C\}\s\{\A, \B, \C, \D\}], \\ &\phantom{{}={}\{} [\{\A\}\s\{\A, \B\}\s\{\A,\C\}\s\{\A, \B, \C\}\s\{\A, \B, \C, \D\}]~\{\A, \B, \C, \D\} \}.
	\end{align*}
	To prove the claim, observe that $\{\A, \B\}$ and $\{\A, \C\}$ are the only incomparable sets appearing in the claimed $\nodeconst_{\rere(\re(\fQ(\Pi)))}$.
	Since the configuration \[\{\A, \B\} \cup \{\A, \C\} \s \{\A, \C\} \cap \{\A, \B\} \s \{\A, \B, \C, \D\} \cap \{\A, \B, \C, \D\}\] is identical to $\{\A\}~\{\A, \B, \C\}~\{\A, \B, \C, \D\}$ and the configuration \[\{\A, \B\} \cup \{\A, \C\} \s \{\A, \C\} \cap \{\A, \B, \C, \D\} \s \{\A, \B, \C, \D\} \cap \{\A, \B\}\] is dominated by $\{\A, \B\}~\{\A, \C\}~\{\A, \B, \C, \D\}$, analogously to the argumentation in the case of $\edgeconst_{\re(\fQ(\Pi))}$ we obtain that $\nodeconst_{\rere(\re(\fQ(\Pi)))}$ is indeed as given above.
	Again, the claim follows by applying~\Cref{obs:easy-exists}.

	By performing the renaming
	\begin{align*}
		\{\A\} &\rightarrow \A_1\\
		\{\A, \B\} &\rightarrow \B\\
		\{\A, \C\} &\rightarrow \C\\
		\{\A, \B, \C\} &\rightarrow \D_1\\
		\{\A, \B, \C, \D\} &\rightarrow \D_2,
	\end{align*}
	we obtain
		\begin{align*}
		\nodeconst_{\rere(\re(\fQ(\Pi)))} &= \{ \B~\C~\D_2, \A_1~\D_1~\D_2 \} \textrm{ and}\\
		\edgeconst_{\rere(\re(\fQ(\Pi)))} &= \{ [\B\s\D_1\s\D_2]~[\C\s\D_1\s\D_2], \quad [\A_1\s\B\s\C\s\D_1\s\D_2]~\D_2 \}.
	\end{align*}
	Observing that
	\[
		 \{ [\B\s\D_1\s\D_2]~[\C\s\D_1\s\D_2], \quad [\A_1\s\B\s\C\s\D_1\s\D_2]~\D_2 \} =  \{ [\B\s\D_1\s\D_2]~[\C\s\D_1\s\D_2], \quad \A_1~\D_2 \}
	\]
	concludes the base case.
	
	For the induction step, fix any $k \geq 2$ and assume the induction hypothesis, i.e., that $\nodeconst_{\fQ^{k - 1}(\Pi)}$ and $\edgeconst_{\fQ^{k - 1}(\Pi)}$ are as given in the lemma.
	We claim that, by applying $\re(\cdot)$ to $\fQ^{k - 1}(\Pi)$, we obtain that the edge constraint $\edgeconst_{\re(\fQ^{k - 1}(\Pi))}$ of $\re(\fQ^{k - 1}(\Pi))$ is given by the following configurations:
	\begin{align*}
		&\{\B,\D_1,\dots,\D_{k - 1}\}~\{\C,\D_1,\dots,\D_{k - 1}\},\\
		&\{\B,\C,\D_1,\dots,\D_{k - 1}\}~\{\D_1,\dots,\D_{k - 1}\},\\
		&\{\A_1, \dots, \A_i, \B, \C, \D_1, \dots, \D_{k - 1}\}~\{\D_{i + 1}, \dots, \D_{k - 1} \} \text{ for each } 1 \leq i \leq k - 2.
	\end{align*}
	Analogously to before (i.e., making use of~\cref{obs:newre,obs:union-non-comparable}), the claim follows from the facts that $\{\B,\D_1,\dots,\D_{k - 1}\}$ and $\{\C,\D_1,\dots,\D_{k - 1}\}$ are the only incomparable listed sets and \[\{\B,\D_1,\dots,\D_{k - 1}\} \cup \{\C,\D_1,\dots,\D_{k - 1}\} \s \{\C,\D_1,\dots,\D_{k - 1}\} \cap \{\B,\D_1,\dots,\D_{k - 1}\}\] is identical to (the listed) $\{\B,\C,\D_1,\dots,\D_{k - 1}\}~\{\D_1,\dots,\D_{k - 1}\}$.

	The claim implies that the output label set for $\re(\fQ^{k - 1}(\Pi))$ is
	\begin{align*}
		\Sigma_{\re(\fQ^{k - 1}(\Pi))} =\ &\{ \{\B,\D_1,\dots,\D_{k - 1}\}, \{\C,\D_1,\dots,\D_{k - 1}\}, \{\B,\C,\D_1,\dots,\D_{k - 1}\} \}\\
		&\cup \{ \{\A_1, \dots, \A_i, \B, \C, \D_1, \dots, \D_{k - 1}\} \mid 1 \leq i \leq k - 2 \}\\
		&\cup \{ \{ \D_i, \dots, \D_{k - 1} \} \mid 1 \leq i \leq k - 1 \}.
	\end{align*}
	By performing the renaming
	\begin{align*}
		\{ \D_i, \dots, \D_{k - 1} \} &\rightarrow \D_i \textrm{ for each } 1 \leq i \leq k - 1\\
		\{\B,\D_1,\dots,\D_{k - 1}\} &\rightarrow \B\\
		\{\C,\D_1,\dots,\D_{k - 1}\} &\rightarrow \C\\
		\{\B,\C,\D_1,\dots,\D_{k - 1}\} &\rightarrow \A_0\\
		\{\A_1, \dots, \A_i, \B, \C, \D_1, \dots, \D_{k - 1}\} &\rightarrow \A_i \textrm{ for each } 1 \leq i \leq k - 2,
	\end{align*}
	we can rewrite the edge constraint and output label set for $\re(\fQ^{k - 1}(\Pi))$ as follows:
	\begin{align*}
		\edgeconst_{\re(\fQ^{k - 1}(\Pi))}\colon\quad &\B~\C\\[-4pt]
		&\A_i~\D_{i + 1} \text{ for each } 0 \leq i \leq k - 2,\\[2pt]
		\Sigma_{\re(\fQ^{k - 1}(\Pi))}\colon\quad &\{ \B, \C \} \cup \{ \A_i \mid 0 \leq i \leq k - 2 \} \cup \{ \D_i \mid 1 \leq i \leq k - 1 \}.
	\end{align*}
	Using the aforementioned renaming, we obtain, by~\Cref{obs:easy-exists}, that the node constraint $\nodeconst_{\re(\fQ^{k - 1}(\Pi))}$ of $\re(\fQ^{k - 1}(\Pi))$ is given by the following condensed configurations:
	\begin{align*}
		&[\A_0\s\dots\s\A_{k - 2}\s\B]~[\A_0\s\dots\s\A_{k - 2}\s\C]~[\A_0\s\dots\s\A_{k - 2}\s\B\s\C\s\D_1\s\dots\s\D_{k - 1}],\\
		&[\A_i\s\dots\s\A_{k - 2}]~[\A_0\s\dots\s\A_{k - 2}\s\B\s\C\s\D_1\s\dots\s\D_i]~[\A_0\s\dots\s\A_{k - 2}\s\B\s\C\s\D_1\s\dots\s\D_{k - 1}] \quad \forall 1 \leq i \leq k - 2.
	\end{align*}
	Finally, we claim that, by applying $\rere(\cdot)$ to $\re(\fQ^{k - 1}(\Pi))$, we obtain that the node constraint $\nodeconst_{\rere(\re(\fQ^{k - 1}(\Pi)))}$ of $\rere(\re(\fQ^{k - 1}(\Pi)))$ is given by the following configurations:
	\begin{align*}
		&\{\A_0,\dots,\A_{k - 2},\B\}~\{\A_0,\dots,\A_{k - 2},\C\}~\{\A_0,\dots,\A_{k - 2},\B,\C,\D_1,\dots,\D_{k - 1}\},\\
		&\{\A_0,\dots,\A_{k - 2}\}~\{\A_0,\dots,\A_{k - 2},\B,\C\}~\{\A_0,\dots,\A_{k - 2},\B,\C,\D_1,\dots,\D_{k - 1}\},\\
		&\{\A_i,\dots,\A_{k - 2}\}~\{\A_0,\dots,\A_{k - 2},\B,\C,\D_1,\dots,\D_i\}~\{\A_0,\dots,\A_{k - 2},\B,\C,\D_1,\dots,\D_{k - 1}\} \quad \forall 1 \leq i \leq k - 2.
	\end{align*}
	Analogously to before, the claim follows from the facts that $\{\A_0,\dots,\A_{k - 2},\B\}$ and $\{\A_0,\dots,\A_{k - 2},\C\}$ are the only incomparable listed sets, the configuration
	\begin{align*}
		&\{\A_0,\dots,\A_{k - 2},\B\} \cup \{\A_0,\dots,\A_{k - 2},\C\} \\
		&\{\A_0,\dots,\A_{k - 2},\C\} \cap \{\A_0,\dots,\A_{k - 2},\B\} \\
		&\{\A_0,\dots,\A_{k - 2},\B,\C,\D_1,\dots,\D_{k - 1}\} \cap \{\A_0,\dots,\A_{k - 2},\B,\C,\D_1,\dots,\D_{k - 1}\}
	\end{align*}
	is identical to $\{\A_0,\dots,\A_{k - 2}\}~\{\A_0,\dots,\A_{k - 2},\B,\C\}~\{\A_0,\dots,\A_{k - 2},\B,\C,\D_1,\dots,\D_{k - 1}\}$ (which is listed above), and the configuration
	\begin{align*}
		&\{\A_0,\dots,\A_{k - 2},\B\} \cup \{\A_0,\dots,\A_{k - 2},\C\} \\
		&\{\A_0,\dots,\A_{k - 2},\C\} \cap \{\A_0,\dots,\A_{k - 2},\B,\C,\D_1,\dots,\D_{k - 1}\} \\
		&\{\A_0,\dots,\A_{k - 2},\B,\C,\D_1,\dots,\D_{k - 1}\} \cap \{\A_0,\dots,\A_{k - 2},\B\}
	\end{align*}
	is dominated by $\{\A_0,\dots,\A_{k - 2},\B\}~\{\A_0,\dots,\A_{k - 2},\C\}~\{\A_0,\dots,\A_{k - 2},\B,\C,\D_1,\dots,\D_{k - 1}\}$ (which is listed above).

	The claim implies that the output label set for $\rere(\re(\fQ^{k - 1}(\Pi)))$ is
	\begin{align*}
		\Sigma_{\rere(\re(\fQ^{k - 1}(\Pi)))} =\ &\{ \{\A_0,\dots,\A_{k - 2},\B\}, \{\A_0,\dots,\A_{k - 2},\C\}, \{\A_0,\dots,\A_{k - 2},\B,\C\} \}\\
		&\cup \{ \{\A_i,\dots,\A_{k - 2}\} \mid 0 \leq i \leq k - 2 \}\\
		&\cup \{ \{\A_0,\dots,\A_{k - 2},\B,\C,\D_1,\dots,\D_i\} \mid 1 \leq i \leq k - 1 \}.
	\end{align*}
	By performing the renaming
	\begin{align*}
		\{\A_i,\dots,\A_{k - 2}\} &\rightarrow \A_{i + 1} \textrm{ for each } 0 \leq i \leq k - 2\\
		\{\A_0,\dots,\A_{k - 2},\B\} &\rightarrow \B\\
		\{\A_0,\dots,\A_{k - 2},\C\} &\rightarrow \C\\
		\{\A_0,\dots,\A_{k - 2},\B,\C\} &\rightarrow \D_1\\
		\{\A_0,\dots,\A_{k - 2},\B,\C,\D_1,\dots,\D_i\} &\rightarrow \D_{i + 1} \textrm{ for each } 1 \leq i \leq k - 1,
	\end{align*}
	we can rewrite the node constraint and output label set for $\rere(\re(\fQ^{k - 1}(\Pi)))$ as follows:
	\begin{align*}
		\nodeconst_{\rere(\re(\fQ^{k - 1}(\Pi)))}\colon\quad &\B~\C~\D_k\\[-4pt]
		&\A_i~\D_i~\D_k \text{ for each } 1 \leq i \leq k - 1,\\[1pt]
		\Sigma_{\rere(\re(\fQ^{k - 1}(\Pi)))}\colon\quad &\{ \B, \C \} \cup \{ \A_i \mid 1 \leq i \leq k - 1 \} \cup \{ \D_i \mid 1 \leq i \leq k \}.
	\end{align*}
	Note that in this last renaming, indices have been ``shifted by $1$''.
	Again making use of this renaming (and~\Cref{obs:easy-exists}), we obtain that the edge constraint $\edgeconst_{\rere(\re(\fQ^{k - 1}(\Pi)))}$ of $\rere(\re(\fQ^{k - 1}(\Pi)))$ is given by the following configurations:
	\begin{align*}
		&[\B\s\D_1\s\dots\s\D_k]~[\C\s\D_1\s\dots\s\D_k]\\
		&[\A_1\s\dots\s\A_i\s\B\s\C\s\D_1\s\dots\s\D_k]~[\D_{i + 1}\s\dots\s\D_k] \text{ for each } 1 \leq i \leq k - 1
	\end{align*}
	Observing that this characterization of $\edgeconst_{\rere(\re(\fQ^{k - 1}(\Pi)))}$ can be equivalently written as
	\begin{align*}
		\edgeconst_{\rere(\re(\fQ^{k - 1}(\Pi)))}\colon\quad &[\B\s\D_1\s\dots\s\D_k]~[\C\s\D_1\s\dots\s\D_k]\\[-4pt]
		&\A_1~\D_j \text{ for each } 1 \leq i < j \leq k
	\end{align*}
	concludes the induction and the entire proof.
\end{proof}

Now we are ready to prove \Cref{thm:sso-with-so}.
\begin{proof}[Proof of \Cref{thm:sso-with-so}]
	We start by showing Property~\ref{prop:lb}.
	By applying~\Cref{thm:newlifting} with $k \in \Theta(\log n)$, resp.\ $k \in \Theta(\log \log n)$, we see that it suffices to show for any $\ell \geq 0$ that $\fQ^{\ell}(\Pi)$ is not solvable in $0$ rounds in the deterministic PN model, even if a solution for sinkless orientation is given as input.
	(Note that we can choose the parameter $L$ in~\Cref{thm:newlifting} to be from $\Theta(\log n)$, resp.\ $\Theta(\log \log n)$, since the characterization of the $\fQ^{\ell}(\Pi)$ given in~\Cref{lem:characterization} guarantees that the number of labels used to describe $\fQ^{\ell}(\Pi)$ is in $O(\ell)$.)
	
	To this end, consider first the case that $\ell \geq 2$, and assume for a contradiction that there is a $0$-round PN-model algorithm $\fA$ that solves $\fQ^{\ell}(\Pi)$ given a solution to sinkless orientation as input.
	Consider a node with input configuration $\lO \s \lI \s \lI$, i.e., two of the half-edges incident to the node are oriented towards the node and the last half-edge is oriented away from the node.
	We consider two cases.

	First consider the case that $\fA$ chooses the node configuration $\A_i \s \D_i \s \D_{\ell} \in \nodeconst_{\fQ^{\ell}(\Pi)}$ for some $1 \leq i \leq \ell$ for such a node.
	If $\fA$ outputs the label $\A_i$ on the outgoing half-edge and therefore outputs label $\D_i$ on one of the incoming half-edges, then on some input graphs, $\fA$ will produce the edge configuration $\A_i \s \D_i$, which is not contained in $\edgeconst_{\fQ^{\ell}(\Pi)}$, yielding a contradiction.
	Hence, $\fA$ necessarily outputs label $\A_i$ on an incoming half-edge.
	Now consider a node with three outgoing half-edges.
	No matter which node configuration $\fA$ chooses for such a node, there is some output label from $\{ \A_1, \dots, \A_{\ell - 1}, \B, \C \}$ that $\fA$ outputs on an outgoing half-edge.
	Since none of the labels contained in this set yields an edge configuration in $\edgeconst_{\fQ^{\ell}(\Pi)}$ when combined with $\A_i$, we conclude that $\fA$ will produce an incorrect output on some input graphs, yielding a contradiction.

	Now consider the complementary case, i.e., that $\fA$ chooses the node configuration $\B \s \C \s \D_{\ell}$ for a node with precisely one incident outgoing half-edge.
	Then $\fA$ outputs one of $\B$ and $\C$ on an incoming half-edge.
	Due to symmetry, we can assume w.l.o.g.\ that $\fA$ outputs $\B$ on an incoming half-edge.
	Now consider again an edge with three outgoing half-edges.
	No matter which node configuration $\fA$ chooses for such a node, there is some output label from $\{ \A_1, \dots, \A_{\ell - 1}, \B \}$ that $\fA$ outputs on an outgoing half-edge.
	Since none of the labels contained in this set yields an edge configuration in $\edgeconst_{\fQ^{\ell}(\Pi)}$ when combined with $\B$, we conclude that $\fA$ will produce an incorrect output on some input graphs, yielding a contradiction.

	We remark that, while we omitted an explicit specification of the ports of the incoming, resp.\ outgoing, edges in the above considerations, it is straightforward to select such a specification.
	Moreover, for the case that $\ell \in \{ 0, 1 \}$, we obtain a contradiction analogously to the second case in the above considerations.
	This concludes the proof that $\Pi$ satisfies Property~\ref{prop:lb}.

	Now we turn to showing Property~\ref{prop:fp}.
	First we prove that there is no nontrivial fixed point relaxation of $\Pi$.
	We will do so by showing that each fixed point relaxation of $\Pi$ can be solved in $0$ rounds in our setting, i.e., given a sinkless orientation as input.
	Let $\hat{\Pi}$ be an arbitrary fixed point relaxation of $\Pi$.
	Let $y$ be an integer satisfying $y \geq |\Sigma_{\hat{\Pi}}| + 2$.
	Consider the problem $\fQ^y(\Pi)$ obtained from $\Pi$ after $y$ recursive applications of $\rere(\re(\cdot))$.

	As $\hat{\Pi}$ is a relaxation of $\Pi$, we obtain by \Cref{lem:commutative} that $\fQ^y(\hat{\Pi})$ is a relaxation of $\fQ^y(\Pi)$.
	Since $\fQ^y(\hat{\Pi}) = \hat{\Pi}$ (due to $\hat{\Pi}$ being a fixed point), it follows that $\hat{\Pi}$ is a relaxation of $\fQ^y(\Pi)$.
	Moreover, by \Cref{lem:characterization}, the definition of $y$ implies that $\nodeconst_{\fQ^y(\hat{\Pi})}$ contains configuration $\A_i~\D_i~\D_y$ for each $1 \leq i \leq |\Sigma_{\hat{\Pi}}| + 1$.

	Recall the definition of a relaxation function given in \Cref{def:relax}, and let $f$ be a relaxation function from $\fQ^y(\Pi)$ to $\hat{\Pi}$.
	By the pigeonhole principle, we obtain that $\nodeconst_{\fQ^y(\hat{\Pi})}$ must contain two configurations $\A_i~\D_i~\D_y$ and $\A_j~\D_j~\D_y$ satisfying $i < j < |\Sigma_{\hat{\Pi}}| + 2$ and $f(\A_i) = f(\A_j)$.
	Moreover, the definition of $f$ ensures that $f(\A_j)~f(\D_j)~f(\D_y)$ is contained in $\hat{\Pi}$.
	We claim that if we select any two (not necessarily distinct) labels in this configuration, the configuration consisting of the selected two labels is contained in $\edgeconst_{\hat{\Pi}}$, except possibly if we select $f(\A_j)$ twice.
	In the following, we prove this claim.

	By \Cref{lem:characterization} and the fact that $j < |\Sigma_{\hat{\Pi}}| + 2 \leq y$, we know that the four configurations $\D_j~\D_j$, $\D_j~\D_y$, $\D_y~\D_y$, and $\A_j~\D_y$ are all contained in $\edgeconst_{\fQ^y(\Pi)}$.
	Hence, by the fact that $f$ is a relaxation function from $\Pi$ to $\hat{\Pi}$, we obtain that the configurations $f(\D_j)~f(\D_j)$, $f(\D_j)~f(\D_y)$, $f(\D_y)~f(\D_y)$, and $f(\A_j)~f(\D_y)$ are all contained in $\edgeconst_{\hat{\Pi}}$.
	Moreover, since $i < j$, \Cref{lem:characterization} also implies that $\A_i~\D_j$ is contained in $\edgeconst_{\fQ^y(\Pi)}$, which in turn implies that $f(\A_i)~f(\D_j)$ is contained in $\edgeconst_{\hat{\Pi}}$.
	As $f(\A_i) = f(\A_j)$, it follows that also $f(\A_j)~f(\D_j)$ is contained in $\edgeconst_{\hat{\Pi}}$, which concludes the proof of the claim.
	
	The proved claim now offers a simple way to solve $\hat{\Pi}$ in $0$ rounds given a sinkless orientation as input: each node $v$ simply selects one of its incident half-edges for which the corresponding edge is oriented away from $v$ in the given sinkless orientation, outputs $f(\A_j)$ on the selected half-edge, outputs $f(\D_j)$ on an arbitrarily chosen half-edge out of the remaining two, and outputs $f(\D_y)$ on the last remaining half-edge.
	It remains to show that the output returned by this $0$-round algorithm is indeed a correct solution for $\hat{\Pi}$.
	From the design of the algorithm it directly follows that, for each node $v$, the configuration consisting of the output labels on the half-edges incident to $v$ is $f(\A_j)~f(\D_j)~f(\D_y)$ and therefore contained in $\nodeconst_{\hat{\Pi}}$.
	For the correctness on the edges, observe that the design of the algorithm ensures that there is no edge $e$ for which the output label on \emph{both} half-edges is $\A_j$ (since $e$ is outgoing for only one of its endpoints in the given sinkless orientation).
	Now, the claim that we proved above guarantees that, for each edge $e$, the configuration consisting of the output labels on the half-edges belonging to $e$ is contained in $\edgeconst_{\hat{\Pi}}$, which concludes the proof for the statement that there is no nontrivial fixed point relaxation of $\Pi$.
	
	Now, applying~\Cref{thm:worthless-input} yields that there is no generalized fixed point $\fF$ (with sinkless orientation as input), as that would imply the existence of a non-generalized fixed point relaxation that is nontrivial in the setting with sinkless orientation as input.
This implies the second part of Property~\ref{prop:fp}.
\end{proof}

\section{Fixed point relaxations for homomorphism problems}\label{sec:homom}

In \cite{brandt-chang-etal-2022-local-problems-on-trees-from-the}, the authors show an interesting connection between two fields that, at a first glance, seem unrelated: distributed computing and descriptive combinatorics. A class of problems that has been widely studied in the context of descriptive combinatorics is the so-called Borel class. In \cite{brandt-chang-etal-2022-local-problems-on-trees-from-the}, the authors show that, if a problem does not belong to the class Borel, then such a problem requires $\Omega(\log n)$ deterministic rounds and $\Omega(\log\log n)$ randomized rounds in the LOCAL model. In the literature of the field of descriptive combinatorics, there is a technique that can be used to show that a problem is not in the Borel class \cite{marks-2016-a-determinacy-approach-to-borel}. The authors of \cite{brandt-chang-etal-2022-local-problems-on-trees-from-the} use such a technique to derive lower bounds in the LOCAL model for a so-called homomorphism problem. They leave as an open question the problem of understanding whether the same lower bounds can be achieved via the round elimination technique, and in particular they ask whether there is a nontrivial fixed point relaxation for the problems they achieve lower bounds for. In this section, we solve this open question by providing a positive answer.
We start by defining the class of problems for which \cite{brandt-chang-etal-2022-local-problems-on-trees-from-the} obtains $\Omega(\log n)$-round deterministic and $\Omega(\log \log n)$-round randomized lower bounds via the aforementioned technique from descriptive combinatorics, which, in the distributed context, is often referred to as \emph{Marks' technique}.

\subsection{A subclass of homomorphism problems}\label{ssec:9-col}

Let $\fP$ denote the class of problems \cite{brandt-chang-etal-2022-local-problems-on-trees-from-the} obtain the aforementioned lower bounds for.
The problems in $\fP$ come from a class of problems called \emph{homomorphism problems}.
In a homomorphism problem the task is to compute a homomorphism from the input graph to a given fixed graph $H$.
This task can be reformulated as a coloring problem: color the nodes of the input graph such that each color is a node of $H$ and any two neighboring nodes $u, v$ in the input graph must output colors $c_u, c_v$ such that $c_u$ and $c_v$ are adjacent nodes in $H$.
To phrase a homomorphism problem as a node-edge-checkable problem, we simply require outputting the same color at each half-edge incident to the same node and that the two outputs on the half-edges belonging to the same edge satisfy the aforementioned coloring property.

The class $\fP$ studied in \cite{brandt-chang-etal-2022-local-problems-on-trees-from-the} is the class of all homomorphism problems where graph $H$ satisfies a certain reasonably natural property, called $\Delta$-(*).
For a formal definition of $\Delta$-(*), see~\cite[Definition 4.8, arXiv version]{brandt-chang-etal-2022-local-problems-on-trees-from-the}.
The authors show further that, for any\footnote{Recall that for all considered lower bound techniques, it suffices to study regular graphs. Here $\Delta$ refers to the maximum degree of the ($\Delta$-regular) input instance, not to the maximum degree of $H$ or $H_{\Delta}$.} $\Delta \geq 3$, a graph $H$ satisfies $\Delta$-(*) if and only if there is a homomorphism from $H$ to a specific graph $H_{\Delta}$~\cite[Proposition 4.10, arXiv version]{brandt-chang-etal-2022-local-problems-on-trees-from-the}.
Observe that the existence of a homomorphism from a graph $H$ to $H_{\Delta}$ implies that there is a relaxation from the homomorphism problem with target graph $H$ to the homomorphism problem with target graph $H_{\Delta}$ as any homomorphism from the input graph $G$ to $H$ can be concatenated with the homomorphism from $H$ to $H_{\Delta}$ to yield a homomorphism from $G$ to $H_{\Delta}$ (and this argumentation translates straightforwardly to the node-edge-checkability formalism).
Hence, to obtain the statement that each problem from $\fP$ has a nontrivial fixed point relaxation, it suffices to show that, for each $\Delta \geq 3$ the homomorphism problem with target graph $H_{\Delta}$ has a nontrivial fixed point relaxation. 

Assume for the remainder of this section that $\Delta \geq 3$ is fixed.
In the following, we state the homomorphism problem with target graph $H_{\Delta}$, introduced in~\cite[Section 4.1, arXiv version]{brandt-chang-etal-2022-local-problems-on-trees-from-the}, in the node-edge-checkability formalism.
We will refer to this problem as $\ninecol$.

As alluded to before, we can regard $\ninecol$ as a coloring problem where only certain color combinations are allowed for neighboring nodes.
The colors of $\ninecol$ are all pairs $(y, z) \in \{1, \dots, \Delta\} \times \{1, \dots, \Delta\}$, i.e., $\Sigma_{\ninecol} := \{1, \dots, \Delta\} \times \{1, \dots, \Delta\}$.
The node constraint $\nodeconst_{\ninecol}$ contains precisely the configurations $i \s \dots \s i$ for all colors $i \in \Sigma$.
The edge constraint $\edgeconst_{\ninecol}$ contains all configurations $(y, z)\s(y', z')$ satisfying
\begin{enumerate}[noitemsep]
	\item $y \neq 1 \neq z$, $y \neq y'$, and $z \neq z'$ (or, dually, $y' \neq 1 \neq z'$, $y \neq y'$, and $z \neq z'$), or
	\item $y = y' = 1$, $z \neq 1 \neq z'$, and $z \neq z'$, or
	\item $z = z' = 1$, $y \neq 1 \neq y'$, and $y \neq y'$.
\end{enumerate}

As an example, we explicitly provide the node and edge constraint of $\ninecol$ for $\Delta = 3$ below.

\begin{equation*}
	\begin{aligned}
		\begin{aligned}
			\nodeconst_{\ninecol}\text{:}\\
			&\lnine{1}{1} &\s& \lnine{1}{1} &~& \lnine{1}{1} \\
			&\lnine{1}{2} &\s& \lnine{1}{2} &~& \lnine{1}{2} \\
			&\lnine{1}{3} &\s& \lnine{1}{3} &~& \lnine{1}{3} \\
			&\lnine{2}{1} &\s& \lnine{2}{1} &~& \lnine{2}{1} \\
			&\lnine{2}{2} &\s& \lnine{2}{2} &~& \lnine{2}{2} \\
			&\lnine{2}{3} &\s& \lnine{2}{3} &~& \lnine{2}{3} \\
			&\lnine{3}{1} &\s& \lnine{3}{1} &~& \lnine{3}{1} \\
			&\lnine{3}{2} &\s& \lnine{3}{2} &~& \lnine{3}{2} \\
			&\lnine{3}{3} &\s& \lnine{3}{3} &~& \lnine{3}{3} \\
		\end{aligned}
		\qquad
		\begin{aligned}
			\edgeconst_{\ninecol}\text{:}\\
			& [\lnine{1}{1}] &~& [\lnine{2}{2} \s \lnine{2}{3} \s \lnine{3}{2} \s \lnine{3}{3}] \\
			& [\lnine{1}{2}] &~& [\lnine{1}{3} \s \lnine{2}{3} \s \lnine{3}{3}] \\
			& [\lnine{1}{3}] &~& [\lnine{1}{2} \s \lnine{2}{2} \s \lnine{3}{2}] \\
			& [\lnine{2}{1}] &~& [\lnine{3}{1} \s \lnine{3}{2} \s \lnine{3}{3}] \\
			& [\lnine{2}{2}] &~& [\lnine{1}{1} \s \lnine{1}{3} \s \lnine{3}{1} \s \lnine{3}{3}] \\
			& [\lnine{2}{3}] &~& [\lnine{1}{1} \s \lnine{1}{2} \s \lnine{3}{1} \s \lnine{3}{2}] \\
			& [\lnine{3}{1}] &~& [\lnine{2}{1} \s \lnine{2}{2} \s \lnine{2}{3}] \\
			& [\lnine{3}{2}] &~& [\lnine{1}{1} \s \lnine{1}{3} \s \lnine{2}{1} \s \lnine{2}{3}] \\
			& [\lnine{3}{3}] &~& [\lnine{1}{1} \s \lnine{1}{2} \s \lnine{2}{1} \s \lnine{2}{2}] \\
		\end{aligned}
	\end{aligned}
\end{equation*}

We devote the rest of this section to proving the following theorem and deriving lower bounds from it.
\begin{theorem}\label{thm:FP-for-ninecol}
	There exists a nontrivial fixed point relaxation for $\ninecol$.
\end{theorem}
As discussed above, \Cref{thm:FP-for-ninecol} implies the following corollary that covers all problems for which lower bounds via Marks' technique have been known but (so far) no lower bounds via round elimination.

\begin{corollary}\label{cor:fpsforall}
	Every problem in $\fP$ has a nontrivial fixed point relaxation.
\end{corollary}

To prove \Cref{thm:FP-for-ninecol}, we make use of \Cref{cor:double-tau-relax} where we choose $\Pi$ to be $\ninecol$.
Then Part \ref{item:double-tau} of \Cref{cor:double-tau-relax} states that there exists a nontrivial fixed point relaxation of $\ninecol$, and hence it suffices to prove part \ref{item:general-input} of \Cref{cor:double-tau-relax} which we will do in the following.

To this end, we introduce a problem $\orcx$, which we will prove to be a round elimination fixed point.
Moreover, we will prove that $\orcx$ is a relaxation of $\ninecol$, given a $\Delta$-edge coloring as input.
Lastly we need to show that $\orcx$ is nontrivial given a $\Delta$-edge coloring.

\paragraph{The ORCX problem.}
In the following we define the aforementioned problem $\orcx$.
We set the output label set of $\orcx$ to be $\Sigma_{\orcx} := \one \cup \two$, where $\one := \{ \lO, \lR, \lC, \lX \}$ and $\two := \{ \lo, \lr, \lc, \lx \}$.
The node constraint $\nodeconst_{\orcx}$ of $\orcx$ contains precisely those configurations $\L_1\s\dots\s\L_{\Delta}$ that satisfy that
\begin{enumerate}
	\item\label{item:onecap} there is precisely one index $1 \leq k \leq \Delta$ such that $\L_k \in \one$ and
	\item\label{item:therest} there are two indices $k, k' \in \{ 1, \dots, \Delta \}$ satisfying $k \neq k'$ such that $\L_{k''} \in \{ \lO, \lo \}$ for each $k'' \in \{ 1, \dots, \Delta \} \setminus \{ k, k' \}$ and
		\begin{enumerate}
			\item either $\L_k \in \{ \lX, \lx \}$ and $\L_{k'} \in \{ \lO, \lo \}$,
			\item or $\L_k \in \{ \lR, \lr \}$ and $\L_{k'} \in \{ \lC, \lc \}$.
		\end{enumerate}
\end{enumerate}
The edge constraint $\edgeconst_{\orcx}$ of $\orcx$ contains precisely the following configurations.
\begin{equation*}
	\begin{aligned}
			& [\lO] &~& [\lO\s\lR\s\lC\s\lX] \\
			& [\lO\s\lR] &~& [\lO\s\lR] \\
			& [\lO\s\lC] &~& [\lO\s\lC] \\
			& [\lo] &~& [\lo\s\lr\s\lc\s\lx] \\
			& [\lo\s\lr] &~& [\lo\s\lc]
	\end{aligned}
\end{equation*}

\subsection{A 0-round transformation}\label{ssec:reduction}

We now show that problem $\orcx$ can be reduced to $\ninecol$ if a $\Delta$-edge coloring (with colors $1, \dots, \Delta$) is given as input.

\begin{lemma}\label{lem:nine-to-orcx}
Given a $\Delta$-edge coloring and a solution to $\ninecol$, it is possible to solve $\orcx$ in $0$ rounds.
\end{lemma}
\begin{proof}
	Each node $v$ which is labeled $\lnine{i}{j}$ (i.e., whose incident half-edges are labeled $\lnine{i}{j}$) in the solution to $\ninecol$ produces a solution for $\orcx$ as follows. First, node $v$ marks the incident edge of color $i$ as ``my row'' and the incident edge of color $j$ as ``my column''. Then, on the incident half-edge of color $1$, node $v$ outputs the following label:
	\begin{itemize}
		\item $\lX$, if the edge is marked both as ``my row'' and as ``my column''. Observe that this output is produced if $v$ is labeled $\lnine{1}{1}$.
		\item $\lR$, if the edge is marked as ``my row'' only. Observe that this output is produced if $v$ is labeled $\lnine{1}{j}$ for some $j > 1$.
		\item $\lC$, if the edge is marked as ``my column'' only. Observe that this output is produced if $v$ is labeled $\lnine{i}{1}$ for some $i > 1$.
		\item $\lO$, if the edge is not marked.
	\end{itemize}
	Then, on any incident half-edge of color $z \in \{2, \dots, \Delta\}$, node $v$ outputs the following label:
	\begin{itemize}
		\item $\lx$, if the edge is marked both as ``my row'' and as ``my column''. Observe that this output is produced if $v$ is labeled $\lnine{z}{z}$.
		\item $\lr$, if the edge is marked as ``my row'' only. Observe that this output is produced if $v$ is labeled $\lnine{z}{z'}$ where $z' \neq z$.
		\item $\lc$, if the edge is marked as ``my column'' only. Observe that this output is produced if $v$ is labeled $\lnine{z'}{z}$, where $z' \neq z$.
		\item $\lo$, if the edge is not marked. 
	\end{itemize}

	Since $v$ outputs a label from $\one$ on the incident half-edge of color $1$ but on no other incident half-edge, the node configuration produced by $v$ satisfies Property~\ref{item:onecap} of the definition of $\nodeconst_{\orcx}$.
	Since $v$ marks exactly one incident edge as ``my row'' and exactly one incident edge as ``my column'', there are $\Delta - 2$ incident half-edges on which $v$ outputs a label from $\{ O, o \}$.
	Moreover, if $v$ marks two different incident edges, then of the two half-edges where $v$ does not output a label from $\{ O, o \}$ one receives a label from $\{ R,r \}$, and the other a label from $\{ C, c\}$, while if $v$ marks the same edge twice, then one of the two ``remaining'' label comes from $\{ X, x \}$, and the other from $\{ O, o\}$.
	Hence, the node configuration produced by $v$ satisfies also Property~\ref{item:therest} of the definition of $\nodeconst_{\orcx}$.

	We now argue that all produced edge configurations are in $\edgeconst_{\orcx}$. 
	Let $u$ and $v$ be two neighboring nodes. Assume that $u$ has color $\lnine{i}{j}$ and $v$ color $\lnine{i'}{j'}$. Let $z$ be the color of the edge connecting $u$ and $v$.
	If $z=1$, the algorithm produces an edge configuration that is contained in $[\lO \lR \lC \lX] \s [\lO \lR \lC \lX]$.
	We show that the produced configuration is not in $\{\lR \s \lC, \lR \s \lX, \lC \s \lX, \lX \s \lX\}$, which are all the configurations from $[\lO \lR \lC \lX] \s [\lO \lR \lC \lX]$ not contained in $\edgeconst_{\orcx}$:
	\begin{itemize}
		\item $\lX \s \lX$ cannot be obtained, since $\lX$ is only produced by a node labeled $\lnine{1}{1}$, but only one node among $u$ and $v$ can have this color.
		\item $\lR \s \lX$ cannot be obtained, since $\lR$ is only produced by nodes labeled $\lnine{1}{j''}$ for some $j'' > 1$, which cannot be neighbors of $\lnine{1}{1}$ by the definition of $\edgeconst_{\ninecol}$.
		\item $\lC \s \lX$ cannot be obtained, since $\lC$ is only produced by nodes labeled $\lnine{i''}{1}$ for some $i'' > 1$, which cannot be neighbors of $\lnine{1}{1}$ by the definition of $\edgeconst_{\ninecol}$.
		\item $\lR \s \lC$ cannot be obtained, since nodes labeled $\lnine{1}{j''}$ for some $j'' > 1$ (which are the only nodes that can produce $\lR$) cannot be neighbors of nodes labeled $\lnine{i''}{1}$ for some $i'' > 1$ (which are the only nodes that can produce $\lC$), by the definition of $\edgeconst_{\ninecol}$.
	\end{itemize}
	Let $z \in \{2,3\}$. On edges colored $z$, the algorithm produces an edge configuration that is contained in $[\lo \lr \lc \lx] \s [\lo \lr \lc \lx]$.
	We show that the produced configuration is not in $\{\lr \s \lr, \lc \s \lc, \lr \s \lx, \lc \s \lx, \lx \s \lx\}$:
	\begin{itemize}
		\item $\lx \s \lx$ cannot be obtained, since $\lx$ is only produced by a node labeled $\lnine{z}{z}$, but only one node among $u$ and $v$ can have this color.
		\item $\lr \s \lx$ cannot be obtained, since $\lr$ is only produced by nodes with a label in $\{\lnine{z}{z'} \mid z' \neq z \}$, which cannot be neighbors of $\lnine{z}{z}$ by the definition of $\edgeconst_{\ninecol}$.
		\item $\lr \s \lr$ cannot be obtained, since $\lr$ is only produced by nodes with a label in $\{\lnine{z}{z'} \mid z' \neq z \}$, and two such nodes cannot be neighbors by the definition of $\edgeconst_{\ninecol}$.
		\item $\lc \s \lx$ cannot be obtained, since $\lc$ is only produced by nodes with a label in $\{\lnine{z'}{z} \mid z' \neq z \}$, which cannot be neighbors of $\lnine{z}{z}$ by the definition of $\edgeconst_{\ninecol}$.
		\item $\lc \s \lc$ cannot be obtained, since $\lc$ is only produced by nodes with a label in $\{\lnine{z'}{z} \mid z' \neq z \}$, and two such nodes cannot be neighbors by the definition of $\edgeconst_{\ninecol}$.
		\qedhere
	\end{itemize}
\end{proof}

\subsection{The ORCX problem is a fixed point}\label{ssec:orcx-fp}

In the following we prove that $\orcx$ is a fixed point under the round elimination framework.
More precisely, we prove the following theorem.
\begin{theorem}\label{lem:orcx-fp}
$\rere(\re(\orcx)) = \orcx$.
\end{theorem}

\paragraph{\boldmath Computing the constraint $\edgeconst'_{\orcx}$.}
We start by computing the set of maximal configurations that satisfy the universal quantifier w.r.t.\ $\edgeconst_{\orcx}$.
\begin{lemma}\label{lem:first-forall}
	Let $\edgeconst'_{\orcx}=\edgeconst_{\re(\orcx)}$ be the set of maximal configurations that satisfy the universal quantifier w.r.t.\ $\edgeconst_{\orcx}$. Then, $\edgeconst'_{\orcx}$ contains exactly the following configurations:
	\begin{itemize}[noitemsep]
			\item $\{\lO\} \s \{\lO,\lR,\lC,\lX\}$
			\item $\{\lO,\lR\} \s \{\lO,\lR\}$
			\item $\{\lO,\lC\} \s \{\lO,\lC\}$
			\item $\{\lo\} \s \{\lo,\lr,\lc,\lx\}$
			\item $\{\lo,\lr\} \s \{\lo,\lc\}$.
	\end{itemize}
\end{lemma}
\begin{proof}
By the definition of $\edgeconst_{\orcx}$, it is clear that all listed configurations satisfy the universal quantifier. Moreover, it is clear that there is no listed configuration that dominates another (different) listed configuration and that each configuration obtainable by taking a configuration from $\edgeconst_{\orcx}$ and replacing each label in the configuration by the singleton set containing the label is dominated by a listed configuration. Hence, what remains to show is that there is no maximal configuration that satisfies the universal quantifier that is not listed above. In order to show this, by \Cref{obs:newre}, it is enough to show that, by combining any two configurations of the list we obtain a new configuration that is dominated by some configuration in the list.

Let $\fC=\L_1\s\L_2$ and $\fC'=\L'_1\s\L'_2$ be two arbitrary (possibly the same) configurations from the list. Let $\sigma$ be a permutation of $\{1,2\}$, and let $u\in\{1,2\}$. Let $\fC''$ be the combination of $\fC$ and $\fC'$ w.r.t.\ $u$ and $\sigma$. We first observe that, in order for $C''$ to be a valid configuration, we must not obtain any empty set, and hence either both $\fC$ and $\fC'$ contain only labels from $\{\lO,\lR,\lC,\lX\}$ or they both contain only labels from $\{\lo,\lr,\lc,\lx\}$. Moreover, by \Cref{obs:union-non-comparable}, it must hold that the union is taken on non-comparable sets.

Hence, in the former case, up to symmetries, the only case to consider is when $\fC = \{\lO,\lR\} \s \{\lO,\lR\}$, $\fC'= \{\lO,\lC\} \s \{\lO,\lC\}$, $u=2$, and $\sigma$ is the identity function. In this case, we get that $\fC''= \{\lO\} \s \{\lO,\lR,\lC\}$, which is dominated by $\{\lO\} \s \{\lO,\lR,\lC,\lX\}$.

In the latter case, up to symmetries, the only case to consider is when $\fC=\fC'=\{\lo,\lr\} \s \{\lo,\lc\}$, $u=2$, $\sigma(1)=2$, and $\sigma(2)=1$. In this case, we get that $\fC''= \{\lo\} \s \{\lo,\lr,\lc\}$, which is dominated by $\{\lo\} \s \{\lo,\lr,\lc,\lx\}$.
\end{proof}

For the sake of readability, we rename the obtained sets of labels as follows.
\begin{itemize}[noitemsep]
	\item $\mybox{\lO} := \{\lO\}$
	\item $\mybox{\lR} := \{\lR,\lO\}$
	\item $\mybox{\lC} := \{\lC,\lO\}$
	\item $\mybox{\lX} := \{\lO,\lR,\lC,\lX\}$
	\item $\mybox{\lo} := \{\lo\}$
	\item $\mybox{\lr} := \{\lr,\lo\}$
	\item $\mybox{\lc} := \{\lc,\lo\}$
	\item $\mybox{\lx} := \{\lo,\lr,\lc,\lx\}$
\end{itemize}
Under such a renaming, we obtain that $\edgeconst'_{\orcx}$ contains exactly the following configurations.
\begin{itemize}[noitemsep]
	\item $\mybox{\lO} \s \mybox{\lX}$
	\item $\mybox{\lR} \s \mybox{\lR}$
	\item $\mybox{\lC} \s \mybox{\lC}$
	\item $\mybox{\lo} \s \mybox{\lx}$
	\item $\mybox{\lr} \s \mybox{\lc}$
\end{itemize}
In the following, let $\Sigma'_{\orcx} = \{\mybox{\lO}, \mybox{\lR}, \mybox{\lC}, \mybox{\lX}, \mybox{\lo}, \mybox{\lr}, \mybox{\lc}, \mybox{\lx}\}$.

\paragraph{\boldmath Computing the constraint $\nodeconst'_{\orcx}$.}
We now compute $\nodeconst'_{\orcx}$, that is, we apply the existential quantifier. 

\begin{lemma}\label{lem:first-exists}
	The constraint $\nodeconst'_{\orcx} = \nodeconst_{\re(\orcx)}$ contains precisely those configurations $\L_1\s\dots\s\L_{\Delta}$ that satisfy that
\begin{enumerate}
	\item there is precisely one index $1 \leq k \leq \Delta$ such that $\L_k \in \{ \mybox{\lO}, \mybox{\lR}, \mybox{\lC}, \mybox{\lX} \}$ and
	\item there are two indices $k, k' \in \{ 1, \dots, \Delta \}$ satisfying $k \neq k'$ such that
		\begin{enumerate}
			\item either $\L_k \in \{ \mybox{\lX}, \mybox{\lx} \}$
			\item or $\L_k \in \{ \mybox{\lR}, \mybox{\lX}, \mybox{\lr}, \mybox{\lx} \}$ and $\L_{k'} \in \{ \mybox{\lC}, \mybox{\lX}, \mybox{\lc}, \mybox{\lx} \}$.
		\end{enumerate}
\end{enumerate}
\end{lemma}
\begin{proof}
	The lemma follows directly by \Cref{obs:easy-exists}.
\end{proof}

\paragraph{\boldmath Computing the constraint $\nodeconst''_{\orcx}$.}
We now compute the set of maximal configurations that satisfy the universal quantifier w.r.t.\ $\nodeconst'_{\orcx}$.
\begin{lemma}\label{lem:second-forall}
	Let $\nodeconst''_{\orcx}=\nodeconst_{\rere(\re(\orcx))}$ be the set of maximal configurations that satisfy the universal quantifier w.r.t.\ $\nodeconst'_{\orcx}$. Then, $\nodeconst''_{\orcx}$ contains precisely those configurations $\L_1\s\dots\s\L_{\Delta}$ that satisfy that
\begin{enumerate}
	\item\label{item:onecapdoubleprime} there is precisely one $1 \leq k \leq \Delta$ such that $\L_k \in \{ \{ \mybox{\lO},\mybox{\lR},\mybox{\lC},\mybox{\lX}\}, \{ \mybox{\lR}, \mybox{\lX}\}, \{ \mybox{\lC},\mybox{\lX}\}, \{  \mybox{\lX}\} \}$ and
	\item there are two indices $k, k' \in \{ 1, \dots, \Delta \}$ satisfying $k \neq k'$ such that $\L_{k''}$ is contained in $\{ \{\mybox{\lO},\mybox{\lR},\mybox{\lC},\mybox{\lX}\}, \{\mybox{\lo},\mybox{\lr},\mybox{\lc},\mybox{\lx}\} \}$ for each $k'' \in \{ 1, \dots, \Delta \} \setminus \{ k, k' \}$ and
		\begin{enumerate}
			\item either $\L_k \in \{ \{\mybox{\lX}\}, \{\mybox{\lx}\} \}$ and $\L_{k'} \in \{ \{\mybox{\lO},\mybox{\lR},\mybox{\lC},\mybox{\lX}\}, \{\mybox{\lo},\mybox{\lr},\mybox{\lc},\mybox{\lx}\} \}$,
			\item or $\L_k \in \{ \{\mybox{\lR}, \mybox{\lX}\}, \{\mybox{\lr}, \mybox{\lx}\} \}$ and $\L_{k'} \in \{ \{\mybox{\lC}, \mybox{\lX}\}, \{\mybox{\lc}, \mybox{\lx}\} \}$.
		\end{enumerate}
\end{enumerate}
\end{lemma}
\begin{proof}
By the definition of $\nodeconst'_{\orcx}$, it is clear that all listed configurations (i.e., all configurations characterized in the lemma) satisfy the universal quantifier. Moreover, it is clear that there is no listed configuration that dominates another (different) listed configuration and that each configuration obtainable by taking a configuration from $\nodeconst'_{\orcx}$ and replacing each label in the configuration by the singleton set containing the label is dominated by a listed configuration. Hence, what remains to show is that there is no maximal configuration that satisfies the universal quantifier that is not listed above. In order to show this, by \Cref{obs:newre}, it is enough to show that, by combining any two configurations of the list we obtain a new configuration that is dominated by some configuration in the list.

Let $\fC=\L_1\s\dots\s L_{\Delta}$ and $\fC'=\L'_1\s\dots\s L'_{\Delta}$ be two arbitrary (possibly the same) configurations from the list. Let $\sigma$ be a permutation of $\{1,\dots,\Delta\}$, and let $u\in\{1,\dots,\Delta\}$. Let $\fC''$ be the combination of $\fC$ and $\fC'$ w.r.t.\ $u$ and $\sigma$. 
We first observe that, in order for $\fC''$ to be a valid configuration, we must not obtain any empty set, and hence $\L_i$ and $\L_{\sigma(i)}$ must either be both subsets of $\{ \mybox{\lO},\mybox{\lR},\mybox{\lC},\mybox{\lX}\}$ or both subsets of $\{ \mybox{\lo},\mybox{\lr},\mybox{\lc},\mybox{\lx}\}$, for all $i \neq u$.
Due to Property~\ref{item:onecapdoubleprime} in the lemma, this behavior holds also for $i = u$.

Moreover, by \Cref{obs:union-non-comparable}, it must hold that $\L_u$ and $\L_{\sigma(u)}$ are non-comparable sets.
We thus get that, up to symmetries, $\L_u = \{\mybox{\lR},\mybox{\lX}\}$ and $\L'_{\sigma(u)} = \{\mybox{\lC},\mybox{\lX}\}$ or that $\L_u = \{\mybox{\lr},\mybox{\lx}\}$ and $\L'_{\sigma(u)} = \{\mybox{\lc},\mybox{\lx}\}$.

In the former case, up to symmetries, we must have
\begin{align*}
	\fC &= \{\mybox{\lR},\mybox{\lX}\} \s \{\mybox{\lc},\mybox{\lx}\} \s \{\mybox{\lo},\mybox{\lr},\mybox{\lc},\mybox{\lx}\} \s \dots \s \{\mybox{\lo},\mybox{\lr},\mybox{\lc},\mybox{\lx}\} \textrm{ and}\\
	\fC' &= \{\mybox{\lC},\mybox{\lX}\} \s \{\mybox{\lr},\mybox{\lx}\} \s \{\mybox{\lo},\mybox{\lr},\mybox{\lc},\mybox{\lx}\} \s \dots \s \{\mybox{\lo},\mybox{\lr},\mybox{\lc},\mybox{\lx}\}.
\end{align*}
Depending on the choice of $\sigma$ we obtain that either
\[
	\fC'' = \{\mybox{\lO},\mybox{\lR},\mybox{\lC},\mybox{\lX}\} \s \{\mybox{\lx}\} \s \{\mybox{\lo},\mybox{\lr},\mybox{\lc},\mybox{\lx}\} \s \dots \s \{\mybox{\lo},\mybox{\lr},\mybox{\lc},\mybox{\lx}\},
\]
which is in the list, or
\[
	\fC'' = \{\mybox{\lO},\mybox{\lR},\mybox{\lC},\mybox{\lX}\} \s \{\mybox{\lc},\mybox{\lx}\} \s \{\mybox{\lr},\mybox{\lx}\} \s \{\mybox{\lo},\mybox{\lr},\mybox{\lc},\mybox{\lx}\} \s \dots \s \{\mybox{\lo},\mybox{\lr},\mybox{\lc},\mybox{\lx}\},
\]
which is also in the list.

In the latter of the two cases discussed above, we have to consider four cases:
\begin{itemize}
	\item We have
	\begin{align*}
		\fC &= \fC'= \{\mybox{\lO},\mybox{\lR},\mybox{\lC},\mybox{\lX}\} \s \{\mybox{\lr},\mybox{\lx}\} \s \{\mybox{\lc},\mybox{\lx}\} \s \{\mybox{\lo},\mybox{\lr},\mybox{\lc},\mybox{\lx}\} \s \dots \s \{\mybox{\lo},\mybox{\lr},\mybox{\lc},\mybox{\lx}\}.
	\end{align*}
	In this case, the resulting configuration  $\fC''$ is dominated by
	\begin{align*}
			&\{\mybox{\lO},\mybox{\lR},\mybox{\lC},\mybox{\lX}\} \s \{\mybox{\lx}\} \s \{\mybox{\lo},\mybox{\lr},\mybox{\lc},\mybox{\lx}\} \s \dots \s \{\mybox{\lo},\mybox{\lr},\mybox{\lc},\mybox{\lx}\} \textrm{ or}\\
	&\{\mybox{\lO},\mybox{\lR},\mybox{\lC},\mybox{\lX}\} \s \{\mybox{\lr}, \mybox{\lx}\} \s \{\mybox{\lc}, \mybox{\lx}\} \s \{\mybox{\lo},\mybox{\lr},\mybox{\lc},\mybox{\lx}\} \s \dots \s \{\mybox{\lo},\mybox{\lr},\mybox{\lc},\mybox{\lx}\}.
	\end{align*}
	\item We have
	\begin{align*}
		\fC &= \{\mybox{\lR},\mybox{\lX}\} \s \{\mybox{\lc},\mybox{\lx}\} \s \{\mybox{\lo},\mybox{\lr},\mybox{\lc},\mybox{\lx}\} \s \dots \s \{\mybox{\lo},\mybox{\lr},\mybox{\lc},\mybox{\lx}\} \textrm{ and}\\
		\fC' &= \{\mybox{\lC},\mybox{\lX}\} \s \{\mybox{\lr},\mybox{\lx}\} \s \{\mybox{\lo},\mybox{\lr},\mybox{\lc},\mybox{\lx}\} \s \dots \s \{\mybox{\lo},\mybox{\lr},\mybox{\lc},\mybox{\lx}\}.
	\end{align*}
	In this case, $\fC''$ is dominated by $\{\mybox{\lX}\} \s \{\mybox{\lo},\mybox{\lr},\mybox{\lc},\mybox{\lx}\} \s \dots \s \{\mybox{\lo},\mybox{\lr},\mybox{\lc},\mybox{\lx}\}$.
	\item We have
	\begin{align*}
		\fC &= \{\mybox{\lR},\mybox{\lX}\} \s \{\mybox{\lc},\mybox{\lx}\} \s \{\mybox{\lo},\mybox{\lr},\mybox{\lc},\mybox{\lx}\} \s \dots \s \{\mybox{\lo},\mybox{\lr},\mybox{\lc},\mybox{\lx}\} \textrm{ and}\\
		\fC' &= \{\mybox{\lO},\mybox{\lR},\mybox{\lC},\mybox{\lX}\} \s \{\mybox{\lr},\mybox{\lx}\} \s \{\mybox{\lc},\mybox{\lx}\} \s \{\mybox{\lo},\mybox{\lr},\mybox{\lc},\mybox{\lx}\} \s \dots \s \{\mybox{\lo},\mybox{\lr},\mybox{\lc},\mybox{\lx}\}.
	\end{align*}
	In this case, $\fC''$ is dominated by $\fC$.
	\item We have
	\begin{align*}
		\fC &= \{\mybox{\lC},\mybox{\lX}\} \s \{\mybox{\lr},\mybox{\lx}\} \s \{\mybox{\lo},\mybox{\lr},\mybox{\lc},\mybox{\lx}\} \s \dots \s \{\mybox{\lo},\mybox{\lr},\mybox{\lc},\mybox{\lx}\} \textrm{ and}\\
		\fC' &= \{\mybox{\lO},\mybox{\lR},\mybox{\lC},\mybox{\lX}\} \s \{\mybox{\lr},\mybox{\lx}\} \s \{\mybox{\lc},\mybox{\lx}\} \s \{\mybox{\lo},\mybox{\lr},\mybox{\lc},\mybox{\lx}\} \s \dots \s \{\mybox{\lo},\mybox{\lr},\mybox{\lc},\mybox{\lx}\}.
	\end{align*}
	This case is symmetric to the previous one, i.e., again $\fC''$ is dominated by $\fC$.
	\qedhere
\end{itemize}
\end{proof}

For the sake of readability, we rename the obtained sets of labels as follows:
\begin{itemize}[noitemsep]
	\item $\mybox{\mybox{\lO}} := \{\mybox{\lO},\mybox{\lR},\mybox{\lC},\mybox{\lX}\}$
	\item $\mybox{\mybox{\lR}} := \{\mybox{\lR},\mybox{\lX}\}$
	\item $\mybox{\mybox{\lC}} := \{\mybox{\lC},\mybox{\lX}\}$
	\item $\mybox{\mybox{\lX}} := \{\mybox{\lX}\}$
	\item $\mybox{\mybox{\lo}} := \{\mybox{\lo},\mybox{\lr},\mybox{\lc},\mybox{\lx}\}$
	\item $\mybox{\mybox{\lr}} := \{\mybox{\lr},\mybox{\lx}\}$
	\item $\mybox{\mybox{\lc}} := \{\mybox{\lc},\mybox{\lx}\}$
	\item $\mybox{\mybox{\lx}} := \{\mybox{\lx}\}$
\end{itemize}
In the following, let $\Sigma''_{\orcx} = \{\mybox{\mybox{\lO}}, \mybox{\mybox{\lR}}, \mybox{\mybox{\lC}}, \mybox{\mybox{\lX}}, \mybox{\mybox{\lo}}, \mybox{\mybox{\lr}}, \mybox{\mybox{\lc}}, \mybox{\mybox{\lx}}\}$.

\paragraph{\boldmath Computing the constraint $\edgeconst''_{\orcx}$.}
We now compute $\edgeconst''_{\orcx}$, that is we apply the existential quantifier.

\begin{lemma}\label{lem:second-exists}
	The constraint $\edgeconst''_{\orcx} = \edgeconst_{\rere(\re(\orcx))}$ contains exactly the following condensed configurations. 
	\begin{itemize}
	\item $[\mybox{\mybox{\lO}}] \s [\mybox{\mybox{\lO}} \mybox{\mybox{\lR}} \mybox{\mybox{\lC}} \mybox{\mybox{\lX}}]$
	\item $[\mybox{\mybox{\lR}} \mybox{\mybox{\lO}}] \s [\mybox{\mybox{\lR}} \mybox{\mybox{\lO}}]$
	\item $[\mybox{\mybox{\lC}} \mybox{\mybox{\lO}}] \s [\mybox{\mybox{\lC}} \mybox{\mybox{\lO}}]$
	\item $[\mybox{\mybox{\lo}}] \s [\mybox{\mybox{\lo}} \mybox{\mybox{\lr}} \mybox{\mybox{\lc}} \mybox{\mybox{\lx}}]$
	\item $[\mybox{\mybox{\lr}} \mybox{\mybox{\lo}}] \s [\mybox{\mybox{\lc}} \mybox{\mybox{\lo}}]$
\end{itemize}
\end{lemma}
\begin{proof}
	The lemma directly follows from \Cref{obs:easy-exists}.
\end{proof}

\begin{proof}[Proof of \Cref{lem:orcx-fp}] In \Cref{lem:first-forall} and \Cref{lem:first-exists} we computed $\re(\orcx)$, while in \Cref{lem:second-forall} and \Cref{lem:second-exists} we computed $\rere(\re(\orcx))$. \Cref{lem:orcx-fp} follows by the fact that the equivalence of $\nodeconst''_{\orcx}$ with $\nodeconst_{\orcx}$ and the equivalence of $\edgeconst''_{\orcx}$ with $\edgeconst_{\orcx}$ is witnessed by the renaming $\mybox{\mybox{\ell}} \mapsto \ell$.
\end{proof}

\subsection{The ORCX problem is non-trivial}\label{ssec:orcx-nontrivial}

We previously showed that $\orcx$ is a fixed point under the round elimination framework. We now show that it is a nontrivial fixed point, even when a $\Delta$-edge coloring is given as input.
\begin{lemma}\label{lem:orcx-nontrivial}
	The problem $\orcx$ cannot be solved in $0$ rounds in the PN model, even if a $\Delta$-edge coloring is given.
\end{lemma}
\begin{proof}
	Consider a graph in which, for each node $v$, the $i$-th port of $v$ is assigned to the edge of color $i$ incident to $v$.
	In such a graph, any $0$-round deterministic PN algorithm must pick a node configuration $\L_1 \s \dots \s \L_{\Delta}$ from $\nodeconst_{\orcx}$ and output it on all nodes,  such that label $\L_i$ is output on the edge of color $\sigma(i)$, for some permutation $\sigma$.
	Observe that, if a node $v$ assigns label $\L$ to an incident edge $\{v,u\}$, then $u$ is also assigning $\L$ to the edge $\{v,u\}$.
	Hence, for this algorithm to be correct, all the edge configurations in $\{\L_1 \s \L_1, \dots, \L_{\Delta} \s \L_{\Delta}\}$ must be contained in $\edgeconst_{\orcx}$.

	Observe that all node configurations containing any of the labels $\lr$, $\lc$, $\lx$ cannot be used by the algorithm, as they would create an edge configuration from $\{\lr \s \lr, \lc \s \lc, \lx \s \lx\}$, neither of which is contained in $\edgeconst_{\orcx}$.
	The only remaining node configuration is $\lX \s \lo \s \dots \s \lo$.
	However, using such a configuration would create the edge configuration $\lX \s \lX$, which is not contained in $\edgeconst_{\orcx}$.
\end{proof}

\subsection{Putting things together}\label{ssec:lifting-orcx}

We now combine the statements proved above to prove~\Cref{thm:FP-for-ninecol} along the lines discussed before.
\begin{proof}[Proof of~\Cref{thm:FP-for-ninecol}]
	By combining \Cref{lem:orcx-fp} (that shows that $\orcx$ is a fixed point) and \Cref{lem:orcx-nontrivial} (that shows that $\orcx$ is nontrivial even under the assumed input), we obtain that $\orcx$ is a nontrivial fixed point, even if a $\Delta$-edge coloring is given.
	By \Cref{lem:nine-to-orcx}, in $\Delta$-edge colored graphs, it is possible to convert a solution for $\ninecol$ into a solution for $\orcx$, in $0$ rounds. Thus, given a $\Delta$-edge coloring, $\orcx$ is a relaxation of $\ninecol$.	
	It follows that, in the setting where a $\Delta$-coloring is given as input, there exists a nontrivial fixed point relaxation of $\ninecol$.
	Now~\Cref{cor:double-tau-relax} implies that there also exists a nontrivial fixed point relaxation of $\ninecol$ in the setting without input.	
\end{proof}

By~\Cref{cor:lifting-fp}, we obtain the desired lower bounds for any problem from $\fP$.

\begin{corollary}
	Each problem $\Pi \in \fP$ requires $\Omega(\log n)$ rounds to be solved by a deterministic algorithm and $\Omega(\log \log n)$ rounds to be solved by a randomized algorithm. In particular, for any $\Delta \geq 3$, problem $\ninecol$ requires $\Omega(\log n)$, resp.\ $\Omega(\log \log n)$, rounds to be solved deterministically, resp.\ randomized.
\end{corollary}

\section{Lifting with inputs}\label{sec:lifting}

In order to obtain a lower bound result using the round elimination technique, one has to first produce a sequence of problems, $\Pi_0, \Pi_1, \Pi_2, \ldots, \Pi_T$, such that $\Pi_0$ is the problem for which we want to prove a lower bound, $\Pi_T$ is nontrivial, and for all $i$ it holds that $\Pi_{i+1}$ is a relaxation of $\rere(\re(\Pi_i))$. Once such a sequence is obtained, a lower bound in the LOCAL model is automatically obtained by applying a so-called lifting theorem \cite{balliu-brandt-etal-2022-distributed-delta-coloring}. The strongest lifting theorem that is currently known works in regular graphs in which either no input is provided or the input comes from some specific class. However, the input that we assumed in \Cref{sec:sso-with-so} does not belong to this class. Thus, in this section, we prove a stronger lifting theorem. We actually prove a lifting theorem that is stronger than what we need for the purposes of this paper: the new lifting theorem that we prove holds also in non-regular graphs and also in the case where the input can be an arbitrary problem expressed in the black-white formalism.

\subsection{Preliminaries}

In this section, we consider a class of problems that is strictly larger than the one defined in \Cref{sec:Preliminaries}. For this reason, we define such a class of problems, and we then define the functions $\re$ and $\rere$ for this setting.

\paragraph{Problems with input in the black-white formalism.}
In more detail, we define problems on bipartite graphs that are not necessarily regular and that have constraints that can depend on some given input.
Note that this setting contains the problems defined in \Cref{sec:Preliminaries}: a graph can be seen as a bipartite $2$-colored graph, where nodes are white nodes and edges are black nodes of degree $2$.

\begin{definition}[Problems with input in the black-white formalism]
	A problem $\Pi$ with input in the black-white formalism is a tuple $(\Sigma_{\mathrm{in}}, \Sigma_{\mathrm{out}}, \nodeconst,\edgeconst,g)$, where:
	\begin{itemize}
		\item $\Sigma_{\mathrm{in}}$ and $\Sigma_{\mathrm{out}}$ are finite sets representing, respectively, the input and output labels of $\Pi$.
		\item $\nodeconst = (\nodeconst^1,\ldots,\nodeconst^\Delta)$ consists of $\Delta$ collections $\nodeconst^i$ of cardinality-$i$ multisets $\{\ell_1,\ldots,\ell_i\}$ with $\ell_1,\ldots,\ell_i$ from $\Sigma_{\mathrm{out}}$. $\nodeconst$ is the \emph{white constraint} of $\Pi$. $\nodeconst^i$ represents the constraint of white nodes of degree $i$.
		\item $\edgeconst = (\edgeconst^1,\ldots,\edgeconst^\delta)$ consists of $\Delta$ collections $\nodeconst^i$ of cardinality-$i$ multisets $\{\ell_1,\ldots,\ell_i\}$ with $\ell_1,\ldots,\ell_i$ from $\Sigma_{\mathrm{out}}$. 
		$\edgeconst$ is the \emph{black constraint} of $\Pi$. $\edgeconst^i$ represents the constraint of black nodes of degree $i$.
		\item $g : \Sigma_{\mathrm{in}} \rightarrow 2^{\Sigma_{\mathrm{out}}}$ is a function that assigns to each label from $\Sigma_{\mathrm{in}}$ a subset of labels from $\Sigma_{\mathrm{out}}$.
	\end{itemize}
	The \emph{degree of $\Pi$} is defined as the maximum between $\Delta$ and $\delta$.
	Let $G = (U,V,E)$ be a $2$-colored bipartite graph. Let $U$ be the set of white nodes and $V$ be the set of black nodes. Assume that white nodes have maximum degree $\Delta$ and black nodes have maximum degree $\delta$, and that to each edge $e$  is assigned a label $f_{\mathrm{in}}(e)$ from $\Sigma_{\mathrm{in}}$.
	A correct solution for $\Pi$ on $G$ is given by an edge labeling $f_{\mathrm{out}} : E(G) \rightarrow \Sigma_{\mathrm{out}}$ such that the following holds: 
	\begin{itemize}
		\item For each white node $u$, the multiset consisting of the labels assigned by $f_{\mathrm{out}}$ to the edges incident to $v$ is contained in $\nodeconst^{\deg(u)}$.
		\item For each black node $v$, the multiset consisting of the labels assigned by $f_{\mathrm{out}}$ to the edges incident to $v$ is contained in $\edgeconst^{\deg(v)}$.
		\item For every edge $e$, the label $f_{\mathrm{out}}(e)$ is contained in the label set $g(f_{\mathrm{in}}(e))$.
	\end{itemize}	
\end{definition}

\paragraph{White and black algorithms.}
In order to apply round elimination for problems in the black-white formalism, we need to introduce the concept of white and black algorithms.
A white algorithm for $\Pi$ running on $G = (U,V,E)$ is an algorithm that is executed by white nodes, and in which black nodes act as passive relayers of messages. In such a setting, the output is only produced by white nodes. The output of a white node $u$ defines $f_{\mathrm{out}}(e)$ for each edge $e$ incident to $u$.
The runtime complexity of an algorithm is defined such that there is a natural mapping between the runtime and the distance at which a node sees on $G$. For this reason, if two white nodes $u$ and $u'$ are incident to the same black node $v$, then we define the number of rounds that it takes for $u$ and $u'$ to communicate as $2$.
A black algorithm is defined in the same way, but by reversing the roles of white and black nodes. 

\paragraph{Round elimination.}
We now define $\re(\Pi)$ and $\rere(\Pi)$ for problems $\Pi$ with input in the black-white formalism. Note that round elimination in the case of problems with input has already been introduced in \cite{grunau-rozhon-brandt-2022-the-landscape-of-distributed}, and we now report its trivial generalization to the case in which the black nodes have a degree that is not necessarily $2$.
\begin{definition}[The function $\re$]
	Let $\Pi = (\Sigma^{\Pi}_{\mathrm{in}}, \Sigma^{\Pi}_{\mathrm{out}}, \nodeconst_{\Pi} = (\nodeconst_{\Pi}^1,\ldots,\nodeconst_{\Pi}^\Delta),\edgeconst_{\Pi} = (\edgeconst_{\Pi}^1,\ldots,\edgeconst_{\Pi}^\delta),g_{\Pi})$ be a problem with input in the black-white formalism.
	Then, we define the problem $\re(\Pi) = (\Sigma^{\re(\Pi)}_{\mathrm{in}}, \Sigma^{\re(\Pi)}_{\mathrm{out}}, \nodeconst_{\re(\Pi)},\edgeconst_{\re(\Pi)},g_{\re(\Pi)})$ as follows:
	\begin{itemize}
		\item $\Sigma^{\re(\Pi)}_{\mathrm{in}} := \Sigma^{\Pi}_{\mathrm{in}}$.
		\item $\edgeconst_{\re(\Pi)}$ is defined as $(\edgeconst_{\re(\Pi)}^1,\ldots,\edgeconst_{\re(\Pi)}^\delta)$, where $\edgeconst_{\re(\Pi)}^i$ contains all the multisets $\{\L_1,\ldots,\L_i\}$, where $\L_i \subseteq 2^{\Sigma^{\Pi}_{\mathrm{out}}}$, such that:
		\begin{itemize}
			\item For any choice $(\ell_1,\ldots,\ell_i) \in \L_1 \times \ldots \times \L_i$, the multiset $\{\ell_1,\ldots,\ell_i\}$ is in $\edgeconst_{\Pi}^i$;
			\item By adding any label in $\Sigma^{\Pi}_{\mathrm{out}}$ to any set $\L_i$, the above condition is not satisfied anymore.
		\end{itemize} 
		\item $\Sigma^{\re(\Pi)}_{\mathrm{out}} := \bigcup_{i=1}^{\delta} \bigcup_{\fC \in \edgeconst_{\re(\Pi)}^i} \fC$. That is, the set of output labels of $\re(\Pi)$ is the set of sets that appear in at least one configuration contained in $\edgeconst_{\re(\Pi)}$.
		\item $\nodeconst_{\re(\Pi)}$ is defined as $(\nodeconst_{\re(\Pi)}^1,\ldots,\nodeconst_{\re(\Pi)}^\Delta)$, where $\nodeconst_{\re(\Pi)}^i$ contains all the multisets $\{\L_1,\ldots,\L_i\}$, where $\L_i \in \Sigma^{\re(\Pi)}_{\mathrm{out}}$, such that there exists a choice $(\ell_1,\ldots,\ell_i) \in \L_1 \times \ldots \times \L_i$ satisfying that the multiset $\{\ell_1,\ldots,\ell_i\}$ is in $\nodeconst_{\Pi}^i$;
		\item For each $\ell \in \Sigma^{\Pi}_{\mathrm{in}}$, $g_{\re(\Pi)}$ is defined as  $g_{\re(\Pi)}(\ell) := 2^{g_{\Pi}(\ell)} \cap \Sigma^{\re(\Pi)}_{\mathrm{out}}$. In other words, a label of $\Sigma^{\re(\Pi)}_{\mathrm{out}}$, which is a subset of $\Sigma^{\Pi}_{\mathrm{out}}$, is allowed on input $\ell$ if and only if all its elements are allowed with $\ell$ according to the function $g_{\Pi}$.
	\end{itemize}
\end{definition}

\begin{definition}[The function $\rere$]
	The function $\rere(\Pi)$ is defined analogously as $\re$, except that the roles of $\nodeconst$ and $\edgeconst$ are reversed.
	In more detail, let $\Pi = (\Sigma^{\Pi}_{\mathrm{in}}, \Sigma^{\Pi}_{\mathrm{out}}, \nodeconst_{\Pi},\edgeconst_{\Pi},g_{\Pi})$, and let $\Pi' =  (\Sigma^{\Pi}_{\mathrm{in}}, \Sigma^{\Pi}_{\mathrm{out}}, \edgeconst_{\Pi},\nodeconst_{\Pi},g_{\Pi})$.
	Let $\re(\Pi') = (\Sigma^{\re(\Pi')}_{\mathrm{in}}, \Sigma^{\re(\Pi')}_{\mathrm{out}}, \edgeconst_{\re(\Pi')},\nodeconst_{\re(\Pi')},g_{\re(\Pi')})$.
	The problem $\rere(\Pi)$ is defined as $(\Sigma^{\re(\Pi')}_{\mathrm{in}}, \Sigma^{\re(\Pi')}_{\mathrm{out}}, \nodeconst_{\re(\Pi')},\edgeconst_{\re(\Pi')},g_{\re(\Pi')})$.
\end{definition}

\paragraph{Relaxations.}
The definition of relaxations, provided in \Cref{sec:Preliminaries}, directly extend to the setting considered in this section. In particular, a problem $\Pi'$ is a white (resp.\ black) relaxation of a problem $\Pi$ if there exists a white (resp.\ black) $0$-round algorithm that solves $\Pi'$ given a solution for $\Pi$, where the $0$-round algorithm can exploit the given input.

\subsection{The extended lifting theorem}

\begin{theorem}\label{thm:newlifting}
	Let $\Pi_0, \Pi_1, \ldots, \Pi_{k}$ be a sequence of problems with input in the black-white formalism, where input labels come from $\Sigma_{\mathrm{in}}$. Let $\Pi_{\mathrm{in}}$ be a problem (without inputs) in the black-white formalism where output labels come from $\Sigma_{\mathrm{in}}$.
	Assume that, for all $0 \le i < k$, and for an integer $L$, the following holds:
	\begin{itemize}[noitemsep]
		\item There exists a problem $\Pi'_i$ that is a black relaxation of $\re(\Pi_i)$;
		\item $\Pi_{i+1}$ is a white relaxation of $\rere(\Pi'_i)$;
		\item The number of input and output labels of $\Pi_i$, and the ones of $\Pi'_i$, are upper bounded by $L$.
	\end{itemize}
	Also, assume that $\Pi_k$ has at most $L$ input and output labels and is not $0$-round solvable in the deterministic PN model, even if a solution for $\Pi_{\mathrm{in}}$ is given as input. Let $\Delta$ be the degree of $\Pi_0$. Then, for all $n$, on trees of $n$ nodes and maximum degree at most $\Delta$, $\Pi_0$ requires at least \[\textstyle\min\{k-1, \frac{1}{16} (\log_\Delta n -  \log_\Delta \log L -624)\}\] rounds in the deterministic LOCAL model and at least \[\textstyle\min\{k-1, \frac{1}{16} (\log_\Delta \log n -  \log_\Delta \log L -5)\}\] rounds in the randomized LOCAL model, even if a solution for $\Pi_{\mathrm{in}}$ is given as input.
\end{theorem}

Before proving the theorem, we discuss how to use such a theorem to obtain lower bounds starting from sequences of problems.
The following corollary shows that a nontrivial fixed point directly implies a lower bound.
\begin{corollary}\label{cor:lifting-fp}
	Let $\Pi$ be a problem in the black-white formalism satisfying $\rere(\re(\Pi)) = \Pi$, and where inputs come from $\Sigma_{\mathrm{in}}$. Let $\Pi_{\mathrm{in}}$ be a problem (without inputs) in the black-white formalism where output labels come from $\Sigma_{\mathrm{in}}$.
	Assume that $\Pi$ is not $0$-round solvable in the deterministic PN model, even if a solution for $\Pi_{\mathrm{in}}$ is given as input. Then, on trees of $n$ nodes and maximum degree at most $\Delta = O(1)$, $\Pi$ requires  $\Omega(\log n)$ rounds in the deterministic LOCAL model and $\Omega(\log \log n)$ rounds in the randomized LOCAL model, even if a solution for $\Pi_{\mathrm{in}}$ is given as input.
\end{corollary}
\begin{proof}
	Since $\Pi$ is a nontrivial fixed point, for any $t$ we can construct a sequence of length $t$ satisfying the requirements of \Cref{thm:newlifting}. Hence, the lower bounds of \Cref{thm:newlifting} are determined by the second terms in the $\min$ operator.
	Let $L$ be the maximum between the number of input and output labels of $\Pi$, and let $\Delta$ be the degree of $\Pi$. Observe that $L$ may depend on $\Delta$ but cannot depend on $n$. Since $\Delta$ is a constant that solely depends on $\Pi$, the claim follows.
\end{proof}

A similar statement holds for infinite sequences, assuming that the number of labels does not grow too fast. The following corollary considers the case in which the number of labels grows linearly, but it can be easily generalized to other cases.
\begin{corollary}\label{cor:lifting-linear-growth}
	Let $\Pi_0, \Pi_1, \ldots$ be an infinite sequence of problems with input in the black-white formalism, where input labels come from $\Sigma_{\mathrm{in}}$. Let $\Pi_{\mathrm{in}}$ be a problem (without inputs) in the black-white formalism where output labels come from $\Sigma_{\mathrm{in}}$.
	Assume that, for all $i \ge 0$, and for some constant $c$, the following holds:
	\begin{itemize}[noitemsep]
		\item there exists a problem $\Pi'_i$ that is a black relaxation of $\re(\Pi_i)$,
		\item $\Pi_{i+1}$ is a white relaxation of $\rere(\Pi'_i)$,
		\item the number of input and output labels of $\Pi_i$, and the ones of $\Pi'_i$, are upper bounded by $c \cdot (i+1)$, and
		\item $\Pi_i$ is not $0$-round solvable in the deterministic PN model, even if a solution for $\Pi_{\mathrm{in}}$ is given as input.
	\end{itemize}
	Let $\Delta$ be the degree of $\Pi_0$. Then, for all $n$, on trees of $n$ nodes and maximum degree at most $\Delta$, $\Pi_0$ requires $\Omega(\log n)$ rounds in the deterministic LOCAL model and $\Omega(\log \log n)$ rounds in the randomized LOCAL model, even if a solution for $\Pi_{\mathrm{in}}$ is given as input.
\end{corollary}
\begin{proof}
	For all $i$, we apply \Cref{thm:newlifting} on the prefix sequence of length $i+1$.
	We obtain that, for all $i$ and for all $n$, $\Pi_0$ requires at least $\min\{i-1, \frac{1}{16} (\log_\Delta n -  \log_\Delta \log (c(i+1)) -624)\}$ rounds in the deterministic LOCAL model and at least $\min\{i-1, \frac{1}{16} (\log_\Delta \log n -  \log_\Delta \log (c(i+1)) -5)\}$ rounds in the randomized LOCAL model, even if a solution for $\Pi_{\mathrm{in}}$ is given as input.
	Hence, for each $n$, we obtain infinite different lower bounds, parameterized by $i$. For each fixed $n$, we can obtain an asymptotically optimal lower bound by considering the case in which $i = \log n$, obtaining a lower bound of $\Omega(\log n)$ for deterministic algorithms and $\Omega(\log \log n)$ for randomized ones. (Note that, similarly as in \Cref{cor:lifting-fp}, $\Delta$ is a constant that depends solely on $\Pi_0$.)
\end{proof}

In some cases, we just know that an infinite sequence exists, but we have no bound on the number of labels. In such a case, we directly obtain a lower bound of $\Omega(\log^* n)$.
A similar statement holds for infinite sequences, assuming that the number of labels does not grow too fast. The following corollary considers the case in which the number of labels grows linearly, but it can be easily generalized to other cases.
\begin{corollary}
	Let $\Pi_0, \Pi_1, \ldots$ be an infinite sequence of problems with input in the black-white formalism, where input labels come from $\Sigma_{\mathrm{in}}$. Let $\Pi_{\mathrm{in}}$ be a problem (without inputs) in the black-white formalism where output labels come from $\Sigma_{\mathrm{in}}$.
	Assume that, for all $i \ge 0$, the following holds:
	\begin{itemize}[noitemsep]
		\item there exists a problem $\Pi'_i$ that is a black relaxation of $\re(\Pi_i)$,
		\item $\Pi_{i+1}$ is a white relaxation of $\rere(\Pi'_i)$, and
		\item $\Pi_i$ is not $0$-round solvable in the deterministic PN model, even if a solution for $\Pi_{\mathrm{in}}$ is given as input.
	\end{itemize}
	Let $\Delta$ be the degree of $\Pi_0$. Then, for all $n$, on trees of $n$ nodes and maximum degree at most $\Delta$, $\Pi_0$ requires $\Omega(\log^* n)$ rounds for both randomized and deterministic LOCAL algorithms, even if a solution for $\Pi_{\mathrm{in}}$ is given as input.
\end{corollary}
\begin{proof}
	By the definition of $\re$ and $\rere$, for all $i$, the number of labels of $\Pi_{i+1}$ is at most doubly exponentially larger than the number of labels of $\Pi_{i}$.
    Hence, an upper bound on the number of labels of problem $\Pi_i$ is roughly a power tower of height $2i$.

	Similarly as in the proof of \Cref{cor:lifting-linear-growth}, we apply \Cref{thm:newlifting} on each prefix sequence of length $i+1$, and for each $n$ we then optimize the lower bound by considering the case $i = \frac{1}{2}\log^* n - O(1)$.
\end{proof}

We now consider the case in which we want to prove a lower bound as a function of $\Delta$. In this case, we usually define a problem $\mathcal{P}$ (e.g., MIS) as a family of problems $\Pi^j$ (e.g., MIS$^j$), one for each degree $j$, and we create a round elimination sequence for each such problem. In the case of inputs, we additionally assume that for each $j$, in graphs of degree $j$, we receive as input a solution for $\Pi^j_{\mathrm{in}}$. The following corollary captures exactly this case.
\begin{corollary}
	Let $\mathcal{P} = (\Pi^j \mid j>0)$ be a family of problems, where $\Pi^j$ is a problem of degree $j$ in the black-white formalism, with inputs from $\Sigma_{\mathrm{in}}^{\Pi^j}$.
	Let $\mathcal{P}_{\mathrm{in}} = (\Pi^j_{\mathrm{in}} \mid j>0)$ be a family of problems (without inputs) in the black-white formalism where output labels come from $\Sigma_{\mathrm{in}}^{\Pi^j}$.
	Suppose that, for each $j$, there exists a sequence of problems starting from $\Pi^j$ satisfying the requirements of \Cref{thm:newlifting} w.r.t.\ input $\Pi^j_{\mathrm{in}}$ and of length $f(j)$, for some $f$. 
	Moreover, suppose that, for each $j$, the number of labels of $\Pi^j$ is at most $2^{j^c}$ for some universal constant $c$.
	Then, $\mathcal{P}$ requires \[\Omega(\min\{f(\Delta), \log_\Delta n\})\] rounds for deterministic algorithms and \[\Omega(\min\{f(\Delta), \log_\Delta \log n\})\] for randomized ones, in $n$-node trees of maximum degree $\Delta$, even if a solution for $\mathcal{P}_{\mathrm{in}}$ is given as input.
\end{corollary}
\begin{proof}
The claim follows by applying \Cref{thm:newlifting} on each sequence independently.
\end{proof}

\subsection{Lifting theorem in a nutshell}

We devote the rest of this section for proving \Cref{thm:newlifting}.
Our proof uses ingredients that are already present in the literature \cite{balliu-brandt-etal-2019-lower-bounds-for-maximal, balliu-brandt-olivetti-2020-distributed-lower-bounds, brandt-olivetti-2020-truly-tight-in-delta-bounds-for, balliu-brandt-etal-2020-classification-of-distributed, balliu-brandt-etal-2022-distributed-delta-coloring}, and the proofs that we provide in this section are essentially an extension of the proofs present in \cite[Appendix~A in the arXiv version]{balliu-brandt-etal-2022-distributed-delta-coloring}. In the following, we summarize the high-level ideas behind the lifting theorem. By \emph{(local) failure probability} of an algorithm we denote the worst-case probability, taken over all nodes, that the algorithm fails on that node, i.e., produces a solution that does not satisfy the (hyper)edge constraint on at least one (hyper)edge incident to that node.

In order to lift a result obtained with round elimination to the LOCAL model, we need to follow the steps listed below.
\begin{enumerate}
	\item Let $\Pi'$ be a relaxation of $\rere(\re(\Pi))$. Prove that, if there exists a randomized $k$-round algorithm for $\Pi$ with failure probability $p$ (which could depend on $n$), then there exists a randomized $(k-1)$-round algorithm for $\Pi'$ with some failure probability $p'$ that is not too large compared to $p$.\label{point:randomre}
	\item Prove that, if a problem cannot be solved in $0$ rounds in the deterministic PN model, then any $0$-rounds randomized PN algorithm for the same problem must fail with some large probability.\label{point:zerorounds}
	\item Apply \Cref{point:randomre} recursively. The number of times that we can apply this recursive procedure before obtaining a failure probability that is larger than the one in \Cref{point:zerorounds} is a lower bound on the runtime for solving the problem in the randomized PN model. This directly gives a lower bound in the randomized LOCAL model as well.\label{point:combine}\label{point:withn}\label{point:randlocal}
	\item Prove that a randomized lower bound in the LOCAL model implies an even stronger deterministic lower bound.\label{point:detlocal}
\end{enumerate}

\subsection{Evolution of the failure probability in a single step of round elimination}

We first observe that, existing results in the literature, imply that \Cref{point:randomre} holds also in the case in which the graph is not necessarily regular and in the presence of inputs. 

\begin{lemma}[Theorem 3.4 of \cite{grunau-rozhon-brandt-2022-the-landscape-of-distributed}]\label{lem:singlestep-orig}
	Let $\mathcal{F}$ be the set containing all forests of maximum degree at most $\Delta$. For any set $N$ of positive integers, let $\mathcal{F}_N$ denote the class of forests with a number of nodes that is contained in $N$.
	Let $\Pi$ be a node-edge-checkable LCL problem and $\mathcal{A}$ a randomized algorithm solving $\Pi$ on $\mathcal{F}$ with runtime $T(n)$ and local failure probability at most $p \le 1$. Let $N$ be the set of all positive integers $n$ satisfying $T(n) + 2 \le \log_\Delta n$. Then, there exists a randomized algorithm $\mathcal{A}'$ solving $\rere(\re(\Pi))$ on $\mathcal{F}_N$ with runtime $\max\{0, T(n)-1\}$ and local failure probability at most $S p^{1/(3\Delta +3)}$, where
	\[
	S = \Bigl(10 \Delta\bigl(|\Sigma_{\mathrm{in}}| + \max \{|\Sigma^{\Pi}_{\mathrm{out}}| ,|\Sigma^{\re(\Pi)}_{\mathrm{out}}| \}\bigr)\Bigr)^{4\Delta^{T(n)+1}}.
	\]
\end{lemma}
In other words, the above lemma says that, if there exists a randomized $k$-rounds algorithm for $\Pi$ that has some failure probability $p$, then we can construct a $(k-1)$-rounds algorithm that solves $\rere(\re(\Pi))$ (and hence also a relaxation of it) with failure probability that is not too large compared to $p$.

While this lemma holds for graphs, in the following we show that it is straightforward to extend it to the case of 2-colored bipartite graphs and problems expressed in the black-white formalism.

In the following, let $\Pi$ be a problem in the black-white formalism, let $\Delta$ be its degree, let $\Sigma_{\mathrm{in}}$ be the set of input labels of $\Pi$, let $\Sigma^{\Pi}_{\mathrm{out}}$ be the set of output labels of $\Pi$, and let $\Pi_{\mathrm{in}}$ be a problem (without inputs) in the black-white formalism where output labels come from $\Sigma_{\mathrm{in}}$.
Let $\mathcal{A}$ be a randomized white algorithm solving $\Pi$ on $\mathcal{F}$ with runtime at most $T(n)$ and local failure probability at most $p \le 1$, when a solution for $\Pi_{\mathrm{in}}$ is given as input\footnote{In the original paper, it is assumed that algorithm $\mathcal{A}$ is able to solve $\Pi$ for any input that comes from $\Sigma_{\mathrm{in}}$.	 However, it is easy to check that the same proof holds in the case in which it is assumed that the input is a solution for a fixed problem $\Pi_{\mathrm{in}}$.}.

To simplify the notation, in the following, we state a version of \Cref{lem:singlestep-orig} for problems in the black-white formalism, for a fixed choice of $n$.
Fix a value of $n$, assume that $T(n)$ is even, and let $2T = T(n)$. Assume that $2T+2 \le \log_\Delta n$.
Moreover, let $\mathcal{F}_{n,\Delta}$ be the set of forests of maximum degree at most $\Delta$ and containing $n$ nodes.  
\begin{lemma}\label{lem:singlestep}
    Let $\Pi'$ be a black relaxation of $\re(\Pi)$, and let $\Sigma^{\Pi'}_{\mathrm{out}}$ be the set of output labels of $\Pi'$. 
	Then, there exists a randomized white algorithm $\mathcal{A}''$ solving $\rere(\Pi')$ on $\mathcal{F}_{n,\Delta}$  when a solution for $\Pi_{\mathrm{in}}$ is given as input,  with runtime $\max\{0, 2T-2\}$ and local failure probability at most \[p'' \le 2 \Delta ( s + |\Sigma^{\Pi'}_{\mathrm{out}}| ) (p')^{1/(\Delta+1)},\] where $s = (3 |\Sigma_{\mathrm{in}}|)^{2\Delta^{2T+2}}$, and  $p' \le 2 \Delta ( s + |\Sigma^{\Pi}_{\mathrm{out}}| ) p^{1/(\Delta+1)}$.
\end{lemma}
\begin{proof}
	While the original proof is phrased on graphs, for the ease of the presentation we now express it in the case of bipartite $2$-colored graphs in which black nodes (which correspond to edges) have degree exactly $2$. The proof of \cite{grunau-rozhon-brandt-2022-the-landscape-of-distributed} works as follows.
	
	Given a white algorithm $\mathcal{A}$ for $\Pi$, they first use $\mathcal{A}$  to construct a black algorithm $\mathcal{A}'$ for $\re(\Pi)$ that has complexity $2T - 1$. Observe that $\mathcal{A}'$  solves $\Pi'$ as well. Then, they use $\mathcal{A}'$ to construct a white algorithm $\mathcal{A}''$ for $\rere(\Pi')$ that has complexity $2T-2$. This construction is done such that the failure probability of $\mathcal{A}''$ is not too large compared to the one of $\mathcal{A}$.

	The black algorithm $\mathcal{A}'$ running on a black node $v$ is constructed as follows: gather the $(2T-1)$-radius neighborhood of $v$, imagine all possible extensions of the gathered neighborhood into a $(2T+1)$-radius neighborhood one, and for each of them compute the output of $\mathcal{A}$ that the white neighbors of $v$ give on the edges incident to $v$. The assumption $2T+2 \le \log_\Delta n$ is used to guarantee that $\mathcal{A}$ does not see more than $n$ nodes, and hence it works correctly. The output of $\mathcal{A}'$ on such edges is the set of labels that  $\mathcal{A}$ outputs on a fraction of the extensions that is at least $K$, for some parameter $0 < K < 1$.
	The authors of \cite{grunau-rozhon-brandt-2022-the-landscape-of-distributed}, in \cite[Lemma 3.7, arXiv version]{grunau-rozhon-brandt-2022-the-landscape-of-distributed}, show that, if the failure probability of $\mathcal{A}$ is $p$, then the failure probability of $\mathcal{A}'$ is at most $2 \Delta ( s + |\Sigma^{\Pi}_{\mathrm{out}}| ) p^{1/3}$, where $s = (3 |\Sigma_{\mathrm{in}}|)^{2\Delta^{2T+1}}$.
	This bound is obtained as follows:
	\begin{itemize}[noitemsep]
		\item On white nodes, the failure probability is upper bounded by $p +  |\Sigma^{\Pi}_{\mathrm{out}}| \Delta K + p s \Delta / K$.
		\item On black nodes, the failure probability is upper bounded by $p  s / K^2$.
	\end{itemize}
	Then, the bound is obtained by picking $K = p^{1/3}$.
	In the above formulas, $s$ is the number of possible extensions, and the exponent $2$ on $K$ on the second bound comes from the fact that a black node has $2$ white neighbors.

	By going through the proof of these lemmas, it is easy to see that they work also in the case in which black nodes have degree at most $\Delta$, by using the following new bound on the failure probability of $\mathcal{A}'$ on black nodes:
	\[
	\frac{p \cdot s}{K^\Delta}, \text{ where } s = (3 |\Sigma_{\mathrm{in}}|)^{2\Delta^{2T+2}}.
	\]
	Indeed, a black node has at most $\Delta$ white neighbors, and the possible extensions of a $(2T-1)$-radius neighborhood into a $(2T+1)$-radius neighborhood are at most $s$:
	\begin{itemize}
		\item There are at most $\Delta^{2T}$ black nodes at distance $2T$ from the center, and at most $\Delta^{2T+1}$ white nodes at distance $2T$ from the center. Hence, there are at most $(\Delta+1)\Delta^{2T} \le 2 \Delta^{2T+1}$ nodes in the extension.
		\item There are at most $2 \Delta^{2T+2}$ edges incident to the nodes in the extension.
		\item In order to describe an extension, for each edge, we need to express its label, which comes from a domain of size  $|\Sigma_{\mathrm{in}}|$, and one of the following three possibilities:
		\begin{itemize}
			\item The edge does not exist.
			\item The edge exists, and connects the node to another node that is nearer to the center.
			\item The edge exists, and connects the node to another node that is further away.
		\end{itemize}
		\item Hence, up to isomorphism, there are at most $s = (3 |\Sigma_{\mathrm{in}}|)^{2\Delta^{2T+2}}$ possible extensions.
	\end{itemize}

By picking $K = p^{\frac{1}{\Delta+1}}$, we obtain that the failure probability of $\mathcal{A}'$ is at most \[p' \le 2 \Delta ( s + |\Sigma^{\Pi}_{\mathrm{out}}| ) p^{1/(\Delta+1)},\] where $s = (3 |\Sigma_{\mathrm{in}}|)^{2\Delta^{2T+2}}$.

The algorithm $\mathcal{A}''$ is constructed analogously (by considering neighborhoods of radius $2T-2$ and extensions of radius $2T$), and its failure probability is at most \[p'' \le 2 \Delta ( s + |\Sigma^{\Pi'}_{\mathrm{out}}| ) (p')^{1/(\Delta+1)},\] where $p'$ is an upper bound on the failure probability of $\mathcal{A}'$.
Hence, the claim follows.
\end{proof}

\subsection{Zero-round solvability}

We now show that \Cref{point:zerorounds} holds also in the case of (possibly) non-regular graphs and for problems with input in the black-white formalism.
In other words, we show that if a problem $\Pi$ cannot be solved in $0$ rounds in the deterministic PN model, then any randomized $0$-rounds PN algorithm for $\Pi$ must fail with some large probability.
\begin{lemma}\label{lem:zerorounds}
	Let $\Pi$ be a problem with input that cannot be solved in $0$ rounds with a deterministic white algorithm in the PN model, even if a solution for $\Pi_{\mathrm{in}}$ is given. Then, any randomized $0$-round algorithm solving $\Pi$ must fail with probability at least $\frac{1}{L^{\Delta^2}}$, even if a solution for $\Pi_{\mathrm{in}}$ is given.
\end{lemma}
\begin{proof}
	In $0$ rounds, an algorithm running on node $v$ only sees the input assigned to $v$, and the ports assigned to $v$ (which includes the degree of $v$).
	Hence, we can see any $0$-round algorithm as a function mapping each possible input assigned to white nodes and a permutation of a subset of $\{1,\ldots,\Delta\}$ into a probability assignment to the white output configurations of $\Pi$.

	Since $\Delta$ is an upper bound on the degree of the nodes, and $L$ is an upper bound on the number of output labels, for each input and port assignment there are at most $L^\Delta$ possible output configurations. Hence, there exists a configuration $\fC$ to which the algorithm assigns probability at least $\frac{1}{L^\Delta}$.  

	Since $\Pi$ is not solvable in $0$ rounds with a white algorithm in the deterministic PN model, then for each (deterministic) mapping from input and port assignment to outputs there exists a way to connect (at most $\Delta$) white nodes to a single black node, and assign inputs and ports to these nodes, such that, by applying the considered mapping, we obtain an invalid configuration on the black node.
    
	In the case of randomized algorithms, for each input and port assignment on a white node, there exists a configuration that is given with probability at least $\frac{1}{L^\Delta}$. Hence, it is possible to connect white nodes to a single black node such that the black node obtains an invalid configuration with probability at least $(\frac{1}{L^\Delta})^\Delta = \frac{1}{L^{\Delta^2}}$.
\end{proof}

\subsection{Evolution of the failure probability in multiple steps of round elimination}

\begin{lemma}\label{lem:multiplesteps}
	Let $\Pi_0, \Pi_1, \dots, \Pi_{T}$ be a sequence of problems satisfying the conditions of \Cref{thm:newlifting}.
	Let $\Delta$ be the degree of $\Pi_0$.
	Let $n$ be an integer, and assume that $2T+2 \le \log_\Delta n$.
	Let $\mathcal{A}$ be a randomized $2T$-round white algorithm for $\Pi_0$ on $\mathcal{F}_{n,\Delta}$  with local failure probability at most $p$ (which may depend on $n$). 
	Then there exists a randomized $(2T-2j)$-round white algorithm $\mathcal{A}'$ for $\Pi_{j}$ with local failure probability at most $(2\Delta (s+L))^2 p^{1/(\Delta+1)^{2j}}$, for all $0<j \le T$, where $s = (3 |\Sigma_{\mathrm{in}}|)^{2\Delta^{2T+2}}$.
\end{lemma}
\begin{proof}
	By assumption, $|\Sigma_{\mathrm{out}}^{\Pi}|$ and $|\Sigma_{\mathrm{out}}^{\Pi'}|$ are upper bounded by $L$. Hence, by \Cref{lem:singlestep}, we obtain: 
	\begin{align*}
	p'' &\le 2 \Delta ( s + |\Sigma^{\Pi'}_{\mathrm{out}}| ) (p')^{1/(\Delta+1)} \\
	    &\le 2 \Delta ( s + L) (p')^{1/(\Delta+1)}.
	\end{align*}
	Similarly,
	\begin{align*}
	p' &\le 2 \Delta ( s + |\Sigma^{\Pi}_{\mathrm{out}}| ) p^{1/(\Delta+1)} \\
	    &\le 2 \Delta ( s + L) p^{1/(\Delta+1)}.
	\end{align*}
	Summarizing, we get that:
	\begin{align*}
	p''&\le 2\Delta (s+L) (p')^{\frac{1}{\Delta+1}}, \text{ where }\\
	p' &\le 2\Delta (s+L) p^{\frac{1}{\Delta+1}}.
	\end{align*}
	By recursively applying \Cref{lem:singlestep}, we get the following:
	\[
	p_j \le  2\Delta (s+L) (p_{j-1})^{\frac{1}{\Delta+1}},
	\]
	where $p_0=p$ and $p_{2j}$, are, respectively, the local failure probability bounds for $\Pi_{0}$ and $\Pi_{j}$. Technically, for different values of $j$, we would have different values for $s$, since for each step we have a different runtime. However, the value of $s$ that we are using upper bounds all of them. We prove by induction that, for all $j\ge 0$,
	\[
	p_{j} \le (2\Delta (s+L))^2 p^{\frac{1}{(\Delta+1)^{j}}}.
	\]
	For the base case where $j=0$, we get that $p \le(2\Delta (s+L))^2 p$, which clearly holds. Let us assume that the claim holds for $j$, and let us prove it for $j+1$. We obtain the following, where the second inequality holds by the inductive hypothesis:
	\begin{align*}
	p_{j+1} &\le 2\Delta (s+L) (p_{j})^{\frac{1}{\Delta+1}} \le 2\Delta (s+L) \left(  (2\Delta (s+L))^2 p^{\frac{1}{(\Delta+1)^{j}}} \right)^{\frac{1}{\Delta+1}} \\
	&\le (2\Delta (s+L))^{1+\frac{2}{\Delta+1}} p^{\frac{1}{(\Delta+1)^{j+1}}} \le  (2\Delta (s+L))^2 p^{\frac{1}{(\Delta+1)^{j+1}}}. \qedhere
	\end{align*}
\end{proof}

We now handle \Cref{point:combine}, by showing that any randomized algorithm that runs in strictly less than $2T$ rounds must fail with large probability.
\begin{lemma}\label{lem:randomizedpnlb}
Let $\Pi_0, \Pi_1, \dots, \Pi_{T}$ be a sequence of problems satisfying the conditions of \Cref{thm:newlifting}.
	Let $\Delta$ be the degree of $\Pi_0$.
	Let $n$ be an integer, and assume that $2T+2 \le \log_\Delta n$.
	Any algorithm for $\Pi_0$ running in strictly less than $2T$ rounds on $\mathcal{F}_{n,\Delta}$ must fail with probability at least $1 / L^{\Delta^{16T}}$, even if a solution for $\Pi_{\mathrm{in}}$ is given as input.
\end{lemma}
\begin{proof}
	For any $T' < T$, by applying \Cref{lem:multiplesteps} on the subsequence containing the first $T'+1$ problems, we get that a white algorithm solving $\Pi_0$ in $2T' < 2T$ rounds with local failure probability at most $p$ implies an algorithm solving $\Pi_{T'}$ in $0$ rounds with local failure probability at most $(2\Delta (s+L))^2 p^{1/(\Delta+1)^{2T'}}$, where $s = (3 |\Sigma_{\mathrm{in}}|)^{2\Delta^{2T'+2}} \le (3 L)^{2\Delta^{2T'+2}}$. Then, since by assumption  $\Pi_{T'}$ is not $0$-round solvable in the deterministic PN model\footnote{Observe that, if $\Pi_T$ is not $0$-rounds solvable in the deterministic PN model, then also $\Pi_{T'}$ is not.}, by applying \Cref{lem:zerorounds} we get that
	\[
	(2\Delta (s+L))^2 p^{1/(\Delta+1)^{2T'}} \ge \frac{1}{L^{\Delta^2}},
	\]
	which implies the following:
	\begin{align*}
	p &\ge \left(\frac{1}{L^{\Delta^2} (2\Delta (s+L))^2 }\right)^{(\Delta+1)^{2T'}}
	\ge \left(\frac{1}{L^{\Delta^2} (2\Delta ((3L)^{2\Delta^{2T'+2}}+L))^2 }\right)^{(\Delta+1)^{2T'}}\\ 
	&\ge \left(\frac{1}{L^{\Delta^2} 4\Delta^2 (4L)^{4\Delta^{2T'+2}} }\right)^{(\Delta+1)^{2T'}} 
	\ge \left(\frac{1}{\Delta^4 L^{\Delta^{2T'+8}} }\right)^{(\Delta+1)^{2T'}}\\ 
	&\ge \left(\frac{1}{L^{\Delta^{2T'+8} + 4 \log \Delta} }\right)^{(\Delta+1)^{2T'}}
	\ge \left(\frac{1}{L^{\Delta^{2T'+11}} }\right)^{(\Delta+1)^{2T'}}\\ 
	&\ge \frac{1}{L^{\Delta^{2T'+11} \cdot (\Delta+1)^{2T'}} }
	\ge \frac{1}{L^{\Delta^{2T'+11} \cdot \Delta^{2T' +1}} }\\ 
	&\ge \frac{1}{L^{\Delta^{4T'+12}} }
	\ge \frac{1}{L^{\Delta^{16T'}} } 
	\ge \frac{1}{L^{\Delta^{16T}} }
	\qedhere
	\end{align*}
\end{proof}

\subsection{A lower bound for randomized LOCAL}

We now prove that a sequence  satisfying the conditions of \Cref{thm:newlifting} implies a lower bound for the LOCAL model on \emph{forests}. We will later show that such a lower bound implies a lower bound for \emph{trees} as well.
\begin{lemma}\label{lem:lb-local-forests}
Let $\Pi_0, \Pi_1, \dots, \Pi_{T}$ be a sequence of problems satisfying the conditions of \Cref{thm:newlifting}.
	Let $\Delta$ be the degree of $\Pi_0$, and let $n$ be an integer.
	Any randomized algorithm running on $\mathcal{F}_{n,\Delta}$ in strictly less than \[\textstyle\min\{2T, \frac{1}{8} (\log_\Delta \log n -  \log_\Delta \log L -3)\}\] rounds must fail with probability $> 1/n$, even if a solution for $\Pi_{\mathrm{in}}$ is given as input.
\end{lemma}
\begin{proof}
    Let $2t = 2\lfloor \min\{2T, \frac{1}{8} (\log_\Delta \log n -  \log_\Delta \log L -2)\} / 2 \rfloor$. Observe that $2t+2 \le \log_\Delta n$.
	We apply \Cref{lem:randomizedpnlb} on the subsequence $\Pi_0, \ldots, \Pi_t$. We obtain that any algorithm for $\Pi_0$ running in strictly less than $2t$ rounds on  $\mathcal{F}_{n,\Delta}$ must fail with probability at least $1 / L^{\Delta^{16t}}$.
	Observe that the following holds.
	\begin{align*}
	& 2t < \frac{1}{8} (\log_\Delta \log n -  \log_\Delta \log L ) \\
	{} \implies {}
	& 16t\log\Delta <  \log \log n - \log \log L \\
	{} \implies {}
	& 16t\log\Delta + \log \log L  < \log \log n \\
	{} \implies {}
	& \Delta^{16t} \log L < \log n \\
	{} \implies {}
	& L^{\Delta^{16t}} < n  \\
	{} \implies {}
	& \frac{1}{L^{\Delta^{16t}}} > 1/n.
	\end{align*}
	Hence, any algorithm for $\Pi_0$ running in strictly less than $2t$ rounds on  $\mathcal{F}_{n,\Delta}$ must fail with probability strictly larger than $1/n$.
\end{proof}

\Cref{lem:lb-local-forests} proves a lower bound for forests. We show that such a lower bound implies a lower bound for trees as well.
For this purpose, we exploit a lemma already proved in \cite{grunau-rozhon-brandt-2022-the-landscape-of-distributed}\footnote{In  \cite{grunau-rozhon-brandt-2022-the-landscape-of-distributed}, this lemma is stated for complexities in $o(\log^* n)$, since it was enough for the purposes of that paper. However it is easy to check that such an assumption is not used in the proof.}.
\begin{lemma}[Lemma 3.3 of \cite{grunau-rozhon-brandt-2022-the-landscape-of-distributed}, rephrased]\label{lem:forests-to-trees}
	Let $\Pi$ be a problem in the black-white formalism that has deterministic (resp.\ randomized) complexity $T(n)$ on trees. Then, the deterministic (resp.\ randomized) complexity of $\Pi$ on forests is at most $2T(n^2) +2$.
\end{lemma}
We are now ready to prove a lower bound for trees, that is, we prove the randomized lower bound of \Cref{thm:newlifting}.  Let $\mathcal{T}_{n,\Delta}$ be the class of trees of $n$ nodes of maximum degree at most $\Delta$.
\begin{lemma}\label{lem:lb-local-trees}
	Let $\Pi_0, \Pi_1, \dots, \Pi_T$ be a sequence of problems satisfying the conditions of \Cref{thm:newlifting}. 
	Let $\Delta$ be the degree of $\Pi_0$, and let $n$ be an integer.
	Any randomized algorithm running on $\mathcal{T}_{n,\Delta}$ in strictly less than \[\textstyle\min\{T-1, \frac{1}{16} (\log_\Delta \log n -  \log_\Delta \log L -5)\}\] rounds must fail with probability $> 1/n$.
\end{lemma}
\begin{proof}
	Let $\Delta$ be the degree of $\Pi$.
Suppose for a contradiction that there exists an algorithm for $\mathcal{T}_{n,\Delta}$ with runtime strictly less than \[\textstyle\frac{1}{2}( \min\{2T, \frac{1}{8} (\log_\Delta \log \sqrt{n} -  \log_\Delta \log L -3)\} -2 ).\] Then, by  \Cref{lem:forests-to-trees}, this implies an algorithm for $\mathcal{F}_{n,\Delta}$ with runtime strictly less than \[\textstyle\min\{2T, \frac{1}{8} (\log_\Delta \log n -  \log_\Delta \log L -3)\}\] rounds, which is a contradiction with \Cref{lem:lb-local-forests}.
Hence, any algorithm for $\mathcal{T}_{n,\Delta}$ requires at least
\begin{align*}
&\tfrac{1}{2}( \min\{2T, \tfrac{1}{8} (\log_\Delta \log \sqrt{n} -  \log_\Delta \log L -3)\} -2 )\\
={} &\min\{T, \tfrac{1}{16} (\log_\Delta \log \sqrt{n} -  \log_\Delta \log L -3)\} -1	\\
\ge{} &\min\{T, \tfrac{1}{16} (\log_\Delta \log n -  \log_\Delta \log L -4)\} -1\\	
\ge{} &\min\{T-1, \tfrac{1}{16} (\log_\Delta \log n -  \log_\Delta \log L -5)\}
\end{align*}
rounds.
\end{proof}

\subsection{A lower bound for deterministic LOCAL}

We now prove the deterministic lower bound of \Cref{thm:newlifting}.
\begin{lemma}\label{lem:lb-local-trees-det}
	Let $\Pi_0, \Pi_1, \dots, \Pi_T$ be a sequence of problems satisfying the conditions of \Cref{thm:newlifting}. 
	Let $\Delta$ be the degree of $\Pi_0$, and let $n$ be an integer. 
	Any deterministic algorithm running on $\mathcal{T}_{n,\Delta}$ requires at least \[\textstyle\min\{T-1, \frac{1}{16} (\log_\Delta \log n -  \log_\Delta \log L -624)\}/13\] rounds.
\end{lemma}
\begin{proof}
Suppose for a contradiction that there exist $n$ and $T$ satisfying the following:
\begin{itemize}[noitemsep]
	\item There exists a sequence of problems $\Pi_0, \Pi_1, \dots, \Pi_T$ satisfying the conditions of \Cref{thm:newlifting}.
	\item There exists a deterministic algorithm $\mathcal{A}$ running on $\mathcal{T}_{n,\Delta}$ in strictly less than \[\textstyle t = \min\{T-1, \frac{1}{16} (\log_\Delta n -  \log_\Delta \log L -624)\}/13\] rounds.
\end{itemize}
We show that, under such assumptions, we can construct an algorithm $\mathcal{A}'$ with runtime strictly less than $\min\{T-1, \frac{1}{16} (\log_\Delta n -  \log_\Delta \log L -624)\}$ on graphs in $\mathcal{T}_{2^n,\Delta}$. By renaming $n = \log N$, this implies an algorithm with runtime strictly less than $\min\{T-1, \frac{1}{16} (\log_\Delta \log N -  \log_\Delta \log L -624)\}$ on graphs in $\mathcal{T}_{N,\Delta}$. Since any deterministic algorithm trivially implies the existence of a randomized algorithm with the same complexity, this contradicts \Cref{lem:lb-local-trees}.

In order to construct $\mathcal{A}'$, we use a standard indistinguishability argument. In more detail, in \cite[Theorem 6]{chang-kopelowitz-pettie-2016-an-exponential-separation}, it is shown that, under some conditions, we can run an algorithm, designed to work on graphs of size $n$ in time $t$, on graphs of different size in the same runtime $t$, such that the algorithm still works correctly. The conditions are the following:
\begin{itemize}[noitemsep]
	\item The IDs in each $(t+1)$-radius neighborhood are unique and from $\{1,2,\dotsc, n\}$;
	\item Each $(t+1)$-radius neighborhood contains at most $n$ nodes.
\end{itemize}
As we will show later, the second condition is guaranteed by the assumed bound on the runtime of $\mathcal{A}$.
In order to guarantee the first condition, it is sufficient to compute a coloring and use it as a new ID assignment.
For this purpose, we use Linial's algorithm, which can convert a given $k$ coloring into a $5\Delta^2\log k$ coloring in $1$ deterministic round \cite[Corollary 4.1]{linial-1992-locality-in-distributed-graph-algorithms}. Observe that, at the beginning of the algorithm, we can use the original IDs of the nodes, which come from $\{1,\ldots,N\}$, as a coloring.
Hence, at first, algorithm $\mathcal{A}'$ runs $3$ rounds of Linial's algorithm on $G^{2t+2}$ (the graph obtained by putting an edge between all nodes that are at distance at most $2t+2$). This requires at most $6t+6 \le 12t$ rounds. Let $\bar{\Delta} = \Delta^{2t+2}$ be an upper bound of the degree of $G^{2t+2}$.
We obtain a coloring with a palette of size at most
\[
	50\bar{\Delta}^2(\log \log \log N + \log \bar{\Delta}) =
	50\Delta^{2(2t+2)}(\log \log n + \log \Delta^{2t+2}).
\]
Since $t \le \frac{1}{16} (\log_\Delta n -624) / 13 = \frac{1}{16\cdot 13} \log_\Delta n -3 \le \frac{1}{8} \log_\Delta n -3$, we obtain the following:
\begin{itemize}
	\item Each $(t+1)$-radius neighborhood contains at most $\Delta^{t+1} < n$ nodes;
	\item The number of colors is at most
	\begin{align*}
		& 50\Delta^{2(2t+2)}(\log \log n + \log \Delta^{2t+2})\\
		\le{}&  50\Delta^{2(2(\frac{1}{8} \log_\Delta n -3)+2)}(\log \log n + \log \Delta^{2(\frac{1}{8} \log_\Delta n -3)+2})\\
		\le{}&  \frac{50}{\Delta^8} \sqrt{n}(\log \log n + (1/4)\log n - 4\log \Delta)\\
		\le{}&  \frac{100}{\Delta^8} \sqrt{n}\log n
		< n.
	\end{align*}
\end{itemize} 

After computing the coloring, $\mathcal{A}'$ runs the original algorithm $\mathcal{A}$, and as argued, it works correctly. In total, $\mathcal{A}'$ requires strictly less than $12t+t=13t$ rounds. Hence the claim follows.
\end{proof}

\section*{Acknowledgements}
Funded by the European Union. Views and opinions expressed are however those of the author(s) only and do not necessarily reflect those of the European Union or the European Research Council. Neither the European Union nor the granting authority can be held responsible for them. This work is supported by ERC grant \href{https://doi.org/10.3030/101162747}{OLA-TOPSENS} (grant agreement number 101162747) under the Horizon Europe funding programme.

\bibliography{da}

\end{document}